\documentclass[fleqn,usenatbib]{mnras}

\usepackage[T1]{fontenc}
\usepackage[utf8]{inputenc}  

\usepackage{newtxtext,newtxmath}
\usepackage{textgreek} 
\usepackage[pagewise, switch]{lineno} 

\usepackage{graphicx}     
\usepackage{amsmath}      
\usepackage{caption}      
\usepackage{subcaption}   
\usepackage{rotating}
\usepackage{longtable}
\usepackage{booktabs}
\usepackage{multirow}
\usepackage[dvipsnames]{xcolor}  

\usepackage{bm}
\usepackage{fix-cm}  
\DeclareUnicodeCharacter{2009}{\,}
\usepackage{newunicodechar}
\newunicodechar{ }{\,}

\DeclareRobustCommand{\VAN}[3]{#2}
\let\VANthebibliography\thebibliography
\def\thebibliography{\DeclareRobustCommand{\VAN}[3]{##3}\VANthebibliography}

\title[GARCIA -- III. Comparative Study of DM Halos]{The GMRT archive atomic gas survey -- III. Comparative Study of Dark Matter Halos in Nearby Galaxies}

\author[Sarkar et al.]{
Sougata Sarkar$^{1,2}$\thanks{E-mail: sougatas@iisc.ac.in},
Prerana Biswas$^{3}$,
Veselina Kalinova$^{4,8}$,
Nirupam Roy$^{2,7}$,
Narendra Nath Patra$^{5}$ 
\newauthor Sushma Kurapati$^{6,9}$ \\
$^{1}$Joint Astronomy Programme, Indian Institute of Science, Bangalore 560012, India \\
$^{2}$Department of Physics, Indian Institute of Science, Bangalore 560012, India \\
$^{3}$Indian Institute of Astrophysics, Bangalore 560034, India \\
$^{4}$Max-Planck-Institut für Radioastronomie (MPIfR), Auf dem Hügel 69, 53121 Bonn, Germany \\
$^{5}$Department of Astronomy, Astrophysics and Space Engineering, Indian Institute of Technology Indore, Indore 453552, India \\
$^{6}$ASTRON – Netherlands Institute for Radio Astronomy, Oude
Hoogeveensedijk 4, 7991 PD Dwingeloo, The Netherlands \\
$^7$New Mexico Institute of Mining and Technology, Socorro, NM 87801, USA \\
$^8$ Institute of Astronomy and National Astronomical Observatory,
Bulgarian Academy of Sciences, 72 Tsarigradsko Chaussee Blvd., 1784 Sofia,
Bulgaria \\
$^9$ National Centre for Radio Astrophysics (NCRA–TIFR), Post Bag 3, Ganeshkhind, Pune 411007, India
}

\date{Accepted XXX. Received YYY; in original form ZZZ}

\pubyear{\the\year{}}

\begin{document}
\label{firstpage}
\pagerange{\pageref{firstpage}--\pageref{lastpage}}
\maketitle

\begin{abstract}
The distribution of dark matter in the inner regions of galaxies poses a key challenge for small-scale ΛCDM cosmology. While cold dark matter simulations predict cuspy inner density profiles, observations of low surface brightness (LSB) and dwarf galaxies often favour cored profiles, an issue known as the cusp-core problem. We investigate this problem by comparing four dark matter halo profiles: NFW (cuspy), Einasto (intermediate), Burkert (cored), and pseudo-isothermal (pISO) (cored) in a pilot sample of $11$ galaxies from the GMRT archive atomic gas survey (GARCIA). We have performed mass modelling using Markov Chain Monte Carlo (MCMC) techniques, utilising rotation curves derived from robust 3D Kinematic modelling. Baryonic contributions from stars derived using stellar kinematics based on $3.6,\mu\mathrm{m}$ or $r$-band photometry via Multi-Gaussian Expansion (MGE) combined with Jeans Anisotropic Model (JAM) and from gas, calculated directly from the gas surface density (H\,\textsc{i} + He) without assuming any predefined functional form, are included. Our mass modelling shows that all halo profiles provide statistically good fits, yielding consistent estimates of halo mass and stellar mass-to-light ratio. To validate our analysis, we examine the stellar-to-halo mass relation and find broad agreement with empirical models. Non-parametric density profiles derived from baryon-subtracted rotation curves show that NFW fits the inner regions best, while all profiles converge in the outskirts. Future studies with a larger sample from GARCIA will be helpful in refining this trend and addressing the cusp–core issue in greater depth. 
    
\end{abstract}




\section{Introduction}

It is widely acknowledged in the literature that a discrepancy exists between the predicted and observed rotation curves of galaxies. The observed rotation curves imply that the amount of mass inferred from stars and gas in galaxies cannot account for the total dynamical mass of galaxies \citep[e.g.][]{1978PhDT.......195B,1987PhDT.......199B,1986RSPTA.320..447V,1990A&ARv...2....1S,1991ApJ...378..496D}. \citet{Zwicky:1933gu} estimated the total mass of the Coma cluster using the virial theorem and found a discrepancy between the mass of luminous matter and the gravitational mass of the system. \citet{1978ApJ...225L.107R} demonstrated, using a sample of 10 spiral galaxies, that their rotation curves remain approximately flat out to large radii ($\sim 50\mathrm{kpc}$), further supporting the presence of additional mass at these extended distances. This discrepancy gave rise to the idea that galaxies are embedded within extended, massive dark matter halos \citep{1997ApJ...490..493N}. However, alternative interpretations are present that do not include extra matter, but
suggests a modification of gravity \citep[e.g.][]{1983ApJ...270..365M,1983ApJ...270..371M}. 
High-resolution $N$-body simulations of collisionless cold dark matter (CDM) have indicated that dark matter halos develop a density structure that rises steeply toward the centre (cuspy) in a hierarchically clustering universe \citep[e.g.][]{1991ApJ...378..496D}. Building on this, \citet{1996NFW,1997ApJ...490..493N} demonstrated that spherically averaged density profiles, now known as the Navarro Frenk White (NFW) profile, provide a good fit to halos over a wide mass range, from dwarf galaxies to galaxy clusters. Further refinement of simulations, incorporating improved resolution, suggested that halos may exhibit even steeper inner density slopes than predicted by the original NFW model, indicating more pronounced cusps \citep{1997astro.ph.11259M,1998Moore}. But observation \citep[e.g.][]{Evi_cusp_1989ApJ...347..760C,Evi_cusp_1994ApJ...427L...1F,Evi_cusp_2001ApJ...552L..23D,Evi_cusp_2002A&A...385..816D,Evi_cusp_2003RMxAC..17...17D,2015AJ....149..180O,2016MNRAS.462.3628R,10.1093/mnras/sty3223,2025A&A...699A.311M} from low surface brightness galaxies, dwarf galaxies (which are known to be dominated by dark matter) and less massive galaxies revealed that the dark matter density profile is flat in the central region, making a core. In the GHASP survey \citep{2008MNRAS.383..297S}, the mass distribution of 36 spiral and irregular galaxies was studied, revealing the presence of a constant density core, independent of the morphological type of the galaxies. Furthermore, studies of some spiral galaxies 
\citep{2001AJ....121.1952B, 2004MNRAS.351..903G, 2013A&A...557A.131M} also supported a constant 
density core over a cuspy density profile. This tension between CDM simulation and observation 
of galaxies' rotation curve, known as the Cusp-Core 
problem \citep{1994Natur.370..629M,1994ApJ...427L...1F}, indicates a modification of the 
$\Lambda$CDM model on small scales. Although there are some 
observations \citep{Swaters2001_LSB_dwarf_does_not_rull_out_cusp,Swaters2003_LSB_dwarf_does_not_rull_out_cusp,2004PhDT........22Hdoes_not_rull_out_cusp} on LSBs and dwarfs showed that most of 
their rotation curves fit consistently with cored as well as cuspy dark matter (DM) profiles, so we can not rule out the cuspy nature of DM Halo. Similarly, the THINGS survey \citep{Walter_2008}, which provides high quality H\,\textsc{i} rotation curves, found that massive, disk-dominated 
galaxies can be fitted equally well by cuspy and cored dark matter halo 
models \citep{2015AJ....149..180O}. However, the situation is further complicated by the significant scatter in the inner density slopes reported across different studies.

Three main types of solutions have been proposed for the cusp-core problem: two are simulation-based, involving the inclusion of baryonic effects in simulations, and the other considers the nature of dark matter itself. The third is observational, focusing on systematics or measurement uncertainties in rotation curve analysis. Various physical processes can substantially alter the inferred inner structure of dark matter halos. Several studies have shown that baryonic processes such as the acquisition of angular momentum during structure formation, central black hole activity \citep{2010A&A...522A..28L,2011MNRAS.413.1633L}, dynamical friction leading to energy transfer from baryons to dark matter \citep{2001ApJ...560..636E}, star formation and supernovae driven gas outflows \citep[e.g.][]{1996MNRAS.283L..72N,2006Natur.442..539M,2008Sci...319..174M,2010Natur.463..203G,2011AJ....142...24O,2012MNRAS.422.1231G} can modify the inner structure of dark matter halo, resulting in a flatter inner density profile than those predicted by dark matter only simulations.

Alternative dark matter models have also been proposed. For instance, self-interacting dark matter (SIDM) can reduce central densities in halos \citep[e.g.][]{2000PhRvL..84.3760S, Burkert_2000,2016PhRvL.116d1302K,2018PhR...730....1T,2019PhRvX...9c1020R,2024eas..conf.1858B,refId0,2025A&A...699A.311M}, and may produce detectable radiation through dark matter annihilation. Other scenarios include decaying dark matter \citep{Cen_2001}, scalar field dark matter \citep{2012MNRAS.422..282R}, and fuzzy dark matter \citep{2000PhRvL..85.1158H}, each offering different mechanisms for modifying the inner structure of halos.

Another possible explanation lies in observational limitations. The mass distribution of galaxies is typically inferred from their rotation curves and systematic errors in measuring these curves can lead to incorrect conclusions about the inner dark matter density profile \citep[e.g.][]{2000AJ....119.1579V,Swaters2001_LSB_dwarf_does_not_rull_out_cusp,Swaters2003_LSB_dwarf_does_not_rull_out_cusp,2004CuspyLSB,10.1093/mnras/stw3004,2019MNRAS.482..821O,2023MNRAS.521.1316R,2023MNRAS.522.3318D,2024arXiv240416247S}. For example, beam smearing a 
result of finite resolution in H\,\textsc{i} observations can smooth out velocity gradients, 
particularly in galaxies with small-scale structures or high inclination. This effect is particularly pronounced in the central regions of galaxies, where the velocity gradient is higher compared to the outskirts. In the case of 
$H\alpha$ rotation curves, inaccurate slit placement may cause the dynamical centre of the 
galaxy to be missed, resulting in an artificially shallow rotation curve. However, Hα rotation curves through IFU observations are not affected by this limitation. Additinally, 
unmodeled non-circular motions can further reduce the observed inner rotation velocities, 
complicating the interpretation of the underlying mass distribution. Many of these 
observational issues have been significantly mitigated with the advent of high-resolution 
H\,\textsc{i} interferometric data. These data sets extend over large radii and are well-suited for the assumption of circular motion due to the cold nature of H\,\textsc{i} gas. Earlier, 
H\,\textsc{i} rotation curves used for mass modelling were typically derived from 2D velocity 
maps by fitting the tilted-ring model \citep{1974ApJ...193..309R}, a method susceptible to beam 
smearing and projection effects. However, recent advances now allow for the direct use of full 
3D H\,\textsc{i} data cubes in tilted-ring modelling \citep{2015MNRAS.452.3139K,2015MNRAS.451.3021D}, which 
helps to overcome these systematic errors and produce accurate rotation curves shown in 
\citep{2015MNRAS.451.3021D,2017MNRAS.466.4159I,2021MNRAS.503.1753S,Deg_SpekkensWALLABY,2025A&A...699A.311M}. For instance, recent 
studies by \citet{2020MNRAS.491.4993K} have demonstrated that the method used to derive the rotation curve 2D versus 3D fitting can significantly impact the inferred inner density slope of dark matter halos. Their results show that while 2D methods favour cored profiles, such as the pseudo-isothermal (pISO) model, the more accurate 3D rotation curves are better fitted by the cuspy NFW model. 

The GMRT ARChIve Atomic gas survey (\href{https://physics.iisc.ac.in/~nroy/garcia_web/about.html}{GARCIA}) is designed to uniformly analyse H\,\textsc{i} interferometric spectral line data for a sample of 515 nearby galaxies from the GMRT archive, enabling a wide range of scientific investigations. As part of this project, the first paper in the series, GARCIA-I 
\citep{biswas2022}, presents data products from a pilot sample of 11 sources. The second paper, 
GARCIA-II \citep{biswas2023} focuses on 3D kinematic and mass modelling using only the NFW dark 
matter halo. In this paper, we extend that analysis by introducing a comparative framework across four DM halo profiles and incorporating a non-parametric reconstruction of the DM distribution. This allows us to test not only which profiles fit the observed rotation curves, but also which are physically consistent with the actual inner DM density profiles, a key question in the cusp-core debate.

Section~\ref{sec:sample} describes the sample and data used for mass modelling. Section~\ref{sec:dm+result} outlines the dark matter halo profiles, methodology and results, including parametric and non-parametric dark matter, followed by a comparison of scaled profiles. Section~\ref{sec:discuss&conclusion} discusses the implications for the $M_{{star}}{-}M_{200}$ relation and the impact of kinematic disturbances, and the comparison of parametric and non-parametric dark matter. Finally, we summarise in section~\ref{sec:summ}, highlighting key findings and prospects.

\section{Sample Selection and Analysis}
\label{sec:sample}
\begin{table*}
\centering
\small   
\caption{Observation sample properties and derived H\,\textsc{i} and stellar masses \citep{biswas2023}.
References used for methods of distance measurements are:
1 -- tip of the red giant branch (TRGB),
2 -- brightest stars,
3 -- cosmological distance,
4 -- sosies.}
\label{tab:sample_properties_masses}

\begin{tabular}{lcccccccc}
\hline
\hline
\textbf{Source} & \textbf{Morph.} & \textbf{$i_{\rm opt}$} &\textbf{$i_{\rm mom0}$} & \textbf{$i_{\rm kin}$} &
\textbf{Dist.} & \textbf{Method} &
\textbf{$M_{\mathrm{H\,I}}$} & \textbf{$M_{star}$} \\
\textbf{name} & \textbf{Type} & (deg) & (deg) & (deg) & (Mpc) & \textbf{used} &
($M_\odot$) & ($M_\odot$) \\
\hline
NGC0784  & SBd  & 81.1 & 82 & 82.22 & 5.45  & 1 &
$(4.22 \pm 0.06)\times 10^{8}$ &
$(8.0 \pm 2.3)\times 10^{8}$ \\
NGC1156  & IB   & 50.0 & 51 & 48.79 & 6.79  & 2 &
$(5.38 \pm 0.07)\times 10^{8}$ &
$(1.11 \pm 0.31)\times 10^{9}$ \\
NGC3027  & Sc   & 73.8 & 63 & 62.60 & 16.54 & 3 &
$(5.14 \pm 0.06)\times 10^{9}$ &
$(4.6 \pm 1.4)\times 10^{9}$ \\
NGC3359  & Sc   & 52.2 & 55 & 56.31 & 16.79 & 3 &
$(9.95 \pm 0.07)\times 10^{9}$ &
$(1.9 \pm 1.8)\times 10^{9}$ \\
NGC4068  & I    & 60.0 & 51 & 57.47 & 4.39  & 1 &
$(1.284 \pm 0.035)\times 10^{8}$ &
$(1.2 \pm 0.4)\times 10^{8}$ \\
NGC4861  & Sm   & 66.5 & 69 & 78.96 & 9.95  & 1 &
$(9.12 \pm 0.22)\times 10^{8}$ &
$(6 \pm 6)\times 10^{7}$ \\
NGC7292  & I    & 66.4 & 23 & 19.16 & 9.60  & 3 &
$(5.49 \pm 0.15)\times 10^{8}$ & $(1.2 \pm 0.5) \times 10^{9}$ \\
NGC7497  & Sc   & 82.8 & 73 & 80.53 & 19.82 & 3 &
$(5.05 \pm 0.06)\times 10^{9}$ &
$(1.32 \pm 0.29)\times 10^{10}$ \\
NGC7610  & SABc & 42.1 & 50 & 52.91 & 12.60 & 4 &
$(2.07 \pm 0.05)\times 10^{10}$ & $(1.90\pm0.26) \times 10^{10}$ \\
NGC7741  & SBc  & 61.6 & 38 & 50.09 & 46.96 & 3 &
$(1.536 \pm 0.031)\times 10^{9}$ &
$(6.2 \pm 0.9)\times 10^{9}$ \\
NGC7800  & IB   & 67.0 & 52 & 49.20 & 20.56 & 3 &
$(3.69 \pm 0.08)\times 10^{9}$ &
$(1.3 \pm 0.8)\times 10^{9}$ \\
\hline
\end{tabular}
\end{table*}

This study investigates the cusp-core problem \citep{1994ApJ...427L...1F,1994Natur.370..629M}, which highlights a key tension between the cuspy inner density profiles of dark matter halos predicted by $\Lambda$CDM simulations and the cored profiles observed in many galaxies. Accurate rotation curves are essential for probing this discrepancy. {H \,\textsc{i}} interferometric data play a critical role by providing both spatial and kinematic information needed to trace the mass distribution.

For this purpose, we utilise the H\,\textsc{i} rotation curves derived by \citet{biswas2023} for a pilot sample, employing 3D kinematic modelling. This sample comprises eleven bright (apparent magnitude, $m_b = 11.03$--$14.10$), low-redshift ($z = 0.0006$--$0.0012$) galaxies, observed at a velocity resolution of $\sim 6.6\ \mathrm{km\,s^{-1}}$. The sample spans a wide range of H\,\textsc{i} masses ($\sim 10^8$--$10^{10}\ M_{\odot}$) and includes both spiral and irregular galaxies, with a mix of barred and unbarred morphologies. The parameter distribution of the sources has been shown in \citet{biswas2022}; observational properties \citep[see][fig. 2]{biswas2022} and H\,\textsc{i} line flux, B-band magnitude, H\,\textsc{i} mass \citep[see][fig. 2]{biswas2022}, distance, inclination angle with the method for determining, H\,\textsc{i} and stellar masses from \citet{biswas2023} are mentioned in Table~\ref{tab:sample_properties_masses}. \citet{biswas2023} derived the rotation curves by applying 3D tilted-ring modelling to the H\,\textsc{i} data cubes using two pipelines: Fully Automated TiRiFiC (FAT) and \textsc{3D-Barolo}; both methods produced acceptable models. As summarised in \citet{biswas2023}, FAT performs more reliably across a wider inclination range, samples the rotation curve to larger radii by fitting more rings for the same radial spacing, and is less sensitive to initial parameter choices due to its automated \textsc{SoFiA}-based setup. FAT also captures structural features such as warps better than \textsc{3D-Barolo}, and it generally produces lower moment 1 residuals. The choice between the rotation curves derived from FAT \citep{2015MNRAS.452.3139K} and \textsc{3D-Barolo} \citep{2015MNRAS.451.3021D} was based on a comparison of the residual velocity fields (\href{https://physics.iisc.ac.in/~nroy/garcia_web/kinematics.html#fig3}{see figure 3 on GARCIA wedsite}). For 10 of the 11 galaxies, FAT yielded residuals typically below $20\ \mathrm{km\,s^{-1}}$ and lower than those obtained from \textsc{3D-Barolo}. However, for NGC4861, the rotation curve generated by FAT showed higher residuals. In this case, the BBarolo-derived curve was preferred, as it yielded a more reliable rotation profile with lower velocity residuals (\href{https://physics.iisc.ac.in/~nroy/garcia_web/kinematics.html#fig4}{see figure 4 on GARCIA website}). All mass modelling in the study was based on these finalised rotation curves from \citet{biswas2023}.

To decompose the mass distribution, we also need to include stellar mass surface density profiles. However, the stellar mass-to-light ratio ($M/L$) varies significantly across photometric bands, introducing uncertainty in estimating the stellar mass distribution. Young, massive stars dominate the light in the optical bands due to their high luminosity, despite contributing little to the total stellar mass \citep{2001ApJ...550..212B}. As a result, recent star formation can significantly boost optical luminosity without substantially affecting stellar mass, making optical $M/L$ ratios highly sensitive to the recent star formation history (SFH). In contrast, near-infrared (NIR) light is dominated by older, low-mass stars that evolve slowly and contribute most of the stellar mass, making NIR $M/L$ less affected by SFH. Several studies \citep{2001ApJ...550..212B,2003ApJS..149..289B,2009MNRAS.400.1181Z,2014AJ....148...77M,2014ApJ...788..144M,2016ApJ...832..198N,10.1093/mnras/sty3223,2025A&A...695L..23M} have shown that NIR data particularly at 3.6\,$\mu$m provides the most stable estimates for $M/L$.

\citet{biswas2023} calculated the stellar mass surface density profiles using Spitzer 3.6\,$\mu$m data, and supplemented these with SDSS $r$-band images for the two galaxies (NGC7610 and NGC7292) with the Multi Gaussian Expansion (MGE) method \citep{2002MNRAS.333..400C}, and the resulting MGE models were converted into stellar mass distributions by assuming a radially constant stellar mass-to-light ratio $M/L$, which was determined via  Markov Chain Monte Carlo (MCMC) together with the halo mass and concentration (see section~\ref{subsubsec: parametric dm}, see section~5.1 in \citet{biswas2023}).

 All mass modelling in the study relies on the finalised rotation curves and stellar mass surface density profiles from \citet{biswas2023}. Further details on the data reduction, 3D kinematic modelling, and rotation curve extraction can be found in \citet{biswas2022} and \citet{biswas2023}.

\section{Distribution of the dark matter: methodology and results}
\label{sec:dm+result}
\subsection{Dark Matter Halo Profiles}
\begin{figure*}
    \centering
    
    \begin{subfigure}{0.49\linewidth}
        \centering
        \includegraphics[width=\linewidth]{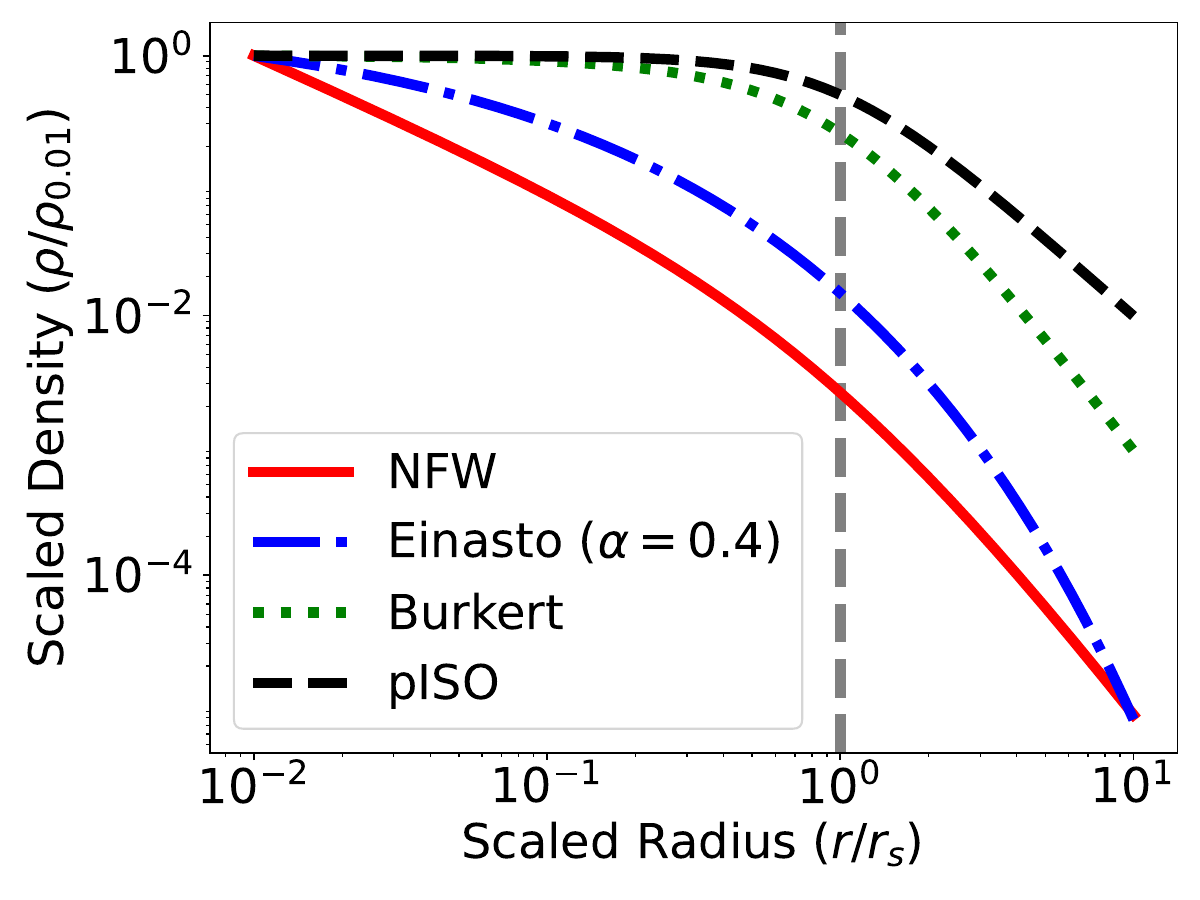}
        \caption{} 
        \label{fig:dm_prof_a}
    \end{subfigure}
    \hfill
    \begin{subfigure}{0.49\linewidth}
        \centering
        \includegraphics[width=\linewidth]{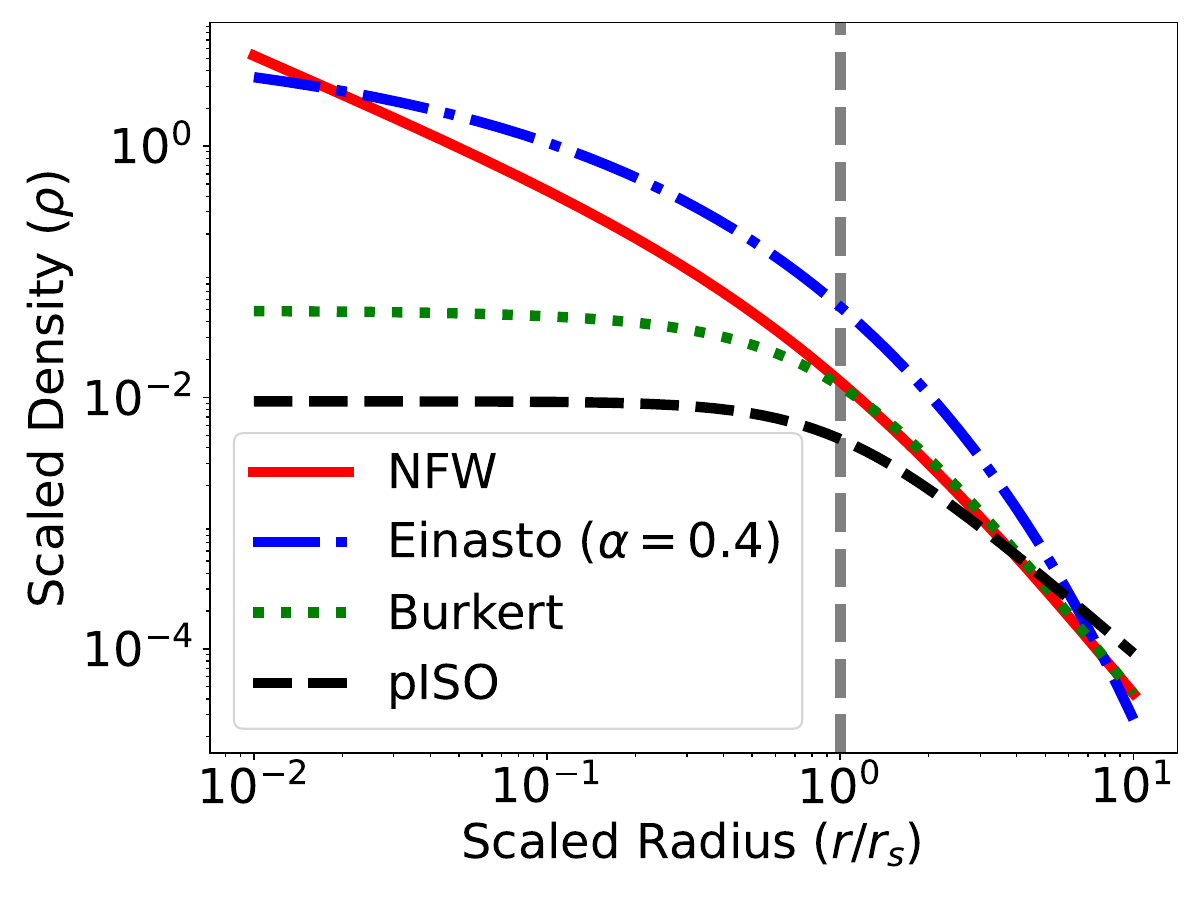}
        \caption{} 
        \label{fig:dm_prof_b}
    \end{subfigure}

    \caption{
    Normalised dark matter density profiles for different halo models.  
    Panel (a): profiles scaled by density, assuming $r_s = 1$ for NFW/Einasto and $r_c = 1$ for pISO/Burkert, with densities normalised at $r = 0.001$ to emphasise the inner slope differences.  
    Panel (b): same profiles scaled such that the enclosed mass at $r = 10$ kpc is matched.  
    }
    \label{fig:dm_prof_combo}
\end{figure*}

In this study, we have considered four dark matter halo profiles (Fig.\ref{fig:dm_prof_combo}) with 
varying halo shape, cuspy, cored and intermediate between cusp and core. A brief overview of these profiles and the theoretical framework for deriving mass and velocity profiles from the density distribution is provided below.

\par
\subsubsection{NFW Profile}
The Navarro-Frenk-White (NFW) arises from N-body simulation of cold dark matter ($\Lambda CDM$) \citet{1996NFW} and describes a cuspy nature of dark matter halo density distribution. It is 
widely used as a model profile for describing dark matter halos and is defined as,

\begin{equation}
    \rho(r) = \frac{\rho_s}{\left(\frac{r}{r_s}\right)\left(1+\frac{r}{r_s}\right)^2}
    \label{eq:rho_NFW}
\end{equation}

where, $\rho_s$ is the characteristic density and $r_s$ is the scale radius. The enclosed mass within radius $r$ is:
\begin{equation}
M_{\mathrm{NFW}}(r) = 4\pi \rho_s r_s^3 \left[ \ln\left(1 + \frac{r}{r_s}\right) - \frac{r/r_s}{1 + r/r_s} \right].
\label{eq:mass_NFW}
\end{equation}

The corresponding velocity profile can be written as:
\begin{align}
    V_{\text{NFW}}(r) &= V_{200} \times \sqrt{ \frac{M_{\mathrm{NFW}}(r)}{M_{200}} \cdot \frac{r_{200}}{r} }
\end{align}

Where $r_{200}$ is the radius within which the average density of a halo is $200$ times the critical density of the universe, the total mass enclosed within this radius is defined as $M_{200}$, and $V_{200}$ is the rotational velocity at $r_{200}$. Substituting the mass expression (equation~\ref{eq:mass_NFW}) and \( V_{200} = 10\, C\, r_s\, H_0 \) with \( H_0 = 72 \) km\,s$^{-1}$\,Mpc$^{-1}$

\begin{align}
    V_{\text{NFW}}(r)
    &= \frac{0.014\, M_{200}^{1/3}}{\sqrt{x}} \times \sqrt{ \frac{ \ln(1 + C x) - \frac{C x}{1 + C x} }{ \ln(1 + C) - \frac{C}{1 + C} } },
    \label{eq:vel_NFW}
\end{align}

where \( x = \frac{r}{20.24\, M_{200}^{1/3}} \) will be same for all halo profiles and \( C = \frac{R_{200}}{r_s} \) is the concentration parameter. We have used the same method to derive the velocity profile for all the dark matter halos.

\subsubsection{pISO Profile}
The density profile of the pseudo-isothermal model \citep{pISO_Begeman1987} is commonly used due to its simplicity and assumes a dark matter halo with a cored structure. The density profile of this model is as follows,
\begin{equation}
    \rho(r) = \frac{\rho_s}{1+\left(\frac{r}{r_c}\right)^2}
    \label{eq:rho_pISO}
\end{equation}
where $\rho_s$ is the characteristic density and $r_c$ is the core radius.

The enclosed mass within radius $r$ is:
\begin{equation}
M_{\mathrm{pISO}}(r) = 4\pi \rho_s r_c^3 \left[ \frac{r}{r_c} - \arctan\left( \frac{r}{r_c} \right) \right]
\end{equation}

The corresponding velocity profile is,
\begin{align}
    V_{\text{pISO}}(r) &= \frac{0.014 {M_{200}}^{\frac{1}{3}}}{\sqrt{x}} \times \sqrt{\frac{{Cx} - \arctan{(Cx)}}{C - \arctan{(C)}}}
    \label{eq:vel_pISO}
\end{align}

where, \( C = \frac{R_{200}}{r_c} \) is the concentration parameter.

\subsubsection{Burkert Profile} 
It is another empirical cored model used here and presented by \citet{burkert1995ApJ...447L..25B} which diverges more slowly than pISO at larger radii. The density profile of the Burkert model is defined as follows.

\begin{align}
    \rho(r) = \frac{\rho_s}{\left(1+\frac{r}{r_c}\right)\left\{1+\left(\frac{r}{r_c}\right)^2\right\}}
    \label{eq:rho_Burkert}
\end{align}
    

where $\rho_s$ is the characteristic density and $r_c$ is the core radius.

The enclosed mass is:


\begin{align}
M_{\mathrm{Burkert}}(r) = 2 \pi \rho_s r_c^3 \Bigg[ 
& \frac{1}{2} \ln \left(1 + \left( \frac{r}{r_c} \right)^2 \right) 
+ \ln \left(1 + \frac{r}{r_c} \right) \nonumber \\
& - \arctan \left( \frac{r}{r_c} \right) 
\Bigg]
\end{align}

The corresponding velocity profile is,
\begin{align}
    V_{\text{Burkert}}(r) & = \frac{0.014 {{M_{200}}^\frac{1}{3}}}{\sqrt{x}}\times \nonumber \\ 
    & \sqrt{\frac{\frac{1}{2}\ln{\big\{1+(Cx)^2\}}+\ln{(1+Cx)}-\arctan{(Cx)}}{\frac{1}{2}\ln{(1+C^2)}+\ln{(1+C)}-\arctan(C)}}
    \label{eq:vel_Burkert}
\end{align}
    
where, \( C = \frac{R_{200}}{r_c} \) is the concentration parameter.

\subsubsection{Einasto Profile}
\citet{NAVARRO_EIN2004MNRAS.349.1039N} explored the mass profiles of cold dark matter halos covering a mass range from dwarf galaxies to galaxy clusters, using numerical simulations and proposed a new profile that better reproduces the logarithmic slope $(- d \ln{\rho} / d \ln{r} )$ than the NFW profile. Later, it was realised that this profile was already introduced by \citet{Einasto_1965TrAlm...5...87E,Einasto_1969Afz.....5..137E} to describe the stellar distribution. The density profile of the Einasto model is described as,

\begin{align}
    \rho_{\text{Einasto}}(r) = {\rho_s} \exp{\left[-\frac{2}{\alpha_e}\left\{\left(\frac{r}{r_s}\right)^{\alpha_e}-1\right\}\right]}
    \label{eq:rho_Einasto}
\end{align}

Where $\rho_s$ is the characteristic density and $r_s$ is the scale radius, this profile introduces an additional parameter called the shape parameter ($\alpha_e$), which controls the curvature of the halo density profile.

The enclosed mass \citep{2005MNRAS.358.1325C,2005MNRAS.362...95M,2006AJ....132.2685M} is,
\begin{align}
M_{\mathrm{Einasto}}(r) =\ & 4\pi\, \rho_s\, r_s^3\, 
\exp\left( \frac{2}{\alpha_\epsilon} \right) 
\left( \frac{2}{\alpha_\epsilon} \right)^{-\frac{3}{\alpha_\epsilon}} 
\frac{1}{\alpha_\epsilon} \nonumber \\
& \times \Gamma\left( \frac{3}{\alpha_\epsilon}, 
\frac{2}{\alpha_\epsilon} \left( \frac{r}{r_s} \right)^{\alpha_\epsilon} \right)
\end{align}

The corresponding velocity profile is,
\begin{align}
    V_{\text{Einasto}}(r) & = \frac{0.014 {{M_{200}}^\frac{1}{3}}}{\sqrt{x}}\times \sqrt{\frac{\Gamma\left(\frac{3}{\alpha_e},\frac{2}{\alpha_e}{(Cx)}^{\alpha_e}\right)}{\Gamma{\left(\frac{3}{\alpha_e},\frac{2}{\alpha_e}{C}^{\alpha_e}\right)}}}
    \label{eq:vel_Einasto}
\end{align}
 \citet{2011AJ....142..109C} showed that the Einasto profile fits the rotation curves of spiral galaxies better than the NFW and pISO models, and this result holds for different choices of the initial mass function (IMF). The choice of IMF affects the stellar mass-to-light ratio ($M/L$), hence it affects how much of the rotation curve is explained by stars versus dark matter. Even though the $M/L$ can change with the IMF, the Einasto model still performs well. They also found that the shape parameter $\alpha$ can vary from about 0.1 to 10 across galaxies of different masses. Though many studies \citep[e.g.][]{2008MNRAS.387..536G,2014MNRAS.441.3359D,10.1093/mnras/sty2968,2016MNRAS.460.1214L,2017MNRAS.465L..84L} showed lower values of Einasto shape parameter. In our study, we test the density profile with different $\alpha$ values and found $\alpha=0.4$ produces dark matter halos that reflect an intermediate structure, with central slopes neither too steep nor too shallow.  We therefore fixed $\alpha = 0.4$ for all galaxies throughout the analysis to represent this transitional behaviour between cuspy and cored profiles and to ensure consistent comparison with other halo models.
 
Figure~\ref{fig:dm_prof_combo} shows different halo models' normalised dark matter density profiles. To allow direct comparison of their inner slopes and structural differences, we computed each profile assuming a scale radius (for NFW and Einasto) or a core radius (for cored models like pISO and Burkert) of unity, i.e., $r_s = 1$ or $r_c = 1$ and the densities were normalised by their respective values at a radius of $r = 0.001$ to highlight the differences in central concentration(see Figure~\ref{fig:dm_prof_a}). This normalisation ensures that all profiles intersect at the same reference point in the innermost region, allowing their inner slope behaviour to be compared more clearly on a log-log scale. As a result, the plot emphasises the intrinsic shape of each profile, particularly in the central regions where the cusp-core differences are most pronounced. Figure~\ref{fig:dm_prof_b} shows the same set of profiles scaled by fixing the enclosed mass within a radius of $r = 10\ \mathrm{kpc}$, illustrating how the relative behaviour of the models changes when matched in total mass rather than scale density. 


 \begin{figure*}  
    \centering
    \begin{subfigure}{0.48\textwidth}
        \centering
        \includegraphics[width=\textwidth]{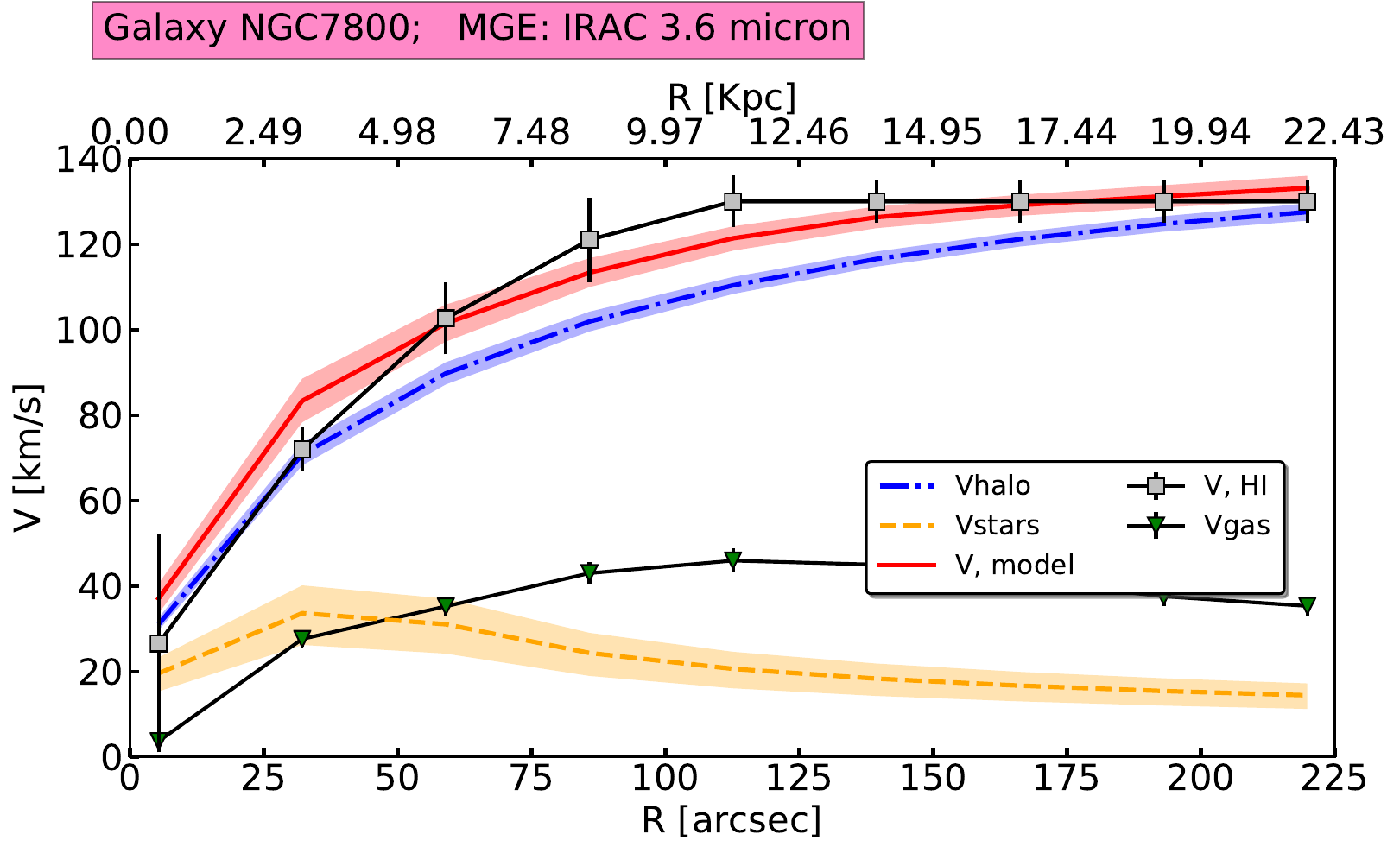}
        \subcaption*{\textbf{NFW profile}}
    \end{subfigure}
    \begin{subfigure}{0.48\textwidth}
        \centering
        \includegraphics[width=\textwidth]{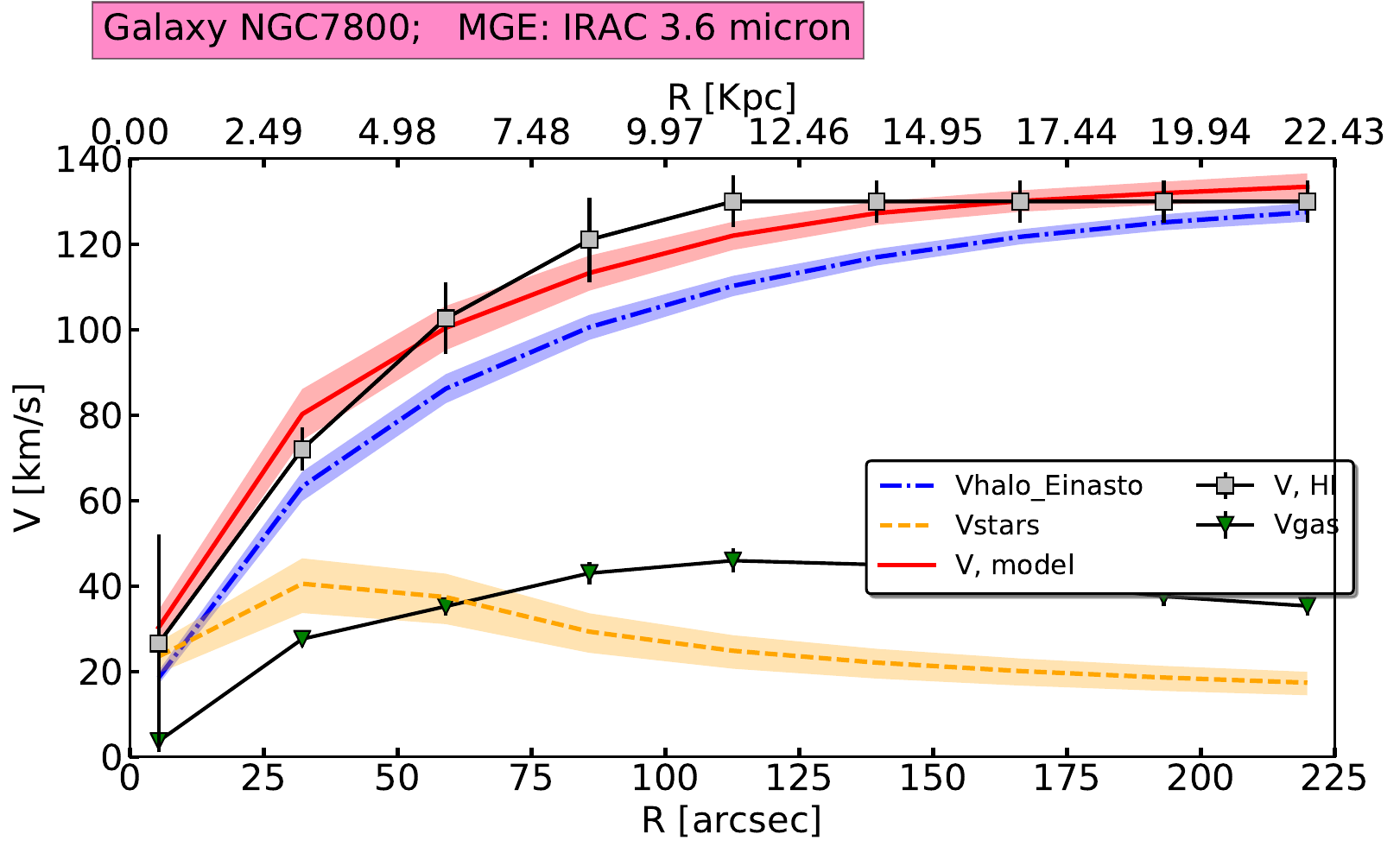}
        \subcaption*{\textbf{Einasto profile}}
    \end{subfigure}
    
    \begin{subfigure}{0.3\textwidth}
        \centering
        \includegraphics[width=\textwidth]{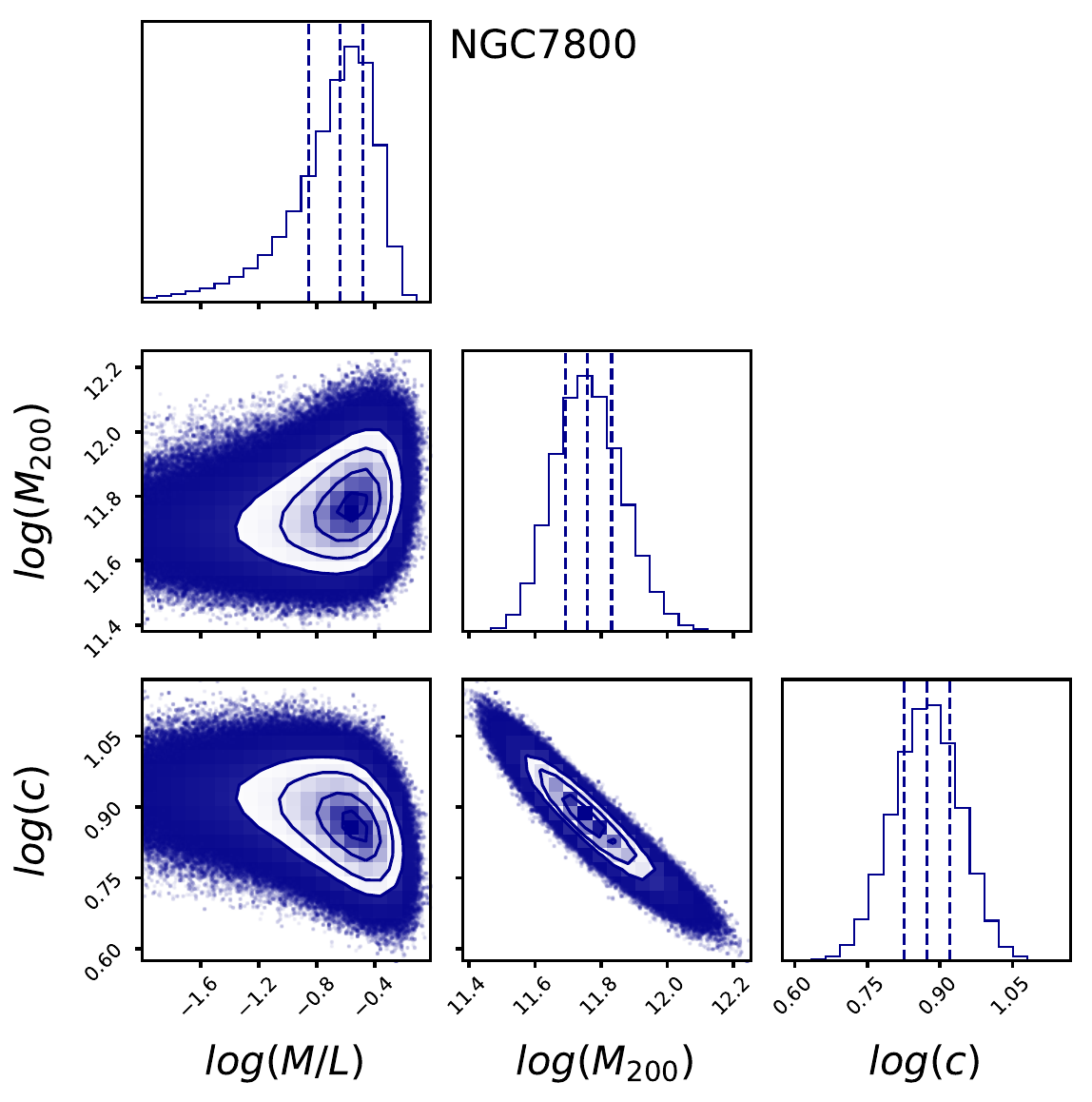}
    \end{subfigure}   
    \hspace*{0.2\textwidth}
    \begin{subfigure}{0.3\textwidth}
        \centering
        \includegraphics[width=\textwidth]{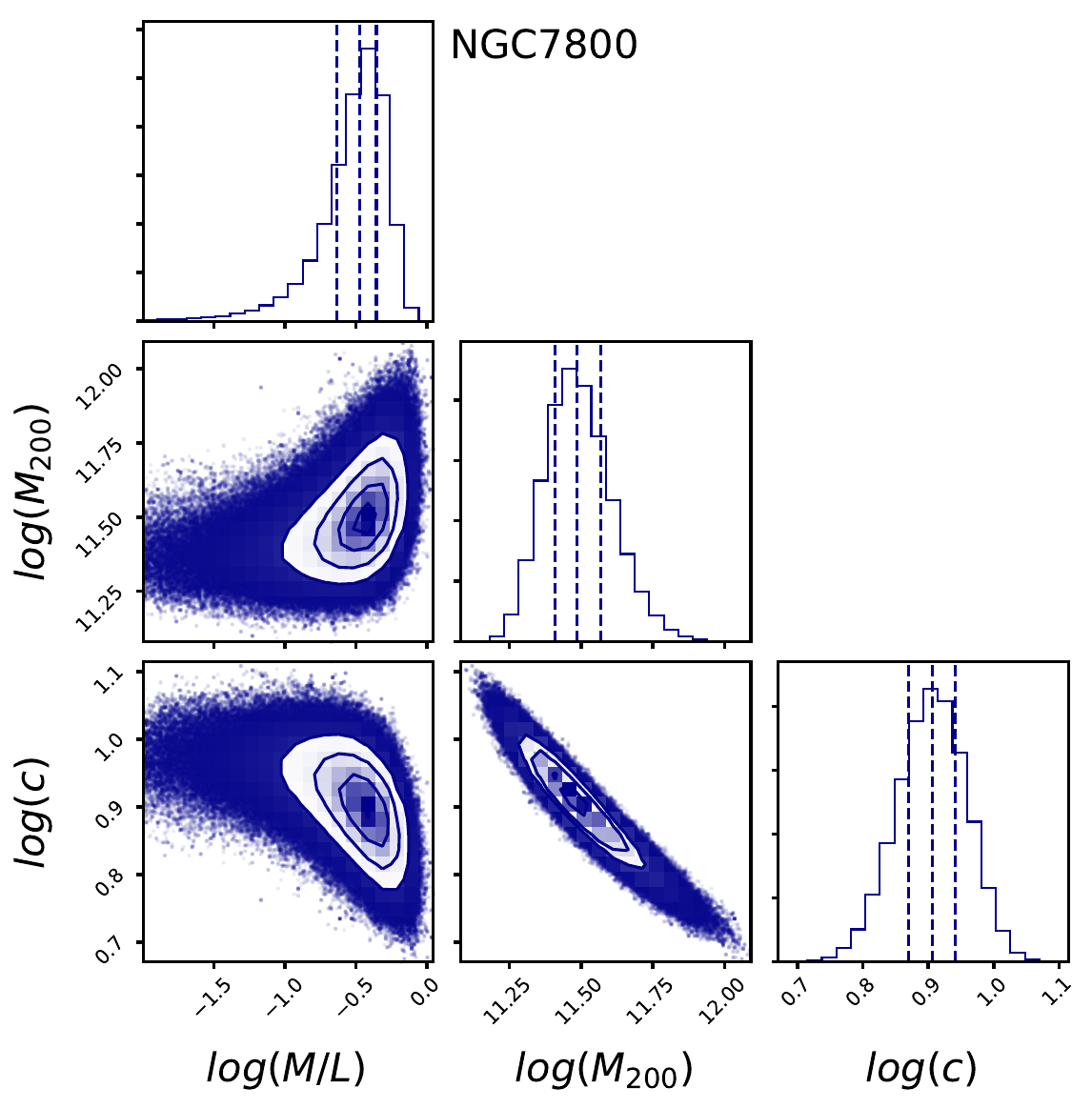}
    \end{subfigure}
 
    \centering
    \begin{subfigure}{0.48\textwidth}
        \centering
        \includegraphics[width=0.9\textwidth]{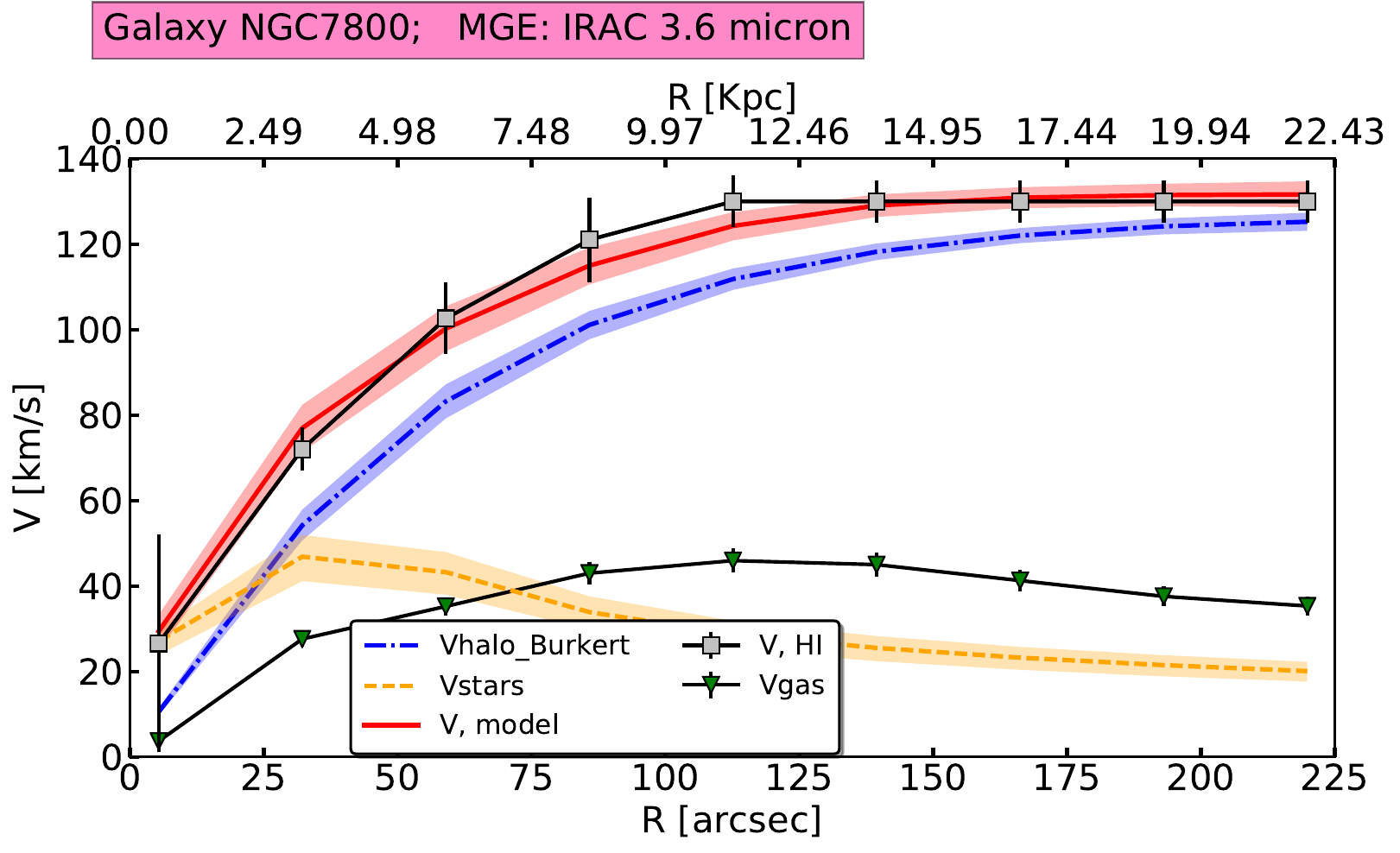}
        \subcaption*{\textbf{Burkert profile}}
    \end{subfigure}
    \begin{subfigure}{0.48\textwidth}
        \centering
        \includegraphics[width=0.9\textwidth]{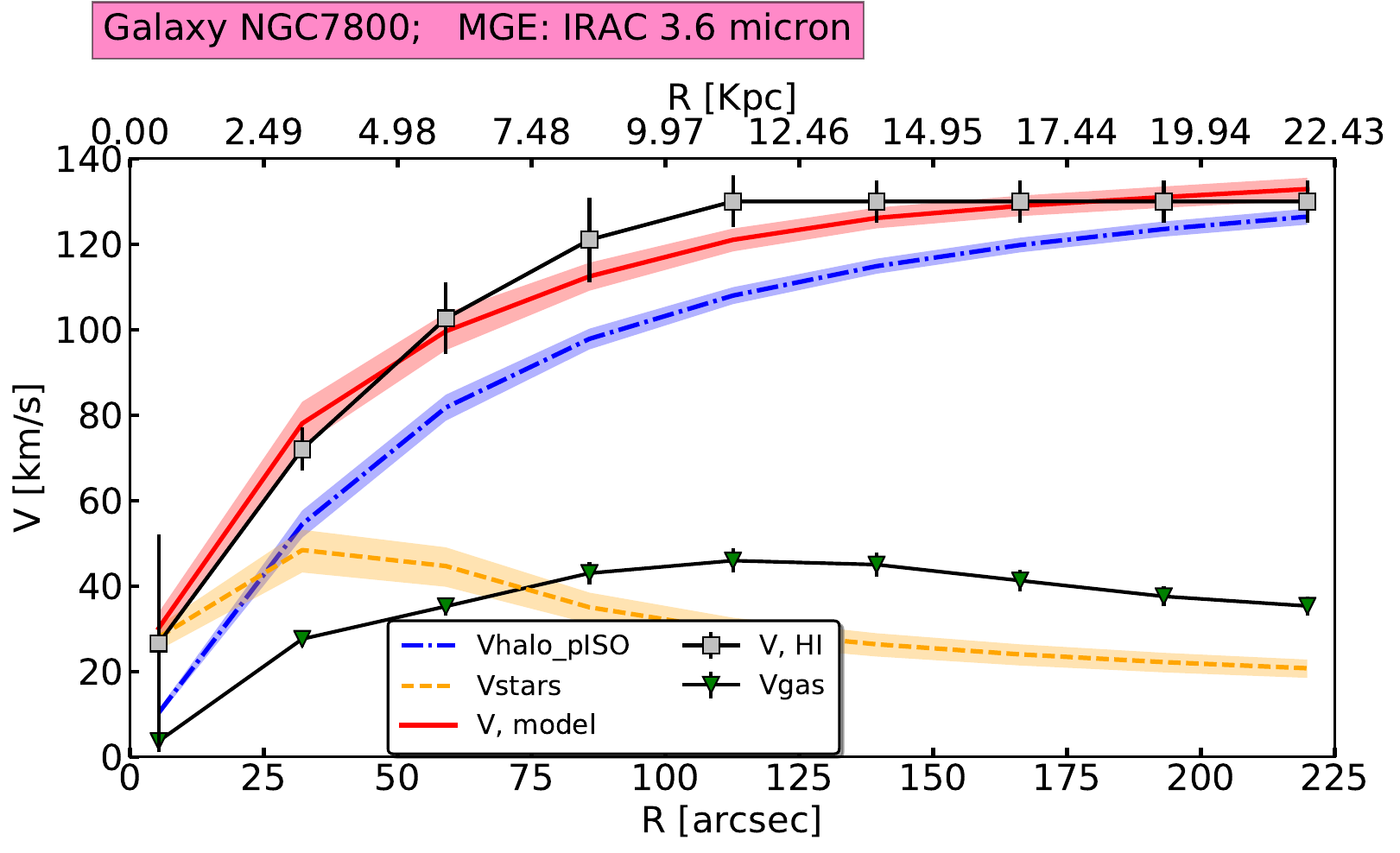}
        \subcaption*{\textbf{pISO profile}}
    \end{subfigure}
    
    \begin{subfigure}{0.3\textwidth}
        \centering
        \includegraphics[width=\textwidth]{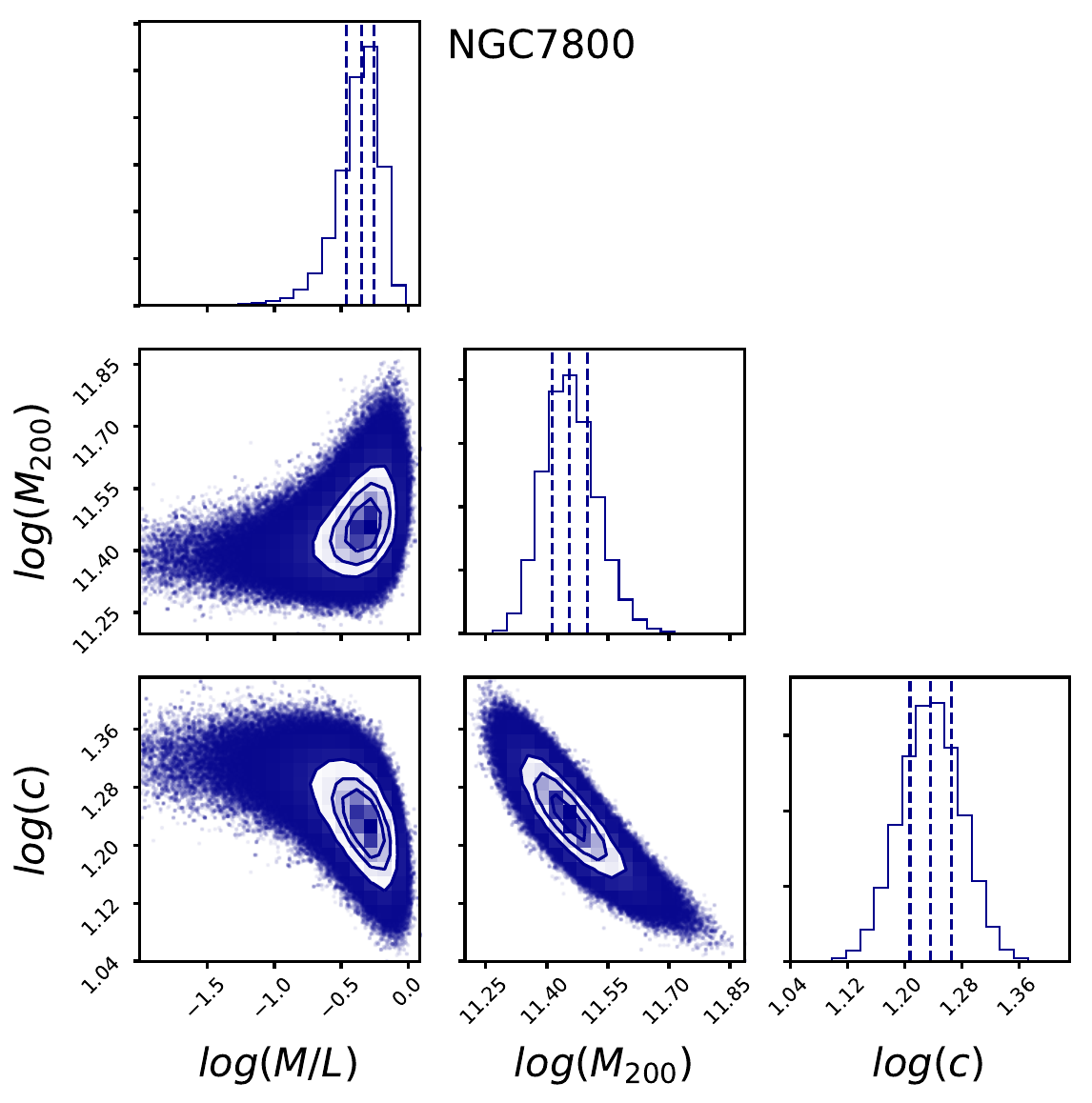}
    \end{subfigure}
     \hspace*{0.2\textwidth}
    \begin{subfigure}{0.3\textwidth}
        \centering
        \includegraphics[width=\textwidth]{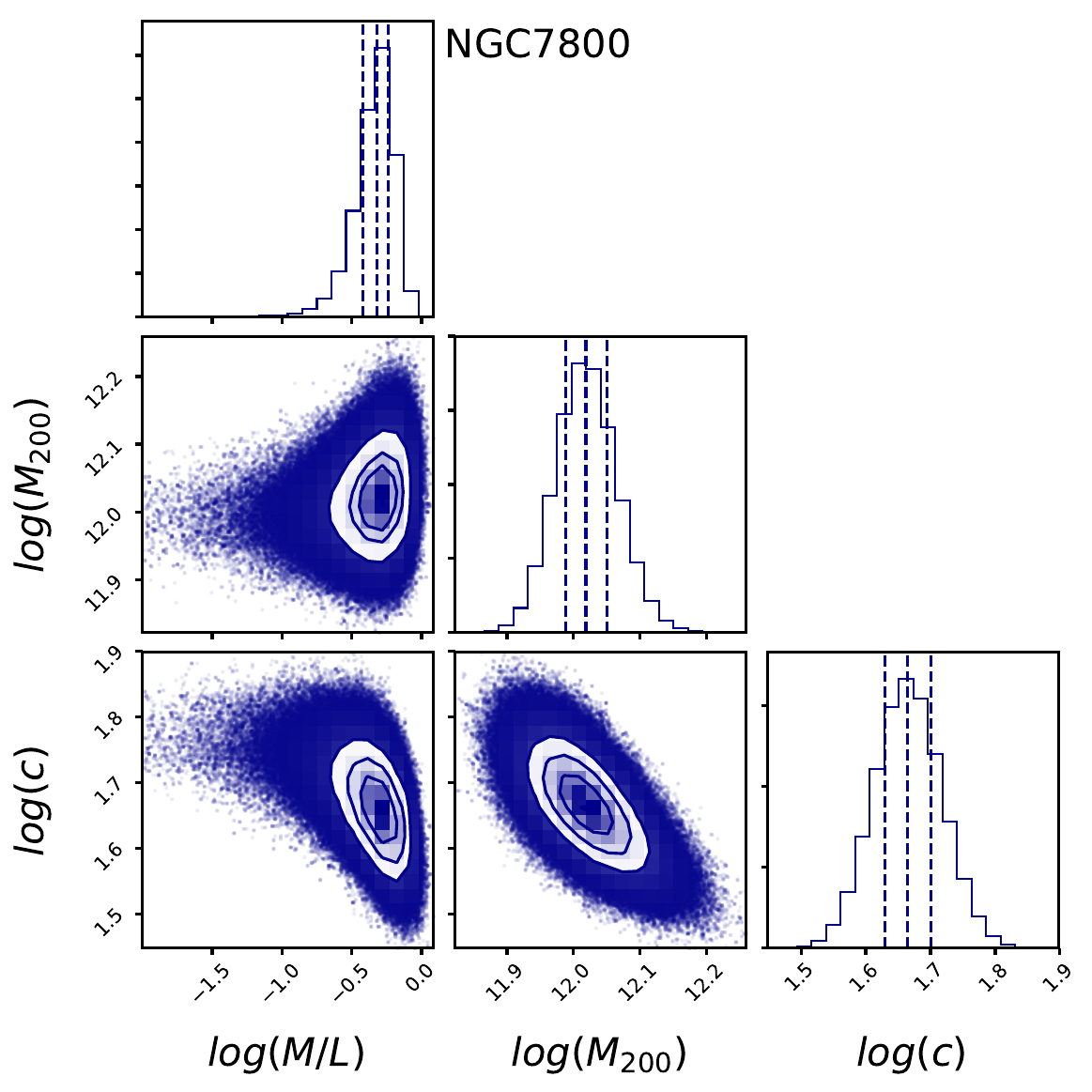}
    \end{subfigure}  
    \caption{ Modelled rotation curves (first and third rows) 
    and the corresponding posterior distributions of the fitted parameters ($M/L$, $M_{200}$, 
    and $C$) from MCMC-based mass modelling (second and fourth rows) for the galaxy NGC 7800. Results are 
    shown for four dark matter halo profiles: NFW, Einasto, Burkert, and pISO. $V_{\mathrm{HI}}$, $V_{\mathrm{model}}$, $V_{\mathrm{halo}}$, $V_{{star}}$, and $V_{\mathrm{gas}}$ represent the observed H\,\textsc{i} rotation velocity, the total modelled velocity (which varies depending on the assumed dark matter halo profile), the dark matter halo contribution, the stellar velocity component, and the gas velocity component, respectively.}
    \label{fig:MCMC RCs}
\end{figure*}

\subsection{Results}
\label{subsec:rasult}
 
\begin{figure*}  
 \centering
    \begin{subfigure}{0.95\textwidth}
        
        \includegraphics[width=\linewidth]{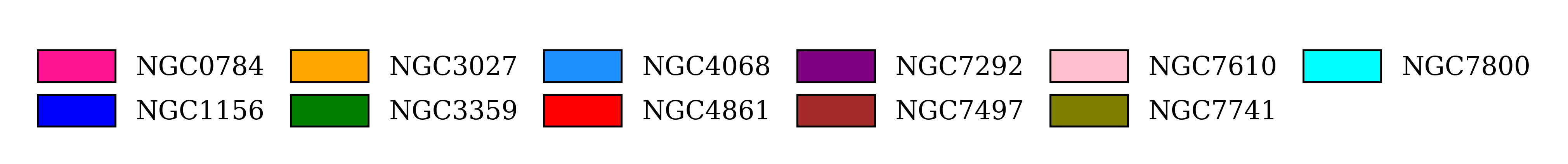}
    \end{subfigure}

    \centering
    \begin{subfigure}{0.412\textwidth}
        \centering
        \includegraphics[width=\textwidth]{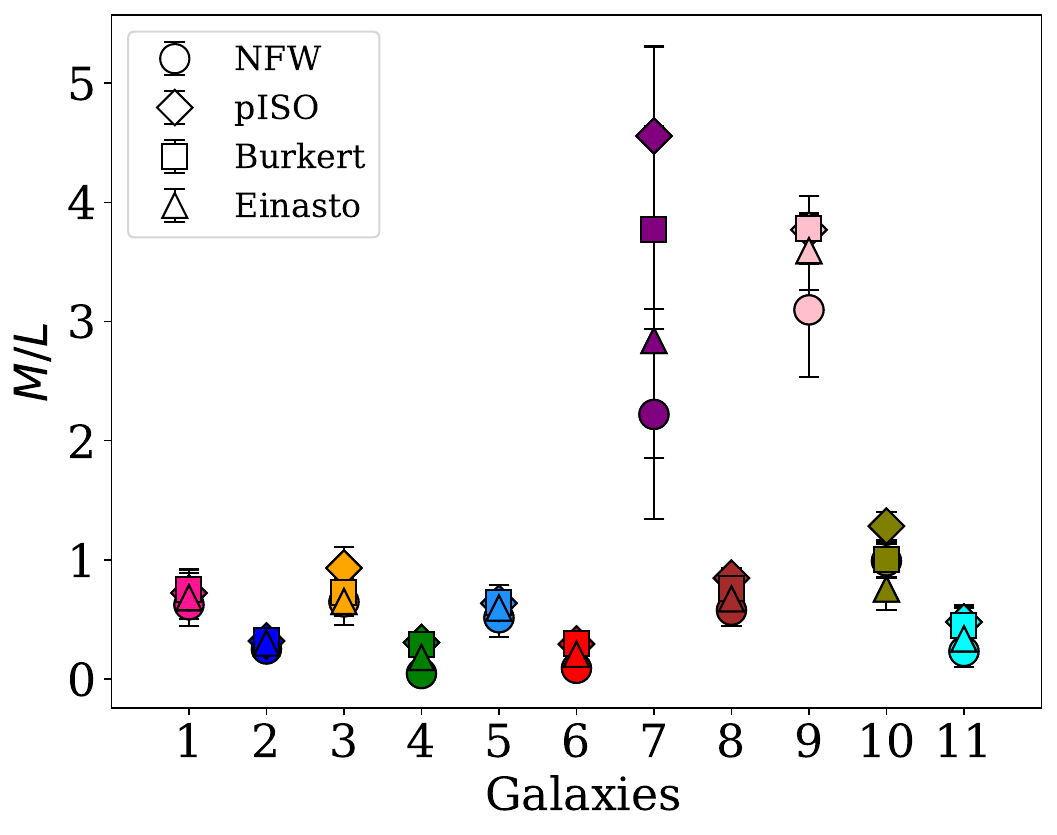}
        \subcaption{\textbf{$M/L$}}
        \label{fig:ml}
    \end{subfigure}
    \begin{subfigure}{0.44\textwidth}
        \centering
        \includegraphics[width=\textwidth]{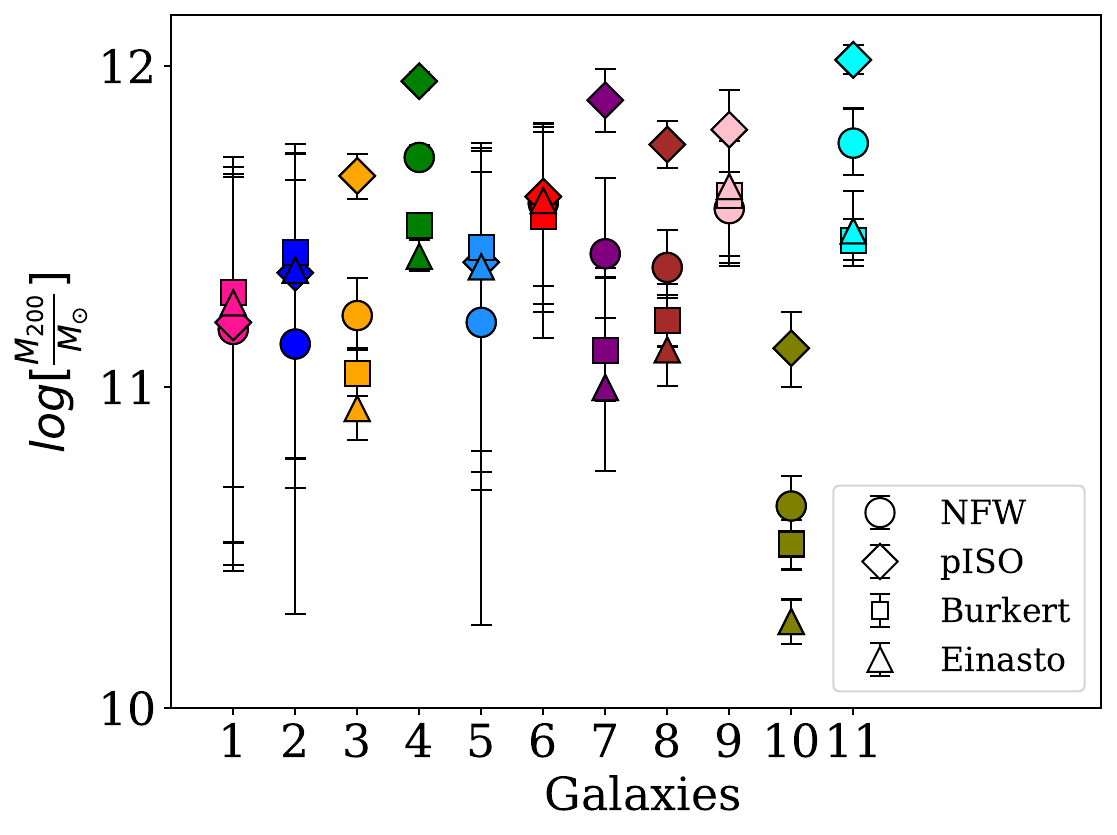}
        \subcaption{\textbf{$M_{200}$} }
        \label{fig:m200}
    \end{subfigure}

    \vspace{0.1cm} 

    \begin{subfigure}{0.43\textwidth}
        \centering
        \includegraphics[width=\textwidth]{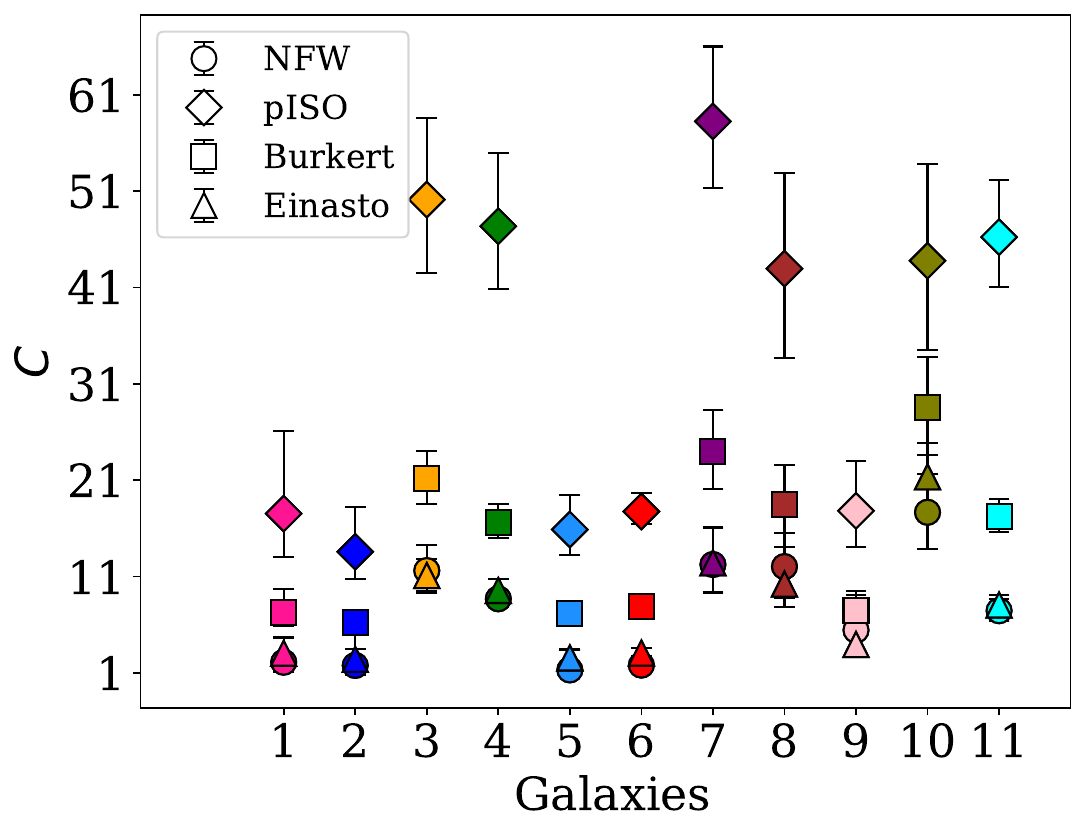}
        \subcaption{\textbf{$C$}}
        \label{fig:C}
    \end{subfigure}
     \hspace{0.3em}
    \begin{subfigure}{0.425\textwidth}
        \centering
        \includegraphics[width=\textwidth]{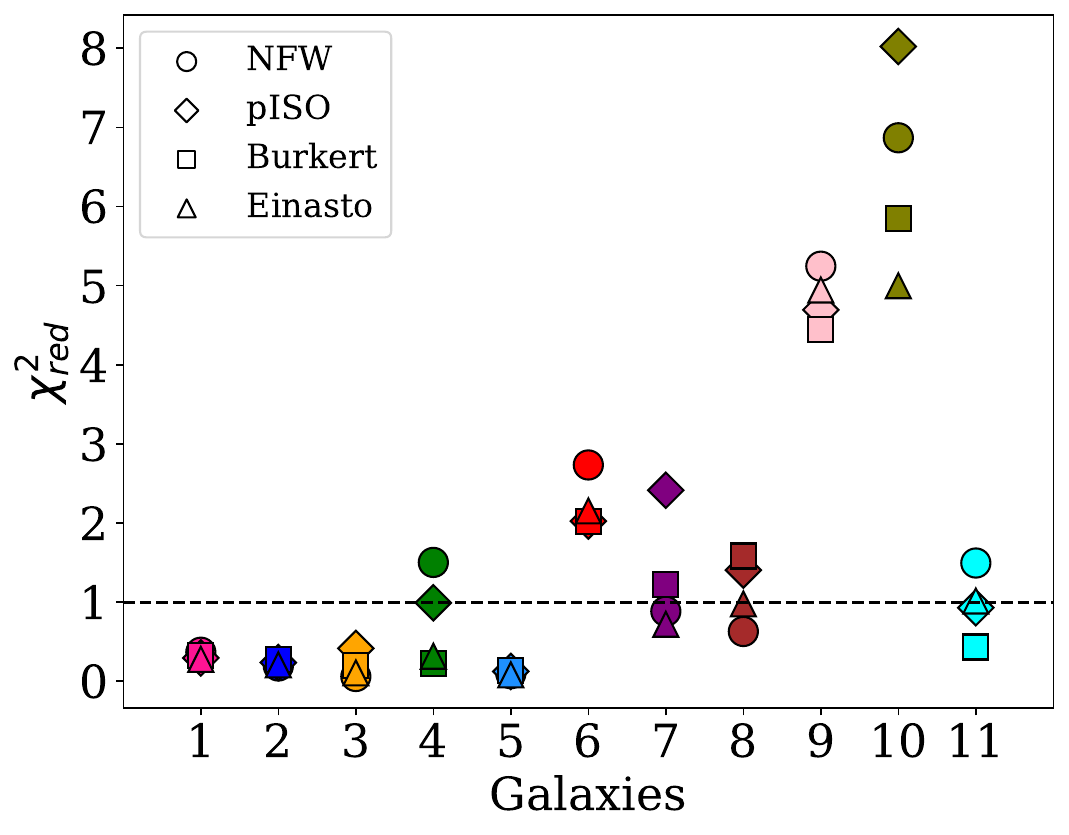}
        \subcaption{\textbf{$\chi^2_{\mathrm{red}}$}}
        \label{fig:chi}
    \end{subfigure}
    \caption{ Distributions of model parameters obtained 
    from the MCMC-based mass modelling. \ref{fig:ml}, \ref{fig:m200} and \ref{fig:C} show the mean values of 
    the posterior distributions for $M/L$, $M_{200}$, and $C$ , respectively. \ref{fig:chi} 
    displays the reduced chi-square ($\chi^2_{\mathrm{red}}$) distribution. Circle, diamond, square and triangle symbols represent the NFW, pISO, Burkert, and Einasto profiles, 
    respectively. Different colours correspond to different galaxies.}
    \label{fig:MCMC result}
\end{figure*}

Rotation curves offer a direct probe of the mass distribution in galaxies, capturing the combined gravitational influence of stars, gas, and dark matter. By analysing these curves, we separate the contributions from these components. We model the total rotation velocity by summing three main components in quadrature: stellar, gas, and dark matter halo. The total velocity follows:
\begin{equation}
{V^{\mathrm{tot}}_{\mathrm{mod}}}^2 = {V_{\mathrm{gas}}}^2 + {V_{\mathrm{star}}}^2 + {V_{\mathrm{halo}}}^2
\label{eq:mod_vel}
\end{equation}

Where ${V_{\mathrm{gas}}}$, ${V_{\mathrm{star}}}$, and ${V_{\mathrm{halo}}}$ represent the contributions from gas, stars, and the dark matter halo, respectively. Here, we are using observed rotational velocity as it closely traces the true circular velocity for cold H\,\textsc{i} gas, i.e. $V_{\mathrm{rot}} \simeq V_{\mathrm{circ}}$.

We adopt the stellar and gas velocity components from \citet{biswas2023}, who performed detailed 3D mass modelling using NFW halo for the GARCIA-I galaxy sample. To model the stellar luminosity distribution \citep[see section 5.1]{biswas2023}, they used the Multi-Gaussian Expansion (MGE) method \citep{2002MNRAS.333..400C}, which accurately reproduces complex light profiles in galaxies with multiple components, such as bulges and disks. The MGE method models the galaxy luminosity distribution by decomposing it into a sum of two-dimensional Gaussian functions and the projected surface brightness. Then, the stellar circular velocity contribution, $V_{star}$, was derived using the MGE components by the Jeans Anisotropic Model (JAM) model \citep{2008MNRAS.390...71C}, which solves the axisymmetric Jeans equations assuming constant anisotropy. JAM is especially effective in converting observed light distributions into stellar kinematic predictions when used with MGE, making it a widely adopted tool for stellar dynamical analysis.

For the gas component, \citet{biswas2023} did not assume a parametric form for the H\,{\sc i} distribution. Instead, they extracted the H\,\textsc{i} surface density profile directly from 3D kinematically modelled H\,\textsc{i} data cubes using a tilted-ring model \citep[see section 5.2 and Appendix A]{biswas2023}. This approach allows for a more realistic representation of the gas distribution and its contribution to the rotation curves. The derived H\,\textsc{i} further multiplied by a factor of 1.33 to include the contribution from helium and the molecular gas (H$_2$) component was not included, as uniform CO data are not available for the full pilot sample. However, previous studies on disk galaxies \citep[e.g.][]{2016AJ....151...94F,2022MNRAS.514.3329M} have shown that neglecting the molecular gas has only a minor impact on the inferred rotation curves and dark-matter halo parameters. The gas velocities were then calculated from these corrected surface densities using the standard relation (assuming axisymmetric razor-thin disk) between surface mass density and gravitational potential \citep[see][Eq.~2.188]{2008gady.book.....B}. At the mass range of our galaxies, the finite thickness and flaring of the H,\textsc{i} disc are expected to introduce only minor corrections to the derived circular velocities and halo parameters \citep[e.g.][]{2022MNRAS.514.3329M,2025A&A...699A.311M}.

All these components $V_{star}$, $V_{\mathrm{gas}}$, and along with the NFW dark matter halo model fitted by \citet{biswas2023} are adopted in our analysis.

The dark matter component was determined using these components $V_{star}$, $V_{\mathrm{gas}}$ in two different approaches. The first is a parametric method (section \ref{subsubsec: parametric dm}), where the dark matter halo is modelled using four profiles: NFW(adopted from \citet{biswas2023}), Einasto, Burkert and pISO. Each profile describes how the dark matter density changes with radius using a set of parameters. These parameters are optimised using a Markov Chain Monte Carlo (MCMC) framework to find the best fit to the observed rotation curve. The second is a non-parametric method (section \ref{subsubsec:non para dm}), no specific shape or equation is assumed for the dark matter profile. Instead, the dark matter contribution is directly calculated by subtracting the known contributions from stars and gas from the total observed rotational velocity.

\subsubsection{Parametric Dark Matter}
\label{subsubsec: parametric dm}

To model the dark matter halo profiles of our galaxy sample, we employ the MCMC-based circular velocity fitting framework developed by \citet{2017MNRAS.464.1903K}. This framework utilises the \texttt{emcee} package \citep{2013PASP..125..306F}, which implements the affine-invariant ensemble sampler for efficient Markov Chain Monte Carlo (MCMC) parameter estimation. The code has previously been adapted to fit the NFW dark matter profile in the works of \citet{tyulneva2021} and \citet{biswas2023}. 

In the present analysis, we employ an extended version of the MCMC-based fitting tool (Kalinova, Tyulneva et al., in preparation) that enables modelling with multiple halo profiles, specifically the pseudo-isothermal (pISO), Burkert, and Einasto models. The framework simultaneously fits the observed rotation curves by varying three free parameters: the halo mass $M_{200}$, the concentration parameter $C$, and the stellar mass-to-light ratio $M/L$. For computational simplicity, we assume a radially constant $M/L$. For the NFW and Einasto profiles, the concentration is defined as $C = r_{200}/r_s$, where $r_s$ is the scale radius, while for the cored profiles (pseudo-isothermal and Burkert) we use the analogous definition $C = r_{200}/r_c$, with $r_c$ representing the core radius. 

However, $M/L$ in real galaxies is not strictly constant; it can vary systematically with radius due to stellar population gradients in age, metallicity, and star formation history \citep[e.g.][]{2001ApJ...550..212B,2009MNRAS.400.1181Z}. Incorporating such radial variations in $M/L$ could yield more physically realistic mass models and tighter constraints on halo parameters.


In previous studies \citep[e.g.]{2017MNRAS.466.1648K, 2022MNRAS.514.3329M, 2023MNRAS.518.6340D,2025A&A...699A.311M}, various scaling relations, such as the Stellar-to-Halo Mass Relation (SHMR) \citep{2013MNRAS.428.3121M} and the $M_{200}$–$C$ relation \citep{2008MNRAS.391.1940M, 2014MNRAS.441.3359D} have been employed to uniform prior selection. These priors help constrain dark matter halo parameters but can influence the results toward expected trends. \citet{2020ApJS..247...31L} found that using priors based on scaling relations improves consistency with large-scale cosmological trends. However, these priors can degrade the quality of the rotation curve fits, indicating that they may bias the inferred parameter estimates. \citet{2022MNRAS.514.3329M} also found that flat priors on the concentration parameter \( C \) led to poor constraints in some galaxies, whereas Gaussian or lognormal priors based on the \( C-M_{200} \) relation significantly improved convergence and the quality of parameter inference. So, \citet{biswas2023} adopts two prior distributions, Uniform and Lognormal, independent of such scaling relations. Since lognormal priors can result in an unbounded exploration of parameter space, a broad uniform prior is used to constrain them.

We use the same uniform priors on the halo mass ($M_{200}$) and the stellar mass-to-light ratio ($M/L$) as adopted in \citet{biswas2023}, since these parameters do not depend on the choice of halo profile. For $M_{200}$, we assume a lognormal prior with a mean value of $10^{11.225}$ and a width of 0.5 dex, based on constraints from the DiskMass Survey (see Fig.~16 in \citealt{2014MNRAS.441.3359D}). The stellar mass-to-light ratio ($M/L$) is best constrained in the 3.6\,$\mu$m infrared band. Stellar population synthesis models \citep{2014ApJ...788..144M, 2014PASA...31...36S, 2014AJ....148...77M} suggest that $M/L$ values in this band typically lie between 0.4 and 0.6, depending on the initial mass function (IMF) and the underlying stellar population. However, bulges can exhibit systematically higher $M/L_{3.6}$ than discs, even in the near-infrared, in some cases by up to a factor of $\sim$2 \citep{Schombert_2022}. To check the impact of $M/L$ in our sample we have done mass modelling for one galaxies with different $M/L$ for disk and bulge for NFW profile mentioned in Appendix~\ref{B}. 
\citet{2016ApJ...827L..19L} used fixed values of $M/L_{3.6} = 0.5$ for discs and $M/L_{3.6} = 0.7$ for bulges to estimate stellar central density. Similarly, \citet{2020ApJS..247...31L} adopted lognormal priors centred at these same values, each with a standard deviation of $0.1$. Following \citet{biswas2023}, we use a lognormal prior on $M/L$, centred at 0.6 with a standard deviation of 0.2. To avoid unbounded sampling, we also impose a uniform prior over the range $[0.01,\,100]$. For the remaining two galaxies, where SDSS \textit{r}-band data is used instead of 3.6\,$\mu$m, we apply a uniform prior on $M/L$ in the range $[0.01,\,10]$. Unlike the 3.6\,$\mu$m band, where previous studies (e.g., \citealt{2014ApJ...788..144M}) have established a relatively narrow and stable range for $M/L$, the \textit{r}-band lacks such strong constraints. Therefore, a wider uniform prior allows more flexibility in exploring the parameter space.


Since the concentration parameter ($C$) depends on the specific form of the dark matter halo density profile, we use different prior ranges for different profiles to ensure accurate and unbiased modelling. For NFW and Einasto profiles, the concentration parameter tracks the assembly history; systems that collapse earlier form denser interiors because their inner structure reflects the higher cosmic critical density at the time of collapse where as for cored profile the concentration parameter is just phenomenological, which help to fit the galaxy kinamatics though they depend on baryonic physics such as feedback process, disc halo coupling. As shown by \citet{2014MNRAS.441.3359D}, the concentration values obtained from fitting the Einasto profile can vary by up to 15\% compared to the NFW profile. To account for this, we follow \citet{biswas2023} and widen the prior range for the concentration parameter $C$ when using the Einasto profile, by 15\% relative to that used for the NFW profile. We use a mean of 10 with a scale of 6.

For the cored profiles, the concentration parameter is \( C = \frac{r_{200}}{r_c} \), where \( r_{200} \) is the virial radius and \( r_c \) is the core radius. We used $(r_c)$ as a reference to guide the range of $(C)$ values explored by the MCMC fitting. We used \( r_c \) values from previous observational studies of galaxies with similar halo masses and morphologies: for the pISO profile, we refer to Table~6 in \citet{2013A&A...557A.131M}, and for the Burkert profile, to Table~4 in \citet{2017MNRAS.470.2410R}. Based on the ranges of \( r_c \) reported in these works, we adopted lognormal priors for \( C \) with a mean of~8 and a scale of~7.99 for the Burkert profile, and a mean of~15 and a scale of~14 for the pISO profile. To ensure flexibility in parameter exploration and to avoid over-constraining the fits, we used a broad uniform prior on \( C \) within the range \([0.01,\,1000]\) for all profiles, allowing walkers of MCMC to explore a larger parameter space.

\begin{table*}
\centering
\renewcommand{\arraystretch}{1.7}
\caption{Best-fit $M_{200}$ (in $10^{11} M_\odot$), $M/L$, and concentration parameter $C$ for the four DM halo profiles.}
\label{tab:MCMC-diffProf}
\resizebox{\textwidth}{!}{%
\begin{tabular}{l|ccc|ccc|ccc|ccc}
\toprule
\multirow{2}{*}{Galaxy} & \multicolumn{3}{c|}{NFW} & \multicolumn{3}{c|}{Einasto} & \multicolumn{3}{c|}{Burkert} & \multicolumn{3}{c}{pISO} \\
 & $M_{200}$ & $M/L$ & $C$ & $M_{200}$ & $M/L$ & $C$ & $M_{200}$ & $M/L$ & $C$ & $M_{200}$ & $M/L$ & $C$ \\
\midrule
NGC0784 &
$1.51^{+3.00}_{-1.23}$ & $0.62^{+0.18}_{-0.18}$ & $2.12^{+2.57}_{-1.02}$ &
$1.83^{+3.03}_{-1.50}$ & $0.68^{+0.17}_{-0.18}$ & $3.05^{+1.59}_{-0.79}$ &
$1.97^{+3.24}_{-1.71}$ & $0.75^{+0.16}_{-0.17}$ & $7.28^{+2.43}_{-1.37}$ &
$1.59^{+3.01}_{-1.10}$ & $0.72^{+0.17}_{-0.17}$ & $17.54^{+8.53}_{-4.47}$ \\

NGC1156 &
$1.36^{+3.07}_{-1.16}$ & $0.25^{+0.07}_{-0.07}$ & $1.79^{+3.29}_{-0.95}$ &
$2.31^{+3.06}_{-1.82}$ & $0.30^{+0.06}_{-0.06}$ & $2.42^{+1.08}_{-0.60}$ &
$2.62^{+3.09}_{-2.02}$ & $0.33^{+0.06}_{-0.06}$ & $6.20^{+1.18}_{-0.99}$ &
$2.27^{+3.07}_{-1.67}$ & $0.32^{+0.06}_{-0.06}$ & $13.58^{+4.67}_{-2.81}$ \\

NGC3027 &
$1.67^{+0.52}_{-0.35}$ & $0.65^{+0.19}_{-0.19}$ & $11.63^{+2.61}_{-2.36}$ &
$0.86^{+0.26}_{-0.17}$ & $0.65^{+0.20}_{-0.20}$ & $11.11^{+1.70}_{-1.61}$ &
$1.10^{+0.20}_{-0.16}$ & $0.72^{+0.19}_{-0.19}$ & $21.18^{+2.86}_{-2.63}$ &
$4.54^{+0.78}_{-0.68}$ & $0.93^{+0.18}_{-0.17}$ & $50.12^{+8.45}_{-7.63}$ \\

NGC3359 &
$5.19^{+0.47}_{-0.39}$ & $0.04^{+0.04}_{-0.02}$ & $8.70^{+0.76}_{-0.85}$ &
$2.55^{+0.32}_{-0.25}$ & $0.18^{+0.05}_{-0.05}$ & $9.56^{+1.16}_{-1.04}$ &
$3.18^{+0.28}_{-0.25}$ & $0.29^{+0.03}_{-0.03}$ & $16.60^{+1.95}_{-1.64}$ &
$8.96^{+0.60}_{-0.54}$ & $0.30^{+0.03}_{-0.03}$ & $47.36^{+7.57}_{-6.52}$ \\

NGC4068 &
$1.59^{+3.09}_{-1.41}$ & $0.51^{+0.16}_{-0.16}$ & $1.32^{+2.15}_{-0.68}$ &
$2.37^{+3.06}_{-1.89}$ & $0.59^{+0.16}_{-0.15}$ & $2.56^{+0.83}_{-0.51}$ &
$2.72^{+3.05}_{-2.09}$ & $0.64^{+0.15}_{-0.15}$ & $7.20^{+1.04}_{-0.95}$ &
$2.45^{+3.10}_{-1.90}$ & $0.63^{+0.15}_{-0.15}$ & $15.89^{+3.58}_{-2.64}$ \\

NGC4861 &
$3.73^{+2.71}_{-1.93}$ & $0.09^{+0.09}_{-0.06}$ & $1.82^{+0.67}_{-0.41}$ &
$3.80^{+2.83}_{-2.10}$ & $0.21^{+0.11}_{-0.10}$ & $3.07^{+0.51}_{-0.31}$ &
$3.40^{+2.84}_{-1.98}$ & $0.30^{+0.10}_{-0.10}$ & $7.91^{+0.58}_{-0.42}$ &
$3.92^{+2.66}_{-1.85}$ & $0.29^{+0.10}_{-0.10}$ & $17.76^{+1.96}_{-1.31}$ \\

NGC7292 &
$2.60^{+1.89}_{-0.96}$ & $2.22^{+0.88}_{-0.88}$ & $12.26^{+3.81}_{-2.95}$ &
$1.00^{+1.35}_{-0.45}$ & $2.84^{+0.97}_{-0.98}$ & $12.43^{+3.68}_{-2.97}$ &
$1.30^{+0.89}_{-0.39}$ & $3.77^{+0.87}_{-0.84}$ & $23.97^{+4.29}_{-3.86}$ &
$7.82^{+1.94}_{-1.59}$ & $4.56^{+0.75}_{-0.72}$ & $58.23^{+7.77}_{-6.88}$ \\

NGC7497 &
$2.36^{+0.73}_{-0.42}$ & $0.58^{+0.13}_{-0.13}$ & $12.05^{+3.49}_{-3.30}$ &
$1.30^{+0.59}_{-0.29}$ & $0.67^{+0.13}_{-0.13}$ & $10.23^{+2.53}_{-2.42}$ &
$1.61^{+0.48}_{-0.28}$ & $0.76^{+0.12}_{-0.11}$ & $18.44^{+4.18}_{-4.32}$ &
$5.69^{+1.06}_{-0.88}$ & $0.85^{+0.09}_{-0.08}$ & $42.96^{+9.95}_{-9.30}$ \\

NGC7610 &
$3.59^{+2.23}_{-1.21}$ & $3.10^{+0.43}_{-0.56}$ & $5.45^{+4.04}_{-2.02}$ &
$4.21^{+2.52}_{-1.77}$ & $3.59^{+0.31}_{-0.33}$ & $3.96^{+1.30}_{-0.84}$ &
$3.95^{+2.21}_{-1.38}$ & $3.78^{+0.28}_{-0.29}$ & $7.44^{+1.65}_{-1.19}$ &
$6.33^{+2.08}_{-1.65}$ & $3.77^{+0.28}_{-0.28}$ & $17.83^{+5.21}_{-3.78}$ \\

NGC7741 &
$0.43^{+0.10}_{-0.08}$ & $0.99^{+0.14}_{-0.14}$ & $17.68^{+3.93}_{-3.79}$ &
$0.19^{+0.03}_{-0.03}$ & $0.75^{+0.17}_{-0.17}$ & $21.35^{+3.51}_{-3.48}$ &
$0.32^{+0.06}_{-0.05}$ & $1.00^{+0.15}_{-0.15}$ & $28.56^{+5.19}_{-4.93}$ &
$1.32^{+0.39}_{-0.32}$ & $1.28^{+0.11}_{-0.11}$ & $43.78^{+10.00}_{-9.28}$ \\

NGC7800 &
$5.75^{+1.62}_{-1.17}$ & $0.23^{+0.15}_{-0.13}$ & $7.47^{+1.26}_{-1.10}$ &
$3.06^{+1.03}_{-0.68}$ & $0.33^{+0.16}_{-0.15}$ & $8.06^{+1.02}_{-0.95}$ &
$2.85^{+0.49}_{-0.37}$ & $0.45^{+0.15}_{-0.15}$ & $17.24^{+1.79}_{-1.63}$ &
$10.44^{+1.18}_{-1.01}$ & $0.48^{+0.14}_{-0.14}$ & $46.23^{+5.91}_{-5.18}$ \\
\bottomrule
\end{tabular}
}
\end{table*}

Figure~\ref{fig:MCMC RCs} presents the mass modelling results for NGC7800 using four different dark matter halo profiles. The corresponding posterior distributions of the fitted parameters are shown below each modelled rotation curve. All four profiles yield comparably good fits to the observed rotation curve. Mass modelling results for all the galaxies have been uploaded as supplementary material, and they will be uploaded to the GARCIA website later. Table~\ref{tab:MCMC-diffProf} and the four panels in Figure~\ref{fig:MCMC result} present the posterior distributions of the free parameters obtained from the MCMC-based mass modelling of 11 galaxies, modelled using four dark matter halo profiles: NFW (adopted from \citet{biswas2023}), pseudo-isothermal (pISO), Burkert, and Einasto. The parameters shown are the stellar mass-to-light ratio ($M/L$), halo mass ($M_{200}$), concentration parameter ($C$), and reduced chi-square ($\chi^2_{\mathrm{red}}$), with distinct markers for each halo model and consistent color coding for each galaxy.

Although the halo mass ($M_{200}$) and stellar mass-to-light ratio ($M/L$) can depend on the assumed halo profile due to differences in the inner mass distribution (e.g. \citealt{2003MNRAS.339..243J,2024MNRAS.528..693O}), our analysis shows that for the pilot GARCIA sample these parameters remain broadly consistent within their uncertainties across the four tested profiles. The inferred $M/L$ (Fig.~\ref{fig:ml}) values are generally consistent across all halo profiles, clustering around $0.5$–$1.0$, in line with expectations from $3.6\mu m$ stellar population models. Although we adopt a single $M/L$ for both bulge and disc components, this is not expected to have a significant impact on our results. At 3.6\,$\mu$m the stellar mass-to-light ratio is only weakly sensitive to stellar population variations \citep{2014ApJ...788..144M}, so radial gradients in $M/L$ are expected to be modest. Moreover, the statistical uncertainties on the $r$-band $M/L$ from our MCMC fits are larger as expected due to radial variation in $r$-band $M/L$ \citep{2019A&A...621A.120G}. A notable outlier in our sample is NGC7610, for which the best-fitting stellar mass-to-light ratio is higher than for the other galaxies. For this galaxy, we lack Spitzer 3.6\,$\mu$m data, so the stellar mass is constrained from optical $r$-band photometry. It is well established that optical stellar mass-to-light ratios, particularly in $g$ and $r$, are more weakly constrained than near-IR or 3.6\,$\mu$m estimates because they are more sensitive to dust attenuation and recent star formation, and therefore exhibit a larger intrinsic scatter at fixed luminosity or colour \citep[e.g.][]{2003MNRAS.341...33K}. The $M_{200}$ estimates (Fig.~\ref{fig:m200}) are broadly consistent across the different halo profiles. The per-galaxy total variations range from 0.21 to 0.58\,dex, with only a few systems showing moderate deviations. However, these differences remain within the same order of magnitude and are consistent with the expected profile-to-profile variations arising from differences in the inner mass distribution. This suggests that, given the quality of the current data, the total halo mass and global $M/L$ are reasonably well constrained, while the concentration parameter (Fig.~\ref{fig:C}) varies significantly across profiles, reflecting sensitivity to inner halo structure. Cored profiles (Burkert, pISO) show higher concentrations, whereas NFW and Einasto yield lower concentrations.

Table~\ref{tab:long_chisq} and Figure~\ref{fig:chi} present the reduced chi-square values for each dark matter profile and galaxy. Values close to 1 indicate a good fit, and we find that for many galaxies, multiple profiles provide statistically acceptable fits. In some cases, cuspy profiles like NFW perform better (e.g., NGC0784,  NGC7292), while in others, cored profiles like Burkert or pISO yield lower chi-square values (e.g. NGC3359, NGC7741). Out of the 11 galaxies, roughly half favour cuspy profiles and half favour cored ones, indicating no clear preference for one type over the other. Moreover, from visual inspection, NFW appears to reproduce the overall shape of the rotation curves well for several galaxies. This suggests that assuming a fixed parametric form may not capture the full diversity of halo structures.  More flexible models, such as the coreNFW profile, have been shown to successfully describe galaxies spanning several orders of magnitude in stellar mass \citep[e.g.][]{2017MNRAS.467.2019R,2025A&A...699A.311M}. To probe the underlying dark-matter distribution without bias from any specific profile shape, we also apply a non-parametric approach, as described in Section \ref{subsubsec:non para dm}.

\begin{table}
\caption{Reduced $\chi^2$ values for the four DM halo models.}
\label{tab:long_chisq}
\centering
\renewcommand{\arraystretch}{1.3}
\begin{tabular}{lcccc}
\hline
Galaxy & NFW & Einasto & Burkert & pISO \\
\hline
NGC0784 & 0.368 & 0.278 & 0.324 & 0.295 \\
NGC1156 & 0.193 & 0.207 & 0.274 & 0.237 \\
NGC3027 & 0.061 & 0.106 & 0.205 & 0.416 \\
NGC3359 & 1.503 & 0.317 & 0.221 & 0.993 \\
NGC4068 & 0.096 & 0.081 & 0.139 & 0.125 \\
NGC4861 & 2.734 & 2.153 & 2.017 & 2.025 \\
NGC7292 & 0.885 & 0.721 & 1.225 & 2.414 \\
NGC7497 & 0.630 & 0.978 & 1.583 & 1.406 \\
NGC7610 & 5.246 & 4.945 & 4.441 & 4.695 \\
NGC7741 & 6.869 & 4.999 & 5.853 & 8.022 \\
NGC7800 & 1.495 & 1.015 & 0.434 & 0.930 \\
\hline
\end{tabular}
\end{table}


\subsubsection{Non--parametric Dark Matter}
\label{subsubsec:non para dm}

Non-parametric modelling provides a model-independent estimate of the dark matter distribution by subtracting baryonic contributions from observed rotation curves \citet{2013A&A...557A.131M}. The dark matter rotational velocity is obtained from

\begin{equation}
    V_{\rm dm}^2 = V_{\rm obs}^2 - V_{\rm gas}^2 - V_{\rm star}^2,
    \label{eq:NP_vel}
\end{equation}

where \( V_{\rm obs} \) is the observed rotational velocity derived from tilted-ring modelling of the H\,\textsc{i} data cube, \( V_{\rm gas} \) is the gas contribution (including both H\,\textsc{i} and helium), and \( V_{\rm star} \) is the stellar contribution. To compute $V_{\mathrm{star}}$ in this approach, a value of $M/L$ is required. We therefore adopted the best-fit $M/L$ value obtained from the MCMC analysis (section~\ref{subsubsec: parametric dm}). Our non-parametric approach does not assume a specific dark matter halo profile; however, the stellar contribution \( V_{\mathrm{star}} \) is derived from MCMC modelling via a mass-to-light ratio, making the method not entirely model-independent. For comparison, we also repeated the non-parametric calculation using a fiducial constant $M/L$ value, as described in \ref{A1}. 

A direct subtraction using equation~\ref{eq:NP_vel} sometimes yields negative values for \( V_{\rm dm}^2 \), particularly in the inner regions of galaxies. This occurs when the modelled velocity \( V_{\rm mod}^2 \), inferred from MCMC, slightly exceeds the observed velocity \( V_{\rm obs}^2 \) in the mean, even though the fit remains within observational error bars. To overcome this issue and ensure physically meaningful (non-negative) dark matter velocities, we adopt the Monte Carlo method. At each radial point, we generate 10,000 random samples of \( V_{\rm obs} \), \( V_{\rm gas} \), and \( V_{\rm star} \), drawing from normal distributions centered on their respective mean values and with standard deviations computed as \( (\text{error}_{\rm up} + \text{error}_{\rm down}) / 2 \). For each sample, we compute \( V_{\rm dm} \) and discard any cases where the result is imaginary (i.e., where \( V_{\rm dm}^2 < 0 \)). The final dark matter velocity profile is then obtained from the median of the accepted \( V_{\rm dm} \) distribution at each radius.

Given that our sample includes galaxies with varying morphologies and rotational velocity ranges, we have unified their rotation curves through a scaling approach. A common method for scaling \citep{2006MNRAS.373.1117H,2011AJ....141..193O,2015AJ....149..180O} involves normalising the rotation curves at a characteristic point, typically situated between the rising and flat regions of the rotation curve. We have taken both the parametric and non-parametric curves for the circular velocity contribution of the dark halo component, and scaled them to a point where
$\frac{d \log V_{dm}}{d \log r} = 0.3$ \citep{2006MNRAS.373.1117H,2011AJ....141..193O,2015AJ....149..180O,2020MNRAS.491.4993K}. 
For seven galaxies, this point is naturally present within the observed data. However, for four
galaxies (NGC0784, NGC1156, NGC4068 and NGC4861), this value is not directly reached in the
available rotation curve data. In these cases, we have estimated the slope $\frac{d \log V_{dm}}
{d \log r}$ vs. r curve using the first-order polynomial fit to determine the corresponding point, and then extrapolate the rotational velocity at that radius.  

\begin{figure*}  
 \centering
    \begin{subfigure}{0.95\textwidth}
        
        \includegraphics[width=\linewidth]{Results_img/galaxy_colorbar_larger_font_and_title_gap.png}
    \end{subfigure}

    \centering
    \begin{subfigure}{0.49\textwidth}
        \centering
        \includegraphics[width=\textwidth]{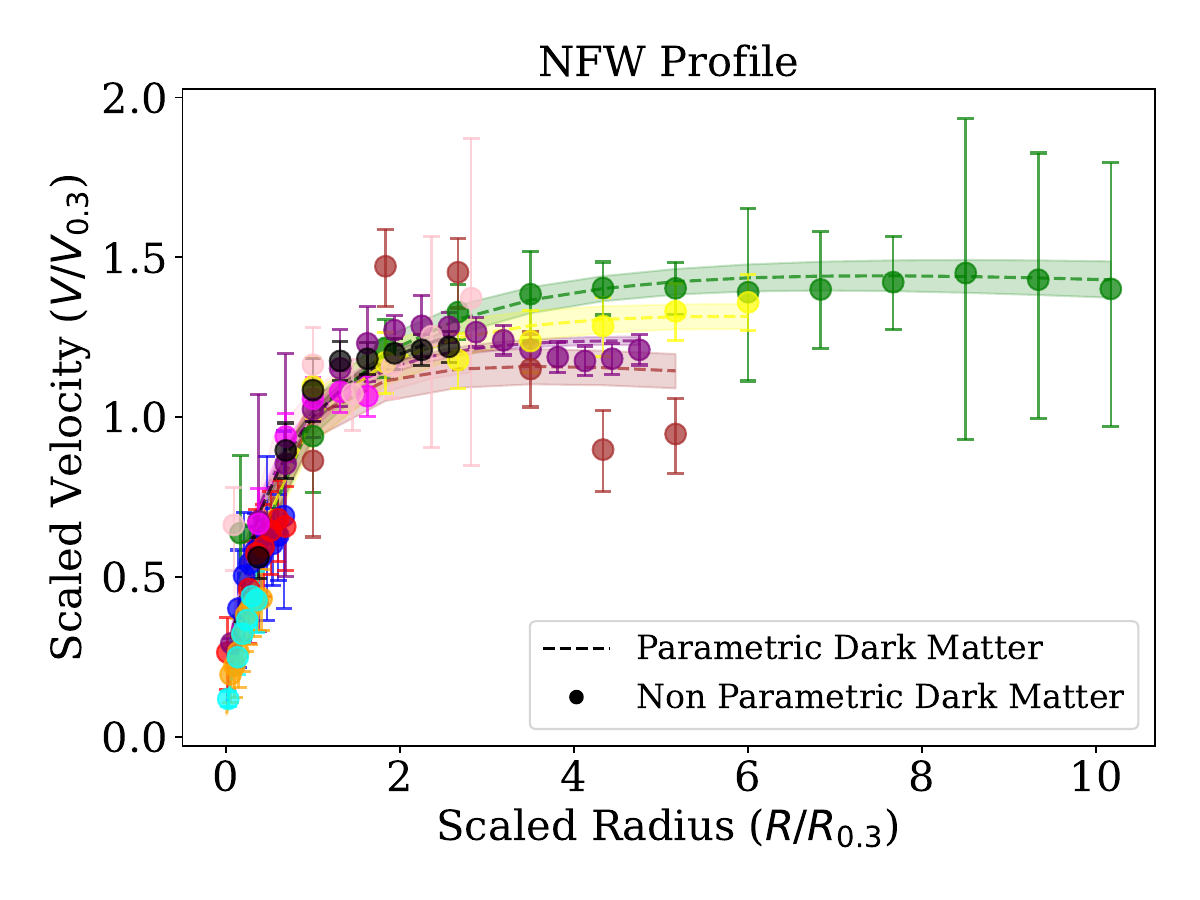}
    \end{subfigure}
    \begin{subfigure}{0.49\textwidth}
        \centering
        \includegraphics[width=\textwidth]{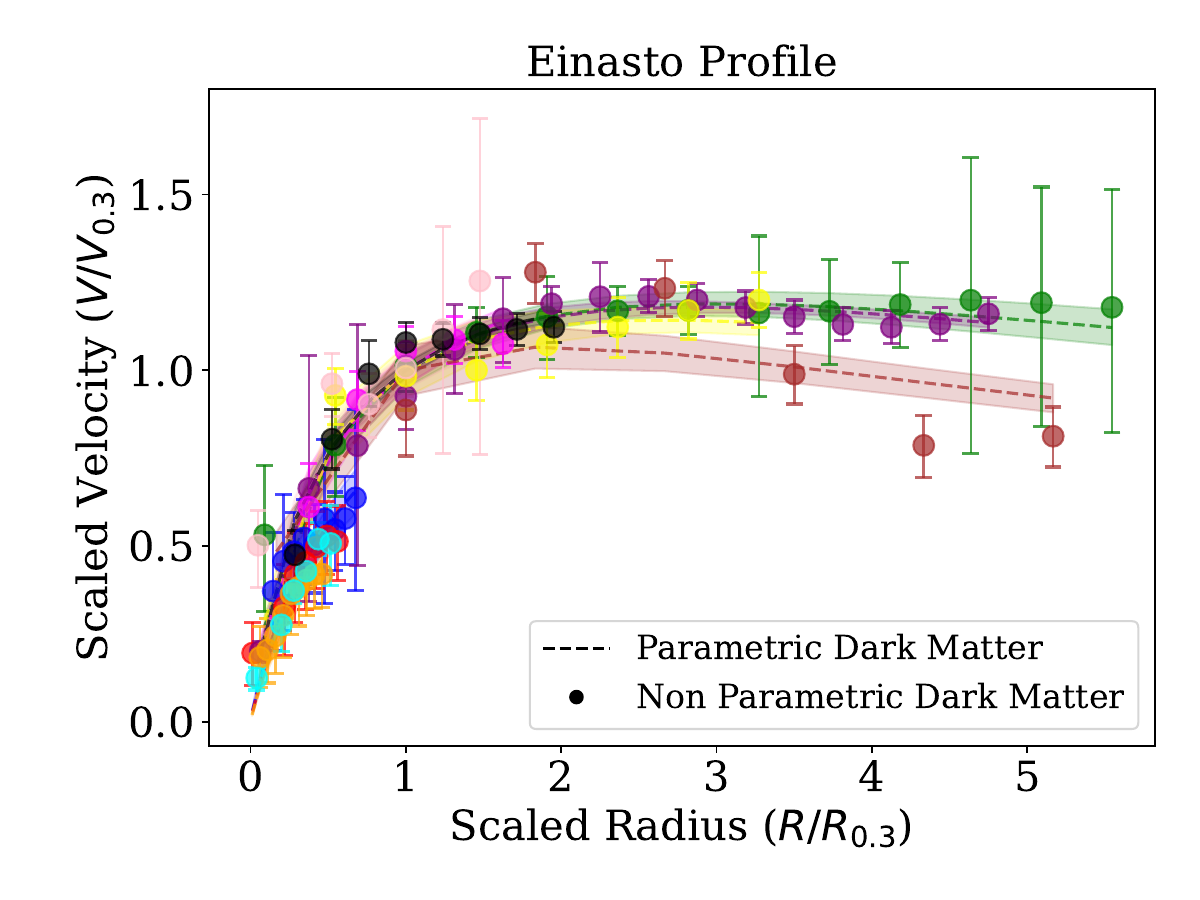}
    \end{subfigure}

    \vspace{-0.05cm} 

    \begin{subfigure}{0.49\textwidth}
        \centering
        \includegraphics[width=\textwidth]{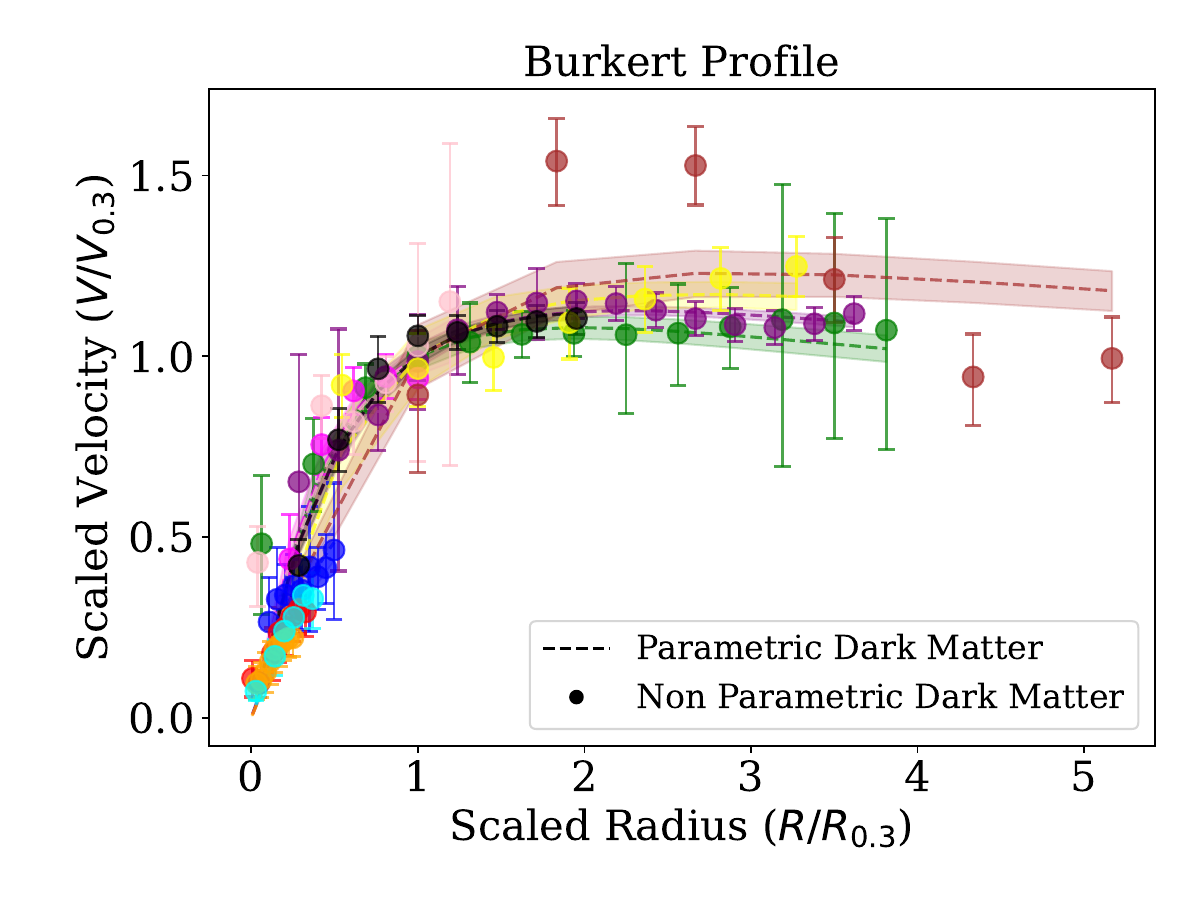}
    \end{subfigure}
    \begin{subfigure}{0.49\textwidth}
        \centering
        \includegraphics[width=\textwidth]{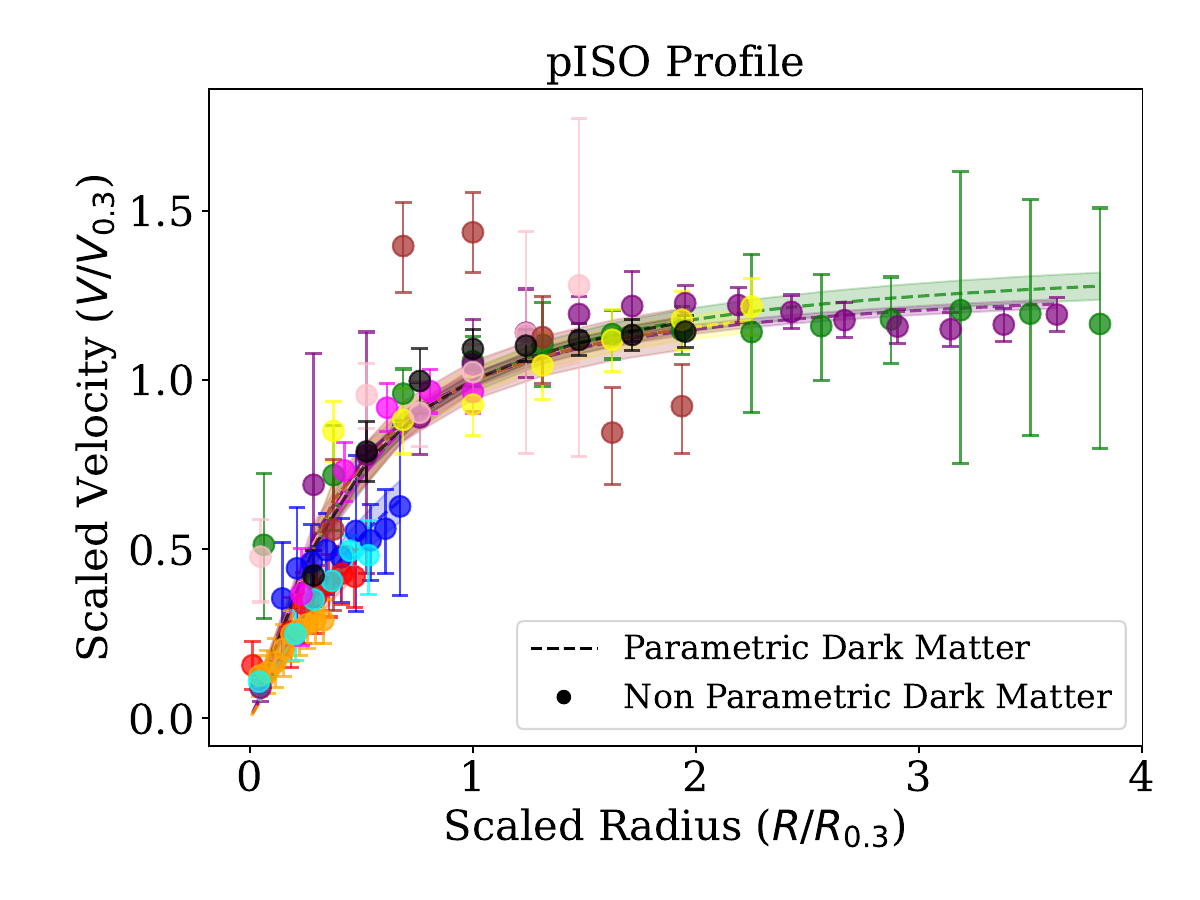}
    \end{subfigure}
    \caption{ Scaled the circular velocity contribution of the dark halo component for parametric and non-parametric dark matter. The dashed line corresponds to parametric dark matter using MCMC-based mass modelling, and each panel represents a different dark matter halo model, with different colours indicating different galaxies.}
    \label{fig:scaled_RC}
\end{figure*}

Figure~\ref{fig:scaled_RC} shows all the circular velocity contributions of the dark halo component derived from parametric and non-parametric approaches. The curves are scaled at the point where $\frac{d \log V}{d \log R} = 0.3$. The dashed line represents the parametric approach extracted from mass modelling via MCMC, and filled circles represent the non-parametric approach. All the velocities are scaled with respect to the rotational velocity $V_{0.3}$ at $R_{0.3}$.

\subsubsection{Scaled Density profiles}
\label{subsubsec:Scaled Density}
\begin{figure*}  
 \centering
    \begin{subfigure}{0.95\textwidth}
        
        \includegraphics[width=\linewidth]{Results_img/galaxy_colorbar_larger_font_and_title_gap.png}
    \end{subfigure}

    \centering
    \begin{subfigure}{0.49\textwidth}
        \centering
        \includegraphics[width=\textwidth]{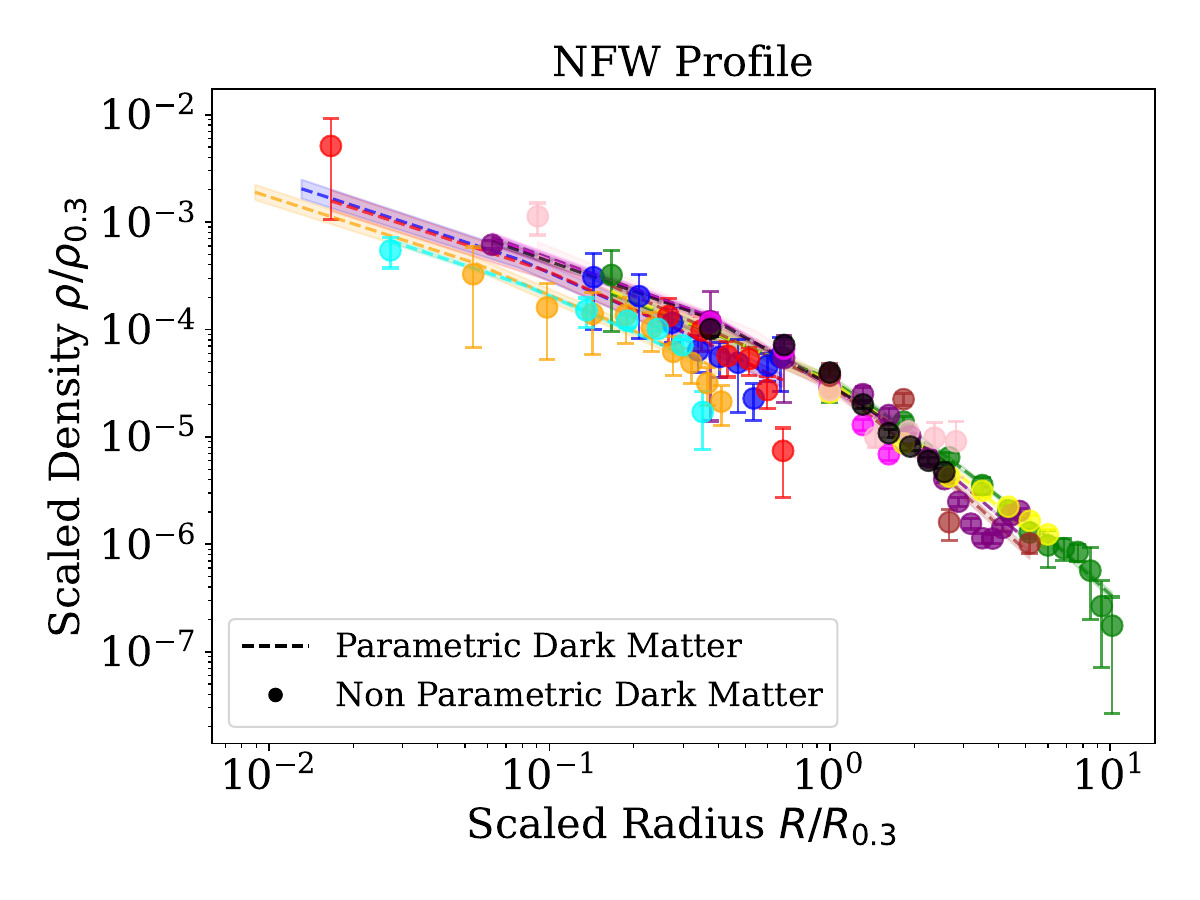}
    \end{subfigure}
    \begin{subfigure}{0.49\textwidth}
        \centering
        \includegraphics[width=\textwidth]{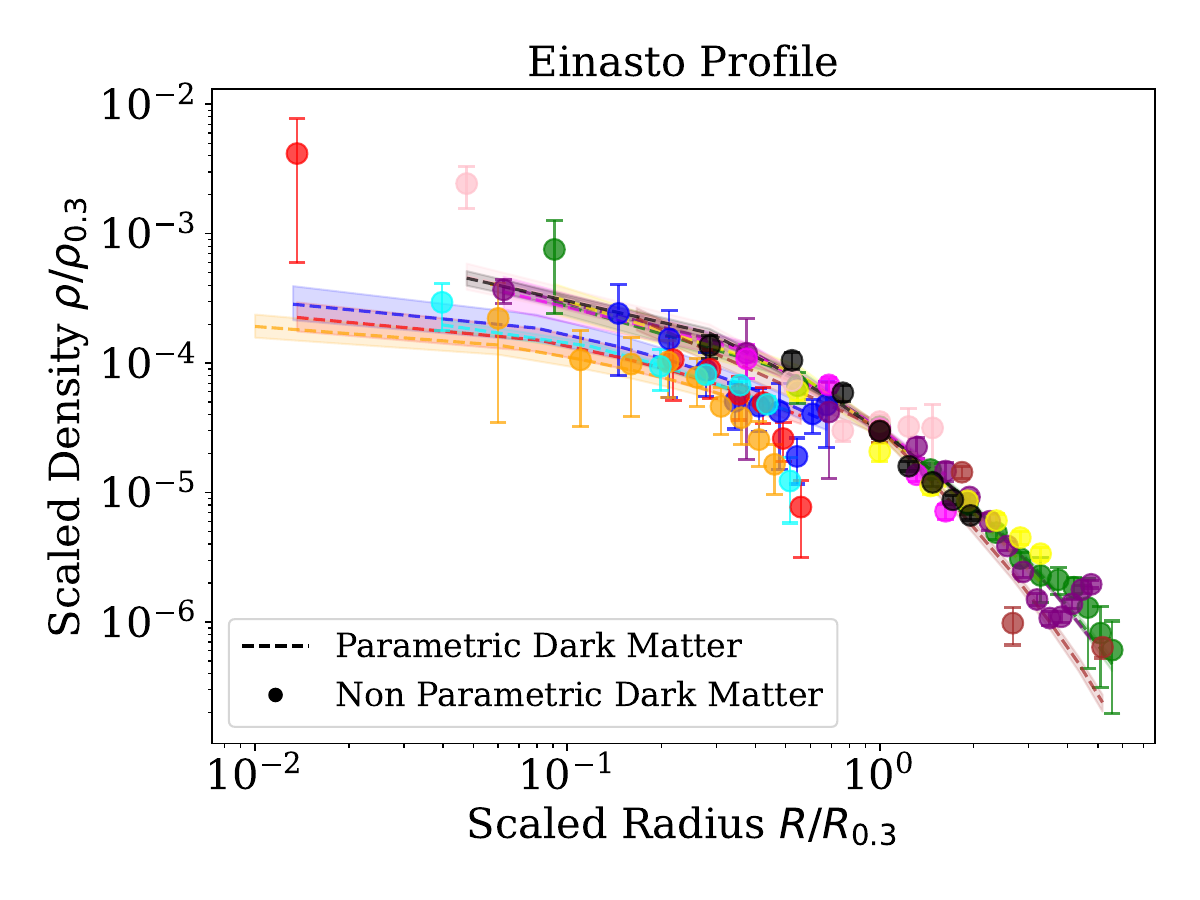}
    \end{subfigure}

    \vspace{0.05cm} 

    \begin{subfigure}{0.49\textwidth}
        \centering
        \includegraphics[width=\textwidth]{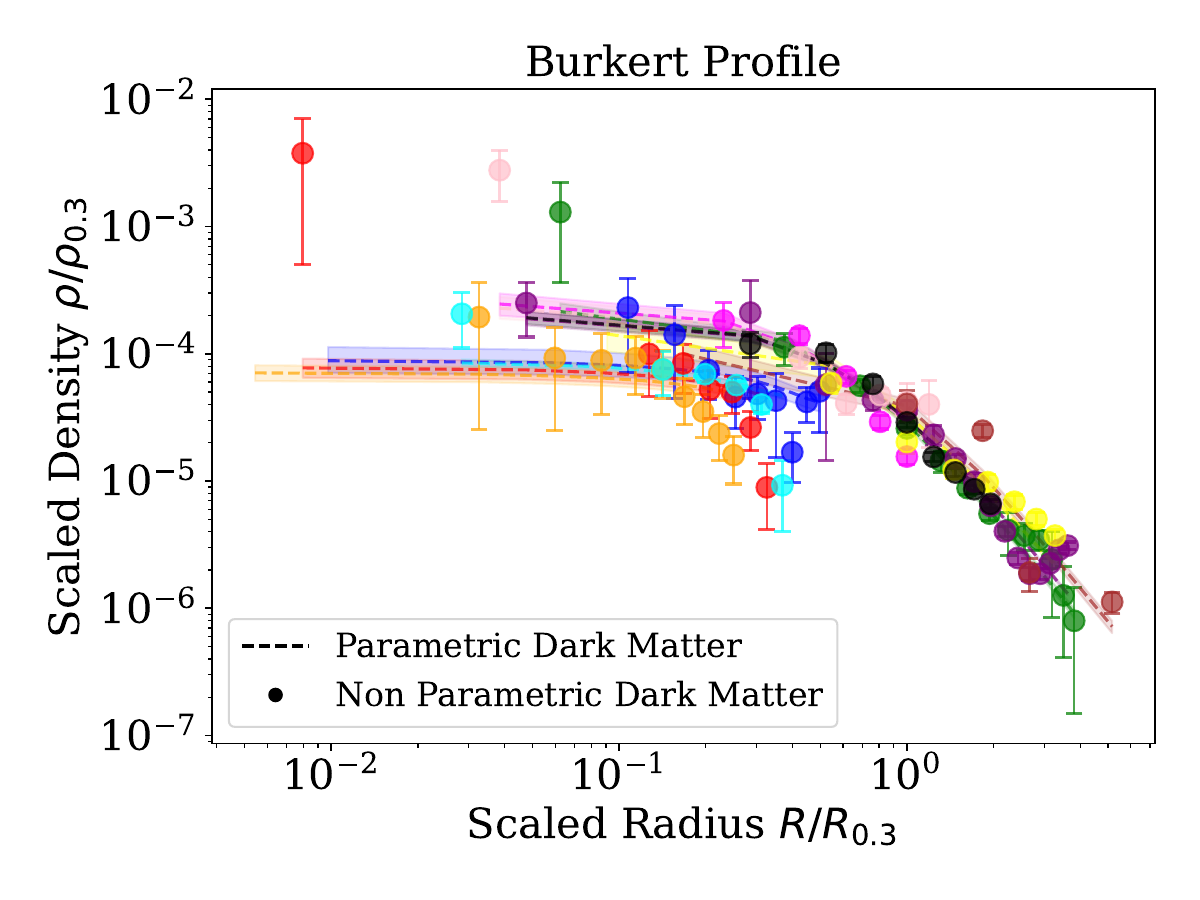}
    \end{subfigure}
    \begin{subfigure}{0.49\textwidth}
        \centering
        \includegraphics[width=\textwidth]{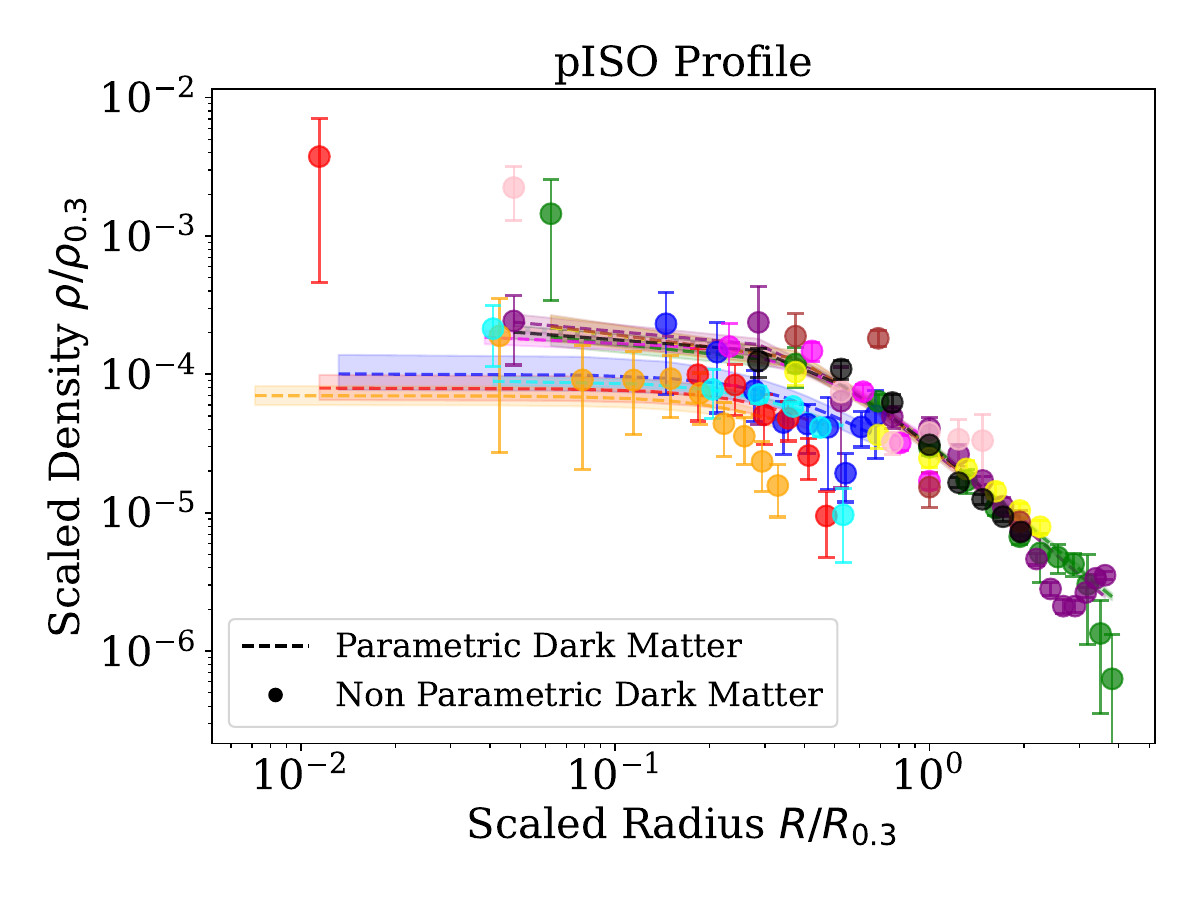}
    \end{subfigure}
    \caption{Scaled dark matter density profiles: parametric (dashed lines; see Section~\ref{subsubsec: parametric dm}) and non-parametric (circles; see Section~\ref{subsubsec:non para dm}). Each panel corresponds to a different dark matter halo model, with different colours representing different galaxies.}
    \label{fig:scaled_den}
\end{figure*}

The mass distribution can be directly inferred from its rotation curve, if the mass distribution is assumed to be spherical. However, this procedure can be unstable in systems with thin disks \citep{Evi_cusp_2001ApJ...552L..23D} because the circular velocity at a given radius in a flattened system receives contributions from mass over a wide range of radii, so small fluctuations or noise in $V_{\rm rot}(r)$ can lead to large and unphysical variations in the recovered mass profile \citep{1997PASA...14...11S}. To avoid this complication, we consider only the dark matter component of the rotation curve, under the assumption that the dark matter forms a spherically symmetric halo. For such a distribution, applying the Poisson equation,  

\begin{equation}
\nabla^2 \Phi = 4\pi G \rho
\end{equation}

along with the gravitational potential \( \Phi = -\frac{GM}{r} \), leads to the following 
expression for the density \citep{Evi_cusp_2001ApJ...552L..23D,2008AJ....136.2648D,2015AJ....149..180O,2020MNRAS.491.4993K}:  

\begin{equation}
\rho(r) = \frac{1}{4\pi G} \left( 2 \frac{V_{dm}}{r} \frac{dV_{dm}}{dr} + \left(\frac{V_{dm}}{r}\right)^2 \right)
\label{eq:scl_den}
\end{equation}
where G is the gravitational constant, \( V_{dm} \) represents the rotational velocity correspond to dark matter and \( r \) is the radius. We derived both parametric and non-parametric dark matter (DM) density profiles for each galaxy. The parametric profiles are obtained from MCMC fits assuming specific halo models, while the non-parametric profiles are computed by subtracting the baryonic contribution from the observed rotation curves. Figure~\ref{fig:scaled_den} compares these two approaches. The figure presents scaled DM density profiles in four panels, each corresponding to a different halo model. The dashed lines represent the parametric DM densities from MCMC modelling (see Section~\ref{subsubsec: parametric dm}), and the circles show the non-parametric DM densities derived directly from the data (see Section~\ref{subsubsec:non para dm}). In the outer regions, all models show good agreement with the non-parametric estimates. However, in the inner regions, only the NFW profile remains consistent with the non-parametric results, while the cored profiles (Burkert and pISO) tend to deviate from them. As we have already mentioned, our non-parametric approach is not entirely model-independent, since the stellar velocity component is derived from MCMC modelling of \( M/L \). To cross-check our results, we repeated the non-parametric analysis using fiducial values of the stellar mass-to-light ratio described in Appendix~\ref{A1}. We find that the overall trends remain unchanged: the outer regions show good agreement between parametric and non-parametric dark matter profiles, whereas cored profiles continue to deviate in the inner regions. This consistency supports the reliability of our findings across different assumptions for \( M/L \).

\section{Discussion and Conclusion}
\label{sec:discuss&conclusion}
\begin{figure*}  
    \centering
    \begin{subfigure}{0.49\textwidth}
        \centering
        \includegraphics[width=\textwidth]{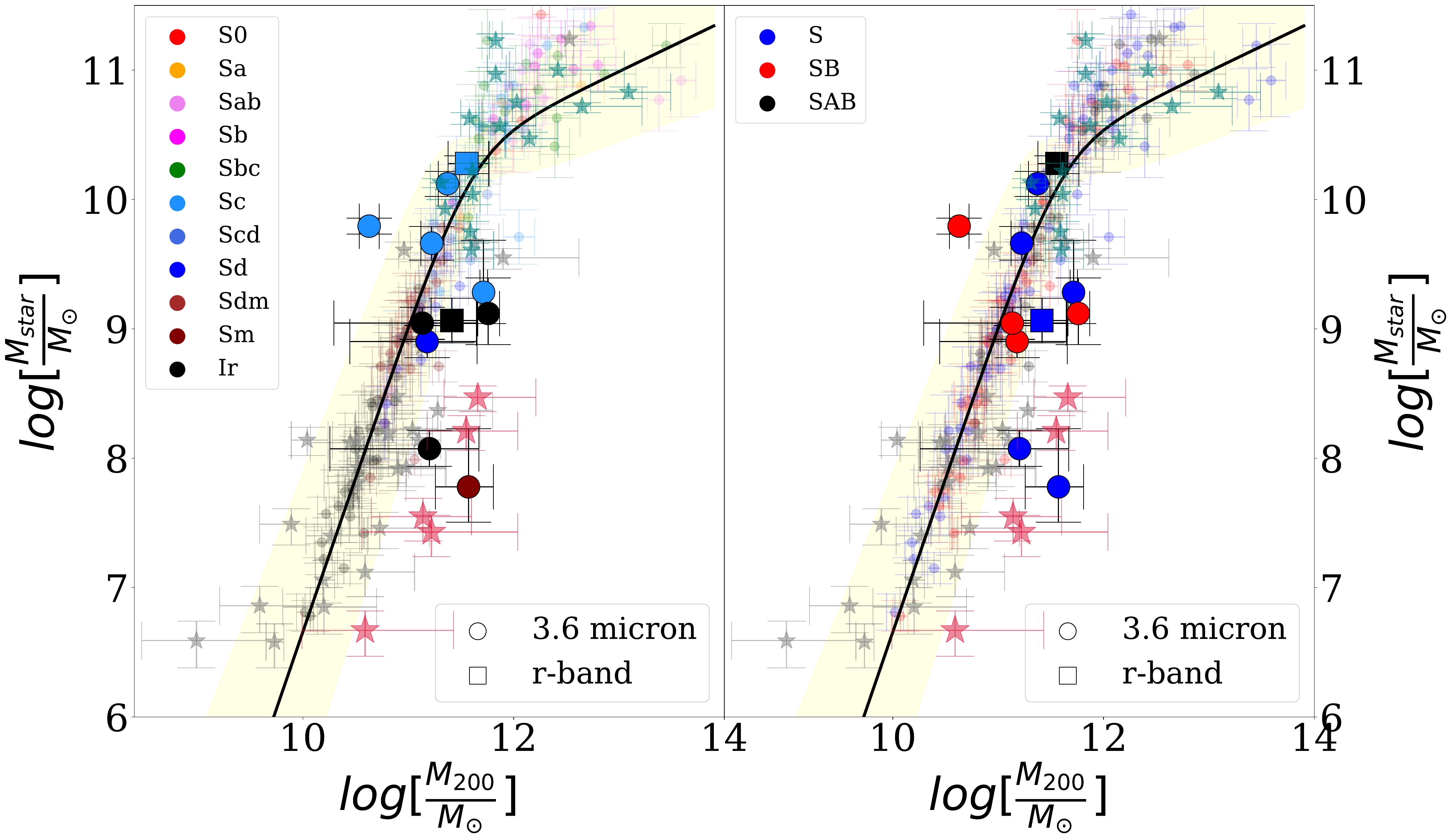}
        \subcaption*{\textbf{NFW profile}}
    \end{subfigure}
    \begin{subfigure}{0.49\textwidth}
        \centering
        \includegraphics[width=\textwidth]{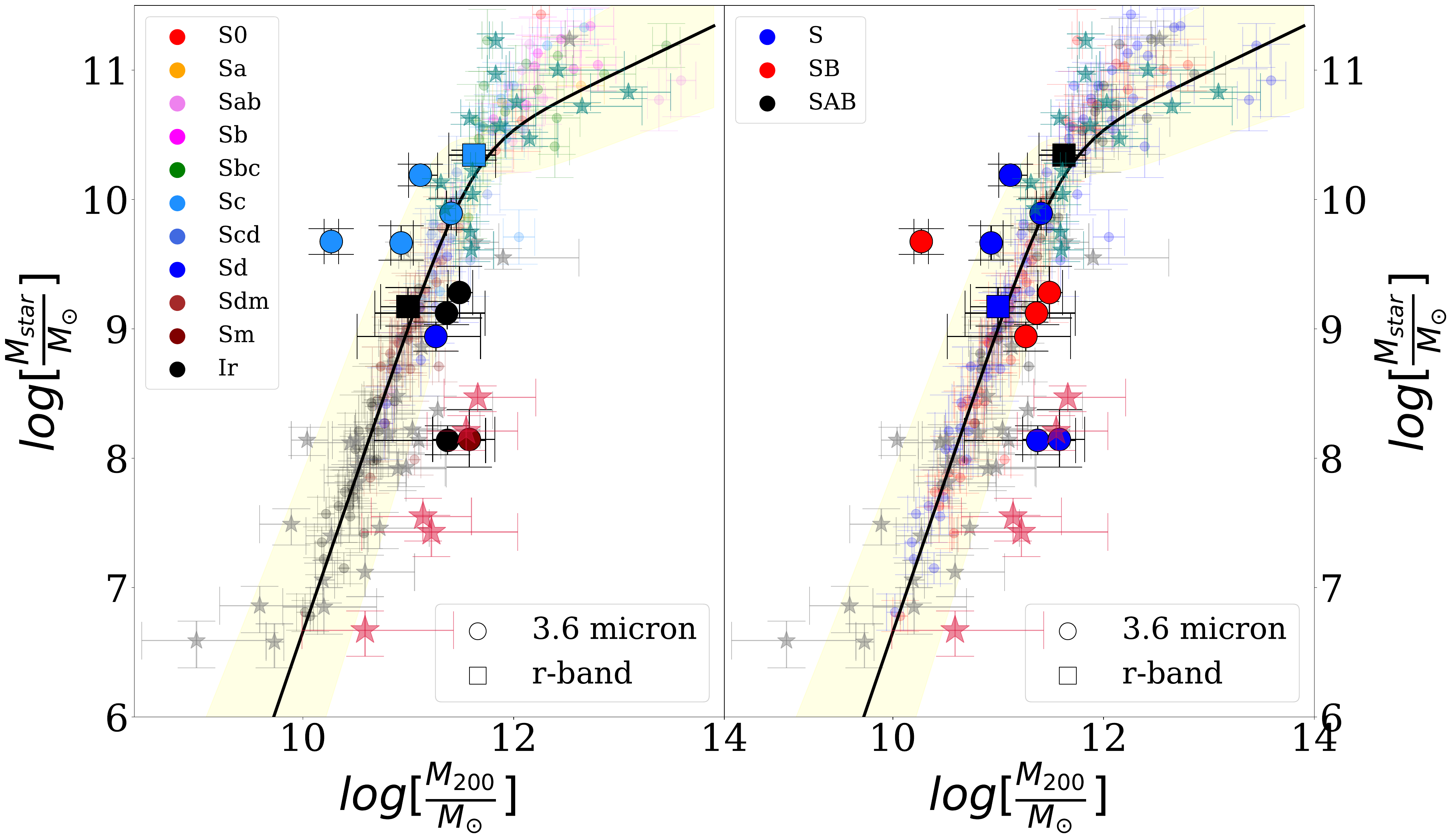}
        \subcaption*{\textbf{Einasto profile}}
    \end{subfigure}

    \vspace{0.1cm} 

    \begin{subfigure}{0.49\textwidth}
        \centering
        \includegraphics[width=\textwidth]{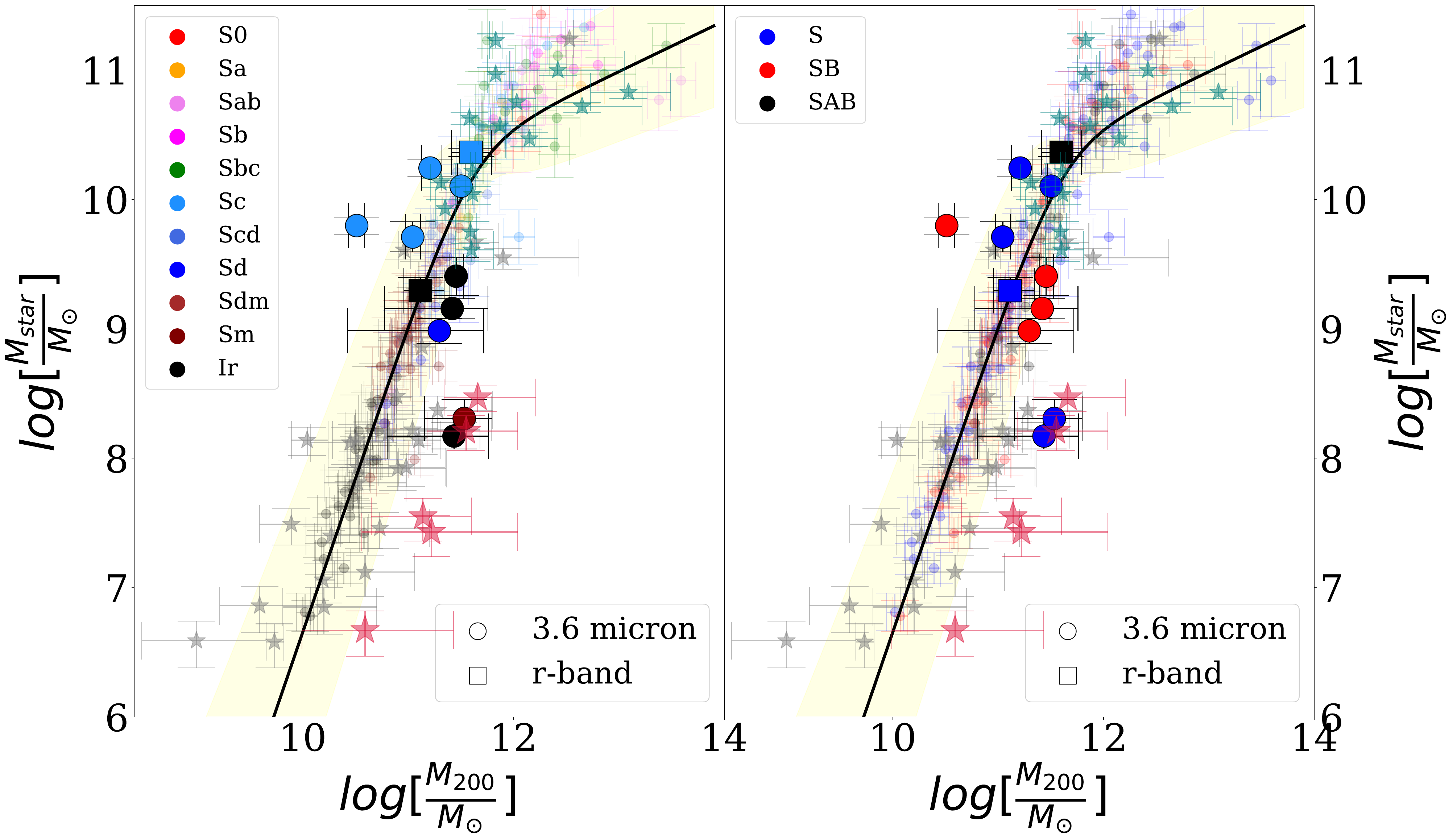}
        \subcaption*{\textbf{Burkert profile}}
    \end{subfigure}
    \begin{subfigure}{0.49\textwidth}
        \centering
            \includegraphics[width=\textwidth]{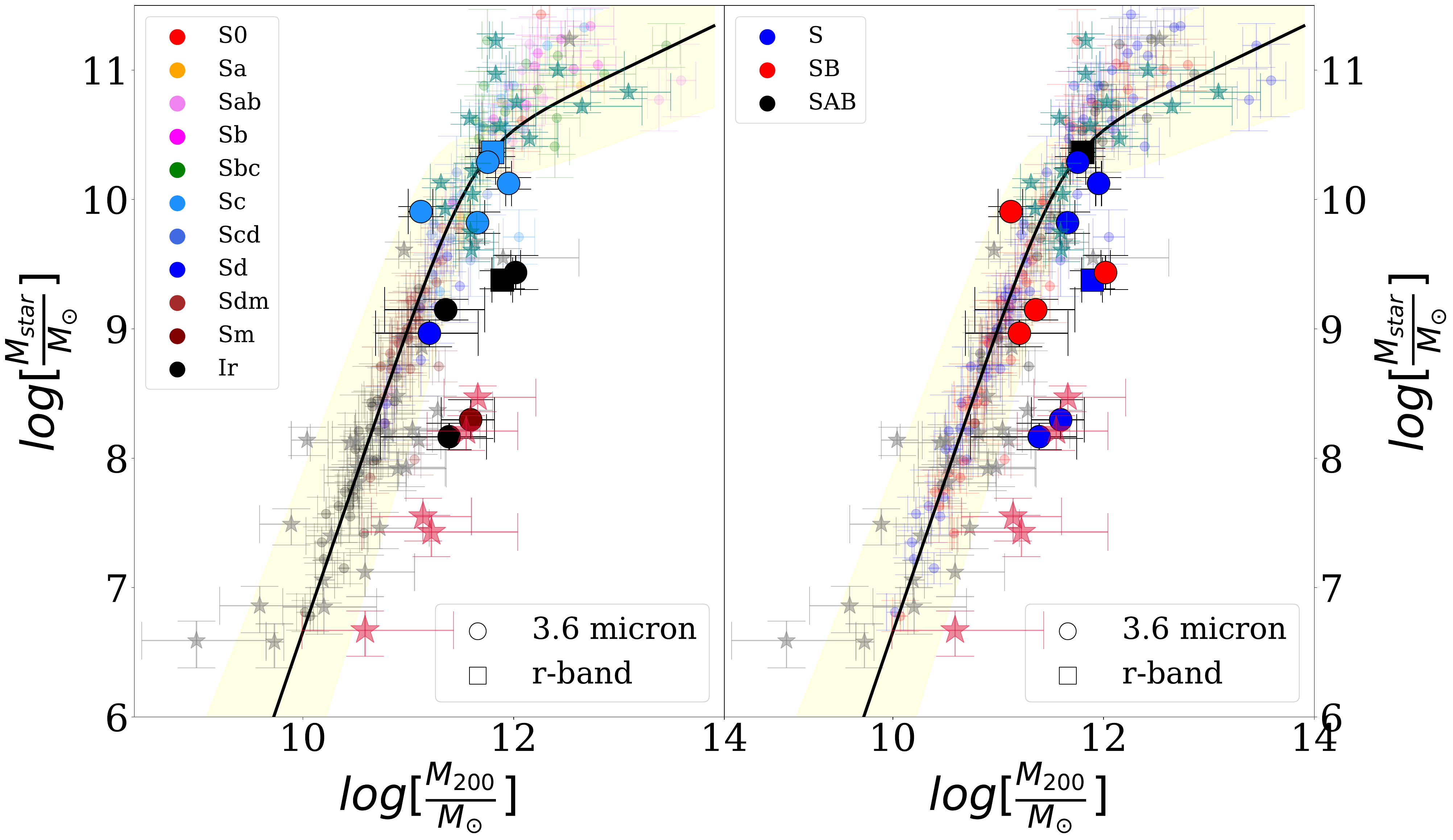}
        \subcaption*{\textbf{pISO profile}}
    \end{subfigure}
      \caption{The $M_{\mathrm{star}}$–$M_{200}$ relation is shown for different dark matter halo profiles. In each figure, the left panel displays galaxies colour-coded by Hubble type, while the right panel shows the same galaxies colour-coded by the presence or absence of bars. Circle and square symbols denote galaxies from the \href{https://physics.iisc.ac.in/~nroy/garcia_web/about.html}{GARCIA} sample, whereas dots represent galaxies from the \href{https://astroweb.cwru.edu/SPARC/} {SPARC} database. Stars represent the galaxies from \citet{2025A&A...699A.311M}: teal stars correspond to massive spirals, grey stars to dwarf galaxies, and pink stars to baryon-deficient dwarfs (BDDs). The black solid line indicates the relation from \citet{2013MNRAS.428.3121M}, and the yellow shaded region shows the corresponding $3\sigma$ uncertainty.}
    \label{fig:Mstar_Mhalo}
\end{figure*} 
This paper presents a comparison of the modelling of dark matter halos using four distinct density profiles, NFW (taken from \citet{biswas2023}), Einasto, Burkert and pISO for a pilot sample of eleven galaxies from the GARCIA survey. The analysis employs both parametric (via MCMC) and non-parametric approaches. We have compared the density profiles from both analyses (Fig.~\ref{fig:scaled_den}) for all galaxies. To check the consistency of the different parameters derived from the MCMC-based mass modelling, we have studied the existing relation and compared our results with previous studies, as discussed in the following subsection \ref{subsec:M_200-M_star_discuss}. 

\subsection{Consistency of different halo profiles}
\label{subsec:M_200-M_star_discuss}
 We studied the $M_{{star}}{-}M_{200}$ relation, which offers the connection between the baryonic and dark matter components of galaxies. Previous studies \citep[e.g.][]{2010ApJ...710..903M,2010ApJ...717..379B,2013MNRAS.428.3121M} have shown that this relation rises steeply at low halo masses, flattens around \( M_{200} \sim 10^{12}~M_{\odot} \), and then declines at higher halo masses. In our analysis, we placed the GARCIA sources on the \(M_{{star}}{-}M_{200} \) plane, using the \( M_{200} \) values derived from different dark matter halo profiles. We also overplotted the corresponding parameters for \href{https://astroweb.cwru.edu/SPARC/}{SPARC}\footnote{\url{https://astroweb.cwru.edu/SPARC/}} (Spitzer Photometry and Accurate Rotation Curves) galaxies for comparison (see Fig.~\ref{fig:Mstar_Mhalo}). The parameter values for the SPARC
 galaxies were taken from \citet{2016AJ....152..157L,2016ApJ...827L..19L,2019MNRAS.484.3267L,2019ApJ...886L..11L,2020ApJS..247...31L}, and morphological classifications were obtained from the \href{http://atlas.obs-hp.fr/hyperleda/}{HyperLeda database}\footnote{\href{http://atlas.obs-hp.fr/hyperleda/}{http://atlas.obs-hp.fr/hyperleda/}} \citep{2014A&A...570A..13M}.
 As shown in Fig.~\ref{fig:Mstar_Mhalo}, the GARCIA sources broadly follow the expected trend of the \(M_{\mathrm{star}}{-}M_{200}\) relation. However, some deviations are evident, consistent with observational studies of massive spirals, discs and gas-rich dwarf galaxies, which also find discrepancies with the abundance-matching SHMR \citep[e.g.][]{2019A&A...626A..56P,2021A&A...649A.119P,2025A&A...699A.311M}. These works suggest a morphological dependence: massive spirals tend to lie above the predicted SHMR, massive discs that grow gradually via gas accretion or minor mergers occupy the rising branch, and massive spheroids that have experienced major mergers or strong AGN feedback follow the falling branch. Possible explanations for deviations of the GARCIA galaxies are discussed below.

NGC4861, a magellanic-type spiral galaxy which falls significantly below the expected \( M_{star} \)–\( M_{200} \) relation from \citet{2013MNRAS.428.3121M}, suggesting a lower stellar mass than expected for its dark matter halo. Similar type of discrepancy also noticed by \citet{2025A&A...699A.311M} for a population of galaxies called baryon deficient dwarfs (BDDs) with halo masses $(log(M_{200}) \sim 10.5-12 M_{\odot})$ and stellar masses as low as $(log(M_{star}) \sim 7-9M_{\odot})$. For NGC 4861, the discrepancy may also be related to its disturbed morphology and clumpy star formation. \citet{1994A&A...283..753D} reported a double nucleus and a countervail structure pointing to a recent merger or tidal interaction that may have disrupted or suppressed star formation. Additionally, a prominent, bright knot in the southeastern part of the stellar disk, identified as a massive star-forming region, has been previously reported and spectroscopically studied \citep{2004MNRAS.355..728F, 2006ApJS..164...52S}. These starburst-driven irregularities likely result from recent or ongoing external perturbations and may suppress or redistribute the stellar content, thereby complicating mass modelling.

NGC7741, a barred spiral galaxy, deviates from the expected \(M_{{star}}{-}M_{200} \) relation when modelled with the NFW, Einasto, and Burkert profiles, all of which yield relatively low \( M_{200} \) values. In contrast, the pISO profile does not exhibit this deviation, although it has the highest reduced \( \chi^2 \) among the models. This discrepancy may stem from the galaxy's distorted rotation curve, which could cause the pISO model to overestimate the halo mass in an attempt to fit the kinematic data. Such a deviation for other profiles may be physically motivated and warrants further investigation, especially given the marginal spatial resolution of our data for this source.

NGC7800, an irregular barred galaxy, deviates from the expected $ M_{{star}}{-}M_{200}$ relation for the NFW and pISO profiles but not for Einasto and Burkert. A possible reason can be induced by its bar and asymmetric morphology. For this, we need to do a detailed analysis of this galaxy.

\subsection{Parametric vs Non-parametric dark matter}
\label{subsec:pmVSnonpm_diss}
Our mass modelling shows that all four dark matter halo profiles: NFW, Burkert, pISO, and Einasto yield broadly consistent fits for most galaxies in the GARCIA-I sample. The total halo mass \( M_{200} \) and stellar mass-to-light ratio \( M/L \) remain stable across different models, except for two galaxies (NGC7292 and NGC7610) where only $r$-band data were available, resulting in less reliable \( M/L \) estimates. In contrast, the concentration parameter \( C \) varies significantly between profiles due to its sensitivity to the inner halo structure. To further probe the nature of the dark matter distribution, we conducted a non-parametric analysis (Sec~\ref{subsubsec:non para dm}) by subtracting the baryonic component from the observed rotation velocity and deriving the corresponding dark matter density profile using equation~\ref{eq:scl_den}. This model-independent method is directly compared with the density profiles inferred from our MCMC fits. Figure~\ref{fig:scaled_den} shows that, in the inner regions of most galaxies, the dark matter density profiles derived from the parametric and non-parametric approaches overlap closely only for the NFW model. The Einasto and cored (Burkert and pISO) profiles reproduce the outer parts of the rotation curves reasonably well, but their inner densities deviate from the non-parametric profiles. This difference arises because the parametric fits depend on the assumed halo shape, whereas the non-parametric method is model-independent. If both methods trace the true mass distribution, their inner density profiles should agree; the closer match for the NFW case therefore suggests that, within our data quality, the NFW profile provides the most consistent description of the inner halo. This result aligns with the findings of \citet{2020MNRAS.491.4993K}, who also reported that, when employing improved 3D modelling techniques for rotation curve derivation, gas-rich void dwarf galaxies exhibit central density slopes consistent with cuspy NFW profiles. The presence of cuspy profiles in both dwarf (weak baryonic influence) and normal galaxies (significant baryonic component present) implies that dark matter halo structure is likely independent of baryonic feedback. Though there are severel studies showed that baryonic feedback can play a crucial role in transforming central dark matter cusps into cores, particularly in dwarf galaxies, as shown in numerous hydrodynamical simulations and observational studies \citep[e.g.][]{2013MNRAS.429.3068T,2019MNRAS.488.2387B}. Our sample encompasses a range of morphologies, including galaxies with a significant baryonic component, yet they exhibit a similar trend to that observed in dwarf systems.  

From the mass modelling of NGC7800 (Fig.~\ref{fig:MCMC RCs}), we observe in the corner plots that the distributions of the concentration parameter \( C \) and halo mass \( M_{200} \) are elongated for the NFW, Einasto, and Burkert profiles, whereas for the pISO profile, the contours are more circular and less correlated. This trend is consistent across most galaxies in our sample, with the elongation in the \( M_{200} - C \) plane increasing from cored to cuspy profiles. The observed elongation suggests a degeneracy between \( M_{200} \) and \( C \) in the inner regions of the halo, reflects the well known concentration–mass  (\( C - M_{200} \) ) relation \citep[e.g.][etc.]{2001MNRAS.321..559B,2003ApJ...597L...9Z,2006MNRAS.368.1931L,2008MNRAS.391.1940M, 2014MNRAS.441.3359D,2015MNRAS.452.1217C,2015ApJ...799..108D,2016MNRAS.457.4340K} in halo fits. In cuspy profiles, the central density rises steeply, providing more flexibility in fitting the inner rotation curve through adjustments in either mass or concentration. In contrast, cored profiles exhibit a flat inner density distribution, requiring a larger total mass to reproduce a steep inner rotation curve. This reduces the flexibility of the concentration parameter, making it more tightly constrained compared to cuspy profiles. 

Our analysis confirms that the galaxies in the GARCIA-I sample broadly follow the expected \(M_{{star}}{-}M_{200} \) relation, supporting the well established empirical link between baryonic and dark matter components in galaxy formation. However, some galaxies show deviations from this relation. These discrepancies may arise from various astrophysical processes. For instance, past merger events or tidal interactions can disrupt the baryonic structure and suppress star formation, resulting in lower stellar masses for a given halo mass \citep{2022ApJ...936L..11S}. Morphological disturbances, such as asymmetries or warps, can also impact mass estimates. In addition to physical effects, some of the observed scatter appears to be artificial, particularly for fits involving the cored pseudo-isothermal (pISO) profile. In several cases, this model tends to overestimate the halo mass \( M_{200} \), leading to an offset from the expected relation. This trend is evident in Fig.~\ref{fig:MCMC result}. The presence of a bar can complicate the derivation of accurate rotation curves, particularly in the inner regions where non-circular motions and the bar’s pattern speed may bias the kinematic modelling. In this study, we did not apply any specific correction for the non-circular motions associated with bar dynamics. In our pilot sample, four galaxies (NGC0784, NGC1156, NGC7741, and NGC7800) exhibit bar-like structures. While these effects likely contribute to the deviations observed, a more detailed investigation would be necessary to quantify the impact of bar-driven perturbations on mass modelling results.

\section{Summary}
\label{sec:summ}
In summary, we have modelled a pilot sample of GARCIA galaxies using four different dark matter halo profiles via MCMC. We found that all profiles yield consistent fits for most galaxies. However, based on the reduced $\chi^2$ values, it is difficult to identify a single preferred halo model across the sample. We subtracted the baryonic contribution from the observed rotation velocity and derived the corresponding dark matter density profile to further investigate the dark matter distribution. We find that the inner mass distribution for the GARCIA sample is more consistent with a cuspy NFW profile. In contrast, previous studies based on 2D rotation curve modelling often favoured cored profiles. These results underscore the importance of employing careful 3D modelling with high-resolution kinematic data to accurately constrain the rotation curve and better understand the nature of dark matter distribution in galaxies. Some recent 3D kinematic modelling \citep[e.g.][]{2017MNRAS.467.2019R,2022MNRAS.514.3329M,2025A&A...699A.311M} report the presence of cores in several galaxies, though this outcome is not universal and appears to depend on galaxy properties and data quality. Our analysis is subject to different limitations that we plan to address in forthcoming work. First, the present study is based on a pilot sample of 11 galaxies, the forthcoming GARCIA-batch II samples will provide a substantially larger ($\sim25$) and more diverse dataset which will provide improved statistics and a better understanding of halo properties across different morphological types. Second, the mass models assume a radially constant stellar mass-to-light ratio; although this is not expected to strongly bias our main conclusions, future papers will relax them by explicitly accounting for exploring radially varying $M/L$ profiles informed by multi-band photometry and stellar population modelling, and by quantifying the impact of non-circular motions, mainly in barred galaxies. We plan to carry out a detailed comparison between bulge–disc decompositions, MGE modelling, and variable $M/L$ for the full GARCIA-II sample in a subsequent paper. Third, our study does not include the molecular gas component because CO observation are not available for the full pilot sample; in future work we will incorporate H$_2$ either from direct CO observations to obtain a more complete baryonic budget. The current findings are based on a relatively small sample and moderate-resolution data from GMRT. With the advent of next-generation radio telescopes such as the Square Kilometre Array (SKA), which will provide unprecedented sensitivity and spatial resolution, future studies will be able to resolve the inner kinematics of galaxies more precisely. This will significantly enhance our ability to distinguish between cored and cuspy halo profiles, refining our understanding of galaxy formation and dark matter physics.

\section*{Acknowledgements}

This research has made use of the \textit{Spitzer Space Telescope} and Sloan Digital Sky Survey (SDSS) databases. The \textit{Spitzer Space Telescope} was operated by the Jet Propulsion Laboratory, California Institute of Technology, under a contract with the National Aeronautics and Space Administration (NASA), with support for this work provided by an award issued by JPL/Caltech. Funding for the Sloan Digital Sky Survey (SDSS) has been provided by the Alfred P.~Sloan Foundation, the participating institutions, NASA, the National Science Foundation, the U.S.~Department of Energy, the Japanese Monbukagakusho, and the Max Planck Society. This research has also made use of the NASA/IPAC Extragalactic Database (NED), operated by the Jet Propulsion Laboratory, California Institute of Technology, under contract with NASA. SS acknowledges the Council of Scientific and Industrial Research
(CSIR), Government of India, for supporting her research
under the CSIR Junior/Senior Research Fellowship program through grant no.~$09/0079(12143)/2021-EMR-I$. NR acknowledges support from the United States–India Educational Foundation through the Fulbright Program.

We also acknowledge the extensive use of archival data from the GMRT online archive. The GMRT is operated by the National Centre for Radio Astrophysics of the Tata Institute of Fundamental Research. We sincerely thank the entire GMRT team for providing and maintaining this invaluable facility.

We thank Md Rashid for his valuable inputs to the discussions related to this study. SS thanks Rachana for her help in obtaining the latest SDSS $g$- and $r$-band magnitudes.

We also thank the anonymous reviewer for the feedback and suggetions, which helped us to improve this manuscript.

\section{Data Availablity}
All the derived quantities and models produced in this study will be shared at the reasonable request of the corresponding author.



\bibliographystyle{mnras}
\bibliography{ref}

@article{Zwicky:1933gu,
    author = "Zwicky, F.",
    title = "{Die Rotverschiebung von extragalaktischen Nebeln}",
    doi = "10.1007/s10714-008-0707-4",
    journal = "Helv. Phys. Acta",
    volume = "6",
    pages = "110--127",
    year = "1933"
}

@ARTICLE{1978ApJ...225L.107R,
       author = {{Rubin}, V.~C. and {Ford}, Jr., W.~K. and {Thonnard}, N.},
        title = "{Extended rotation curves of high-luminosity spiral galaxies. IV. Systematic dynamical properties, Sa -> Sc.}",
      journal = {\apjl},
         year = 1978,
        month = nov,
       volume = {225},
        pages = {L107-L111},
          doi = {10.1086/182804},
       adsurl = {https://ui.adsabs.harvard.edu/abs/1978ApJ...225L.107R}
}

@PHDTHESIS{1978PhDT.......195B,
       author = {{Bosma}, A.},
        title = "{The distribution and kinematics of neutral hydrogen in spiral galaxies of various morphological types}",
       school = {University of Groningen, Netherlands},
         year = 1978,
        month = mar,
       adsurl = {https://ui.adsabs.harvard.edu/abs/1978PhDT.......195B}
}

@PHDTHESIS{1987PhDT.......199B,
       author = {{Begeman}, K.~G.},
        title = "{HI rotation curves of spiral galaxies}",
       school = {University of Groningen, Kapteyn Astronomical Institute},
         year = 1987,
        month = dec,
       adsurl = {https://ui.adsabs.harvard.edu/abs/1987PhDT.......199B}
}

@ARTICLE{1986RSPTA.320..447V,
       author = {{van Albada}, T.~S. and {Sancisi}, R.},
        title = "{Dark Matter in Spiral Galaxies}",
      journal = {Philosophical Transactions of the Royal Society of London Series A},
         year = 1986,
        month = dec,
       volume = {320},
       number = {1556},
        pages = {447-464},
          doi = {10.1098/rsta.1986.0128},
       adsurl = {https://ui.adsabs.harvard.edu/abs/1986RSPTA.320..447V}
}

@ARTICLE{1990A&ARv...2....1S,
       author = {{Sanders}, R.~H.},
        title = "{Mass discrepancies in galaxies: dark matter and alternatives}",
      journal = {\aapr},
         year = 1990,
        month = jan,
       volume = {2},
       number = {1},
        pages = {1-28},
          doi = {10.1007/BF00873540},
       adsurl = {https://ui.adsabs.harvard.edu/abs/1990A&ARv...2....1S}
}

@ARTICLE{1991ApJ...378..496D,
       author = {{Dubinski}, John and {Carlberg}, R.~G.},
        title = "{The Structure of Cold Dark Matter Halos}",
      journal = {\apj},
         year = 1991,
        month = sep,
       volume = {378},
        pages = {496},
          doi = {10.1086/170451},
       adsurl = {https://ui.adsabs.harvard.edu/abs/1991ApJ...378..4Evidence against dissipation-less dark matter from observations of galaxy haloes96D}
}

@ARTICLE{1994Natur.370..629M,
       author = {{Moore}, Ben},
        title = "{Evidence against dissipation-less dark matter from observations of galaxy haloes}",
      journal = {\nat},
         year = 1994,
        month = aug,
       volume = {370},
       number = {6491},
        pages = {629-631},
          doi = {10.1038/370629a0},
       adsurl = {https://ui.adsabs.harvard.edu/abs/1994Natur.370..629M}
}

@ARTICLE{1994ApJ...427L...1F,
       author = {{Flores}, Ricardo A. and {Primack}, Joel R.},
        title = "{Observational and Theoretical Constraints on Singular Dark Matter Halos}",
      journal = {\apjl},
         year = 1994,
        month = may,
       volume = {427},
        pages = {L1},
          doi = {10.1086/187350},
archivePrefix = {arXiv},
       eprint = {astro-ph/9402004},
 primaryClass = {astro-ph},
       adsurl = {https://ui.adsabs.harvard.edu/abs/1994ApJ...427L...1F}
}

@ARTICLE{1996NFW,
       author = {{Navarro}, Julio F. and {Frenk}, Carlos S. and {White}, Simon D.~M.},
        title = "{The Structure of Cold Dark Matter Halos}",
      journal = {\apj},
         year = 1996,
        month = may,
       volume = {462},
        pages = {563},
          doi = {10.1086/177173},
archivePrefix = {arXiv},
       eprint = {astro-ph/9508025},
 primaryClass = {astro-ph},
       adsurl = {https://ui.adsabs.harvard.edu/abs/1996ApJ...462..563N}
}

@ARTICLE{1997ApJ...490..493N,
       author = {{Navarro}, Julio F. and {Frenk}, Carlos S. and {White}, Simon D.~M.},
        title = "{A Universal Density Profile from Hierarchical Clustering}",
      journal = {\apj},
         year = 1997,
        month = dec,
       volume = {490},
       number = {2},
        pages = {493-508},
          doi = {10.1086/304888},
archivePrefix = {arXiv},
       eprint = {astro-ph/9611107},
 primaryClass = {astro-ph},
       adsurl = {https://ui.adsabs.harvard.edu/abs/1997ApJ...490..493N}
}

@ARTICLE{1997astro.ph.11259M,
       author = {{Moore}, Ben and {Ghigna}, Sebastiano and {Governato}, Fabio and {Lake}, George and {Quinn}, Tom and {Stadel}, Joachim},
        title = "{The Structure and Dynamics of Cold Dark Matter Halos}",
      journal = {arXiv e-prints},
         year = 1997,
        month = nov,
          eid = {astro-ph/9711259},
        pages = {astro-ph/9711259},
          doi = {10.48550/arXiv.astro-ph/9711259},
archivePrefix = {arXiv},
       eprint = {astro-ph/9711259},
 primaryClass = {astro-ph},
       adsurl = {https://ui.adsabs.harvard.edu/abs/1997astro.ph.11259M}
}

@ARTICLE{1998Moore,
       author = {{Moore}, B. and {Governato}, F. and {Quinn}, T. and {Stadel}, J. and {Lake}, G.},
        title = "{Resolving the Structure of Cold Dark Matter Halos}",
      journal = {\apjl},
         year = 1998,
        month = may,
       volume = {499},
       number = {1},
        pages = {L5-L8},
          doi = {10.1086/311333},
archivePrefix = {arXiv},
       eprint = {astro-ph/9709051},
 primaryClass = {astro-ph},
       adsurl = {https://ui.adsabs.harvard.edu/abs/1998ApJ...499L...5M}
}

@ARTICLE{2004CuspyLSB,
       author = {{Hayashi}, E. and {Navarro}, J.~F. and {Power}, C. and {Jenkins}, A. and {Frenk}, C.~S. and {White}, S.~D.~M. and {Springel}, V. and {Stadel}, J. and {Quinn}, T.~R.},
        title = "{The inner structure of {\ensuremath{\Lambda}}CDM haloes - II. Halo mass profiles and low surface brightness galaxy rotation curves}",
      journal = {\mnras},
         year = 2004,
        month = dec,
       volume = {355},
       number = {3},
        pages = {794-812},
          doi = {10.1111/j.1365-2966.2004.08359.x},
archivePrefix = {arXiv},
       eprint = {astro-ph/0310576},
 primaryClass = {astro-ph},
       adsurl = {https://ui.adsabs.harvard.edu/abs/2004MNRAS.355..794H}
}

@ARTICLE{Evi_cusp_1989ApJ...347..760C,
       author = {{Carignan}, Claude and {Beaulieu}, Sylvie},
        title = "{Optical and H i Studies of the ``Gas-rich'' Dwarf Irregular Galaxy DDO 154}",
      journal = {\apj},
         year = 1989,
        month = dec,
       volume = {347},
        pages = {760},
          doi = {10.1086/168167},
       adsurl = {https://ui.adsabs.harvard.edu/abs/1989ApJ...347..760C}
}

@ARTICLE{Evi_cusp_1994ApJ...427L...1F,
       author = {{Flores}, Ricardo A. and {Primack}, Joel R.},
        title = "{Observational and Theoretical Constraints on Singular Dark Matter Halos}",
      journal = {\apjl},
         year = 1994,
        month = may,
       volume = {427},
        pages = {L1},
          doi = {10.1086/187350},
archivePrefix = {arXiv},
       eprint = {astro-ph/9402004},
 primaryClass = {astro-ph},
       adsurl = {https://ui.adsabs.harvard.edu/abs/1994ApJ...427L...1F}
}

@ARTICLE{Evi_cusp_2001ApJ...552L..23D,
       author = {{de Blok}, W.~J.~G. and {McGaugh}, Stacy S. and {Bosma}, Albert and {Rubin}, Vera C.},
        title = "{Mass Density Profiles of Low Surface Brightness Galaxies}",
      journal = {\apjl},
         year = 2001,
        month = may,
       volume = {552},
       number = {1},
        pages = {L23-L26},
          doi = {10.1086/320262},
archivePrefix = {arXiv},
       eprint = {astro-ph/0103102},
 primaryClass = {astro-ph},
       adsurl = {https://ui.adsabs.harvard.edu/abs/2001ApJ...552L..23D}
}

@ARTICLE{Evi_cusp_2002A&A...385..816D,
       author = {{de Blok}, W.~J.~G. and {Bosma}, A.},
        title = "{High-resolution rotation curves of low surface brightness galaxies}",
      journal = {\aap},
         year = 2002,
        month = apr,
       volume = {385},
        pages = {816-846},
          doi = {10.1051/0004-6361:20020080},
archivePrefix = {arXiv},
       eprint = {astro-ph/0201276},
 primaryClass = {astro-ph},
       adsurl = {https://ui.adsabs.harvard.edu/abs/2002A&A...385..816D}
}

@INPROCEEDINGS{Evi_cusp_2003RMxAC..17...17D,
       author = {{de Blok}, W.~J.~G.},
        title = "{The Mass-Density Profiles of Low Surface Brightness Galaxies}",
    booktitle = {Revista Mexicana de Astronomia y Astrofisica Conference Series},
         year = 2003,
       editor = {{Avila-Reese}, Vladimir and {Firmani}, Claudio and {Frenk}, Carlos S. and {Allen}, Christine},
       series = {Revista Mexicana de Astronomia y Astrofisica Conference Series},
       volume = {17},
        month = jun,
        pages = {17-18},
       adsurl = {https://ui.adsabs.harvard.edu/abs/2003RMxAC..17...17D}
}

@article{Swaters2001_LSB_dwarf_does_not_rull_out_cusp,
    author = {van den Bosch, Frank C. and Swaters, Rob A.},
    title = {Dwarf galaxy rotation curves and the core problem of dark matter haloes},
    journal = {Monthly Notices of the Royal Astronomical Society},
    volume = {325},
    number = {3},
    pages = {1017-1038},
    year = {2001},
    month = {08},
    issn = {0035-8711},
    doi = {10.1046/j.1365-8711.2001.04456.x},
    url = {https://doi.org/10.1046/j.1365-8711.2001.04456.x},
    eprint = {https://academic.oup.com/mnras/article-pdf/325/3/1017/2839453/325-3-1017.pdf},
}

@ARTICLE{Swaters2003_LSB_dwarf_does_not_rull_out_cusp,
       author = {{Swaters}, R.~A. and {Madore}, B.~F. and {van den Bosch}, Frank C. and {Balcells}, M.},
        title = "{The Central Mass Distribution in Dwarf and Low Surface Brightness Galaxies}",
      journal = {\apj},
         year = 2003,
        month = feb,
       volume = {583},
       number = {2},
        pages = {732-751},
          doi = {10.1086/345426},
archivePrefix = {arXiv},
       eprint = {astro-ph/0210152},
 primaryClass = {astro-ph},
       adsurl = {https://ui.adsabs.harvard.edu/abs/2003ApJ...583..732S}
}

@PHDTHESIS{2004PhDT........22Hdoes_not_rull_out_cusp,
       author = {{Hayashi}, Eric},
        title = "{The structure of dark matter halos and disk galaxy rotation curves}",
       school = {University of Victoria, Canada},
         year = 2004,
        month = jan,
       adsurl = {https://ui.adsabs.harvard.edu/abs/2004PhDT........22H}
}

@ARTICLE{2015AJ....149..180O,
       author = {{Oh}, Se-Heon and {Hunter}, Deidre A. and {Brinks}, Elias and {Elmegreen}, Bruce G. and {Schruba}, Andreas and {Walter}, Fabian and {Rupen}, Michael P. and {Young}, Lisa M. and {Simpson}, Caroline E. and {Johnson}, Megan C. and {Herrmann}, Kimberly A. and {Ficut-Vicas}, Dana and {Cigan}, Phil and {Heesen}, Volker and {Ashley}, Trisha and {Zhang}, Hong-Xin},
        title = "{High-resolution Mass Models of Dwarf Galaxies from LITTLE THINGS}",
      journal = {\aj},
         year = 2015,
        month = jun,
       volume = {149},
       number = {6},
          eid = {180},
        pages = {180},
          doi = {10.1088/0004-6256/149/6/180},
archivePrefix = {arXiv},
       eprint = {1502.01281},
 primaryClass = {astro-ph.GA},
       adsurl = {https://ui.adsabs.harvard.edu/abs/2015AJ....149..180O}
}

@ARTICLE{2016MNRAS.462.3628R,
       author = {{Read}, J.~I. and {Iorio}, G. and {Agertz}, O. and {Fraternali}, F.},
        title = "{Understanding the shape and diversity of dwarf galaxy rotation curves in {\ensuremath{\Lambda}}CDM}",
      journal = {\mnras},
         year = 2016,
        month = nov,
       volume = {462},
       number = {4},
        pages = {3628-3645},
          doi = {10.1093/mnras/stw1876},
archivePrefix = {arXiv},
       eprint = {1601.05821},
 primaryClass = {astro-ph.GA},
       adsurl = {https://ui.adsabs.harvard.edu/abs/2016MNRAS.462.3628R}
}

@ARTICLE{2025A&A...699A.311M,
       author = {{Mancera Pi{\~n}a}, Pavel E. and {Read}, Justin I. and {Kim}, Stacy and {Marasco}, Antonino and {Benavides}, Jos{\'e} A. and {Glowacki}, Marcin and {Pezzulli}, Gabriele and {Lagos}, Claudia del P.},
        title = "{The galaxy-halo connection of disc galaxies over six orders of magnitude in stellar mass}",
      journal = {\aap},
         year = 2025,
        month = jul,
       volume = {699},
          eid = {A311},
        pages = {A311},
          doi = {10.1051/0004-6361/202554381},
archivePrefix = {arXiv},
       eprint = {2505.22727},
 primaryClass = {astro-ph.GA},
       adsurl = {https://ui.adsabs.harvard.edu/abs/2025A&A...699A.311M}
}

@article{Walter_2008,
doi = {10.1088/0004-6256/136/6/2563},
url = {https://dx.doi.org/10.1088/0004-6256/136/6/2563},
year = {2008},
month = {nov},
publisher = {The American Astronomical Society},
volume = {136},
number = {6},
pages = {2563},
author = {Walter, Fabian and Brinks, Elias and de Blok, W. J. G. and Bigiel, Frank and Kennicutt, Robert C. and Thornley, Michele D. and Leroy, Adam},
title = {THINGS: THE H i NEARBY GALAXY SURVEY},
journal = {The Astronomical Journal}
}

@ARTICLE{2001AJ....121.1952B,
       author = {{Blais-Ouellette}, S{\'e}bastien and {Amram}, Philippe and {Carignan}, Claude},
        title = "{Accurate Determination of the Mass Distribution in Spiral Galaxies. II. Testing the Shape of Dark Halos}",
      journal = {\aj},
         year = 2001,
        month = apr,
       volume = {121},
       number = {4},
        pages = {1952-1964},
          doi = {10.1086/319944},
archivePrefix = {arXiv},
       eprint = {astro-ph/0006449},
 primaryClass = {astro-ph},
       adsurl = {https://ui.adsabs.harvard.edu/abs/2001AJ....121.1952B}
}

@ARTICLE{2004MNRAS.351..903G,
       author = {{Gentile}, G. and {Salucci}, P. and {Klein}, U. and {Vergani}, D. and {Kalberla}, P.},
        title = "{The cored distribution of dark matter in spiral galaxies}",
      journal = {\mnras},
         year = 2004,
        month = jul,
       volume = {351},
       number = {3},
        pages = {903-922},
          doi = {10.1111/j.1365-2966.2004.07836.x},
archivePrefix = {arXiv},
       eprint = {astro-ph/0403154},
 primaryClass = {astro-ph},
       adsurl = {https://ui.adsabs.harvard.edu/abs/2004MNRAS.351..903G}
}

@ARTICLE{2008MNRAS.383..297S,
       author = {{Spano}, M. and {Marcelin}, M. and {Amram}, P. and {Carignan}, C. and {Epinat}, B. and {Hernandez}, O.},
        title = "{GHASP: an H{\ensuremath{\alpha}} kinematic survey of spiral and irregular galaxies - V. Dark matter distribution in 36 nearby spiral galaxies}",
      journal = {\mnras},
         year = 2008,
        month = jan,
       volume = {383},
       number = {1},
        pages = {297-316},
          doi = {10.1111/j.1365-2966.2007.12545.x},
archivePrefix = {arXiv},
       eprint = {0710.1345},
 primaryClass = {astro-ph},
       adsurl = {https://ui.adsabs.harvard.edu/abs/2008MNRAS.383..297S}
}

@ARTICLE{2013A&A...557A.131M,
       author = {{Martinsson}, Thomas P.~K. and {Verheijen}, Marc A.~W. and {Westfall}, Kyle B. and {Bershady}, Matthew A. and {Andersen}, David R. and {Swaters}, Rob A.},
        title = "{The DiskMass Survey. VII. The distribution of luminous and dark matter in spiral galaxies}",
      journal = {\aap},
         year = 2013,
        month = sep,
       volume = {557},
          eid = {A131},
        pages = {A131},
          doi = {10.1051/0004-6361/201321390},
archivePrefix = {arXiv},
       eprint = {1308.0336},
 primaryClass = {astro-ph.CO},
       adsurl = {https://ui.adsabs.harvard.edu/abs/2013A&A...557A.131M}
}

@ARTICLE{2011MNRAS.413.1633L,
       author = {{Le Delliou}, M. and {Henriksen}, R.~N. and {MacMillan}, J.~D.},
        title = "{Black holes and galactic density cusps - I. Radial orbit cusps and bulges}",
      journal = {\mnras},
         year = 2011,
        month = may,
       volume = {413},
       number = {3},
        pages = {1633-1642},
          doi = {10.1111/j.1365-2966.2011.18236.x},
archivePrefix = {arXiv},
       eprint = {0911.2232},
 primaryClass = {astro-ph.GA},
       adsurl = {https://ui.adsabs.harvard.edu/abs/2011MNRAS.413.1633L}
}

@ARTICLE{2010A&A...522A..28L,
       author = {{Le Delliou}, M. and {Henriksen}, R.~N. and {MacMillan}, J.~D.},
        title = "{Black holes and galactic density cusps. Spherically symmetric anisotropic cusps}",
      journal = {\aap},
         year = 2010,
        month = nov,
       volume = {522},
          eid = {A28},
        pages = {A28},
          doi = {10.1051/0004-6361/200913648},
archivePrefix = {arXiv},
       eprint = {0911.2234},
 primaryClass = {astro-ph.GA},
       adsurl = {https://ui.adsabs.harvard.edu/abs/2010A&A...522A..28L}
}

@ARTICLE{2001ApJ...560..636E,
       author = {{El-Zant}, Amr and {Shlosman}, Isaac and {Hoffman}, Yehuda},
        title = "{Dark Halos: The Flattening of the Density Cusp by Dynamical Friction}",
      journal = {\apj},
         year = 2001,
        month = oct,
       volume = {560},
       number = {2},
        pages = {636-643},
          doi = {10.1086/322516},
archivePrefix = {arXiv},
       eprint = {astro-ph/0103386},
 primaryClass = {astro-ph},
       adsurl = {https://ui.adsabs.harvard.edu/abs/2001ApJ...560..636E}
}

@ARTICLE{1996MNRAS.283L..72N,
       author = {{Navarro}, Julio F. and {Eke}, Vincent R. and {Frenk}, Carlos S.},
        title = "{The cores of dwarf galaxy haloes}",
      journal = {\mnras},
         year = 1996,
        month = dec,
       volume = {283},
       number = {3},
        pages = {L72-L78},
          doi = {10.1093/mnras/283.3.L72},
archivePrefix = {arXiv},
       eprint = {astro-ph/9610187},
 primaryClass = {astro-ph},
       adsurl = {https://ui.adsabs.harvard.edu/abs/1996MNRAS.283L..72N}
}

@ARTICLE{2006Natur.442..539M,
       author = {{Mashchenko}, Sergey and {Couchman}, H.~M.~P. and {Wadsley}, James},
        title = "{The removal of cusps from galaxy centres by stellar feedback in the early Universe}",
      journal = {\nat},
         year = 2006,
        month = aug,
       volume = {442},
       number = {7102},
        pages = {539-542},
          doi = {10.1038/nature04944},
archivePrefix = {arXiv},
       eprint = {astro-ph/0605672},
 primaryClass = {astro-ph},
       adsurl = {https://ui.adsabs.harvard.edu/abs/2006Natur.442..539M}
}

@ARTICLE{2008Sci...319..174M,
       author = {{Mashchenko}, Sergey and {Wadsley}, James and {Couchman}, H.~M.~P.},
        title = "{Stellar Feedback in Dwarf Galaxy Formation}",
      journal = {Science},
         year = 2008,
        month = jan,
       volume = {319},
       number = {5860},
        pages = {174},
          doi = {10.1126/science.1148666},
archivePrefix = {arXiv},
       eprint = {0711.4803},
 primaryClass = {astro-ph},
       adsurl = {https://ui.adsabs.harvard.edu/abs/2008Sci...319..174M}
}

@ARTICLE{2010Natur.463..203G,
       author = {{Governato}, F. and {Brook}, C. and {Mayer}, L. and {Brooks}, A. and {Rhee}, G. and {Wadsley}, J. and {Jonsson}, P. and {Willman}, B. and {Stinson}, G. and {Quinn}, T. and {Madau}, P.},
        title = "{Bulgeless dwarf galaxies and dark matter cores from supernova-driven outflows}",
      journal = {\nat},
         year = 2010,
        month = jan,
       volume = {463},
       number = {7278},
        pages = {203-206},
          doi = {10.1038/nature08640},
archivePrefix = {arXiv},
       eprint = {0911.2237},
 primaryClass = {astro-ph.CO},
       adsurl = {https://ui.adsabs.harvard.edu/abs/2010Natur.463..203G}
}

@ARTICLE{2011AJ....142...24O,
       author = {{Oh}, Se-Heon and {Brook}, Chris and {Governato}, Fabio and {Brinks}, Elias and {Mayer}, Lucio and {de Blok}, W.~J.~G. and {Brooks}, Alyson and {Walter}, Fabian},
        title = "{The Central Slope of Dark Matter Cores in Dwarf Galaxies: Simulations versus THINGS}",
      journal = {\aj},
         year = 2011,
        month = jul,
       volume = {142},
       number = {1},
          eid = {24},
        pages = {24},
          doi = {10.1088/0004-6256/142/1/24},
archivePrefix = {arXiv},
       eprint = {1011.2777},
 primaryClass = {astro-ph.CO},
       adsurl = {https://ui.adsabs.harvard.edu/abs/2011AJ....142...24O}
}

@ARTICLE{2012MNRAS.422.1231G,
       author = {{Governato}, F. and {Zolotov}, A. and {Pontzen}, A. and {Christensen}, C. and {Oh}, S.~H. and {Brooks}, A.~M. and {Quinn}, T. and {Shen}, S. and {Wadsley}, J.},
        title = "{Cuspy no more: how outflows affect the central dark matter and baryon distribution in {\ensuremath{\Lambda}} cold dark matter galaxies}",
      journal = {\mnras},
         year = 2012,
        month = may,
       volume = {422},
       number = {2},
        pages = {1231-1240},
          doi = {10.1111/j.1365-2966.2012.20696.x},
archivePrefix = {arXiv},
       eprint = {1202.0554},
 primaryClass = {astro-ph.CO},
       adsurl = {https://ui.adsabs.harvard.edu/abs/2012MNRAS.422.1231G}
}

@ARTICLE{2000PhRvL..84.3760S,
       author = {{Spergel}, David N. and {Steinhardt}, Paul J.},
        title = "{Observational Evidence for Self-Interacting Cold Dark Matter}",
      journal = {\prl},
         year = 2000,
        month = apr,
       volume = {84},
       number = {17},
        pages = {3760-3763},
          doi = {10.1103/PhysRevLett.84.3760},
archivePrefix = {arXiv},
       eprint = {astro-ph/9909386},
 primaryClass = {astro-ph},
       adsurl = {https://ui.adsabs.harvard.edu/abs/2000PhRvL..84.3760S}
}

@article{Burkert_2000,
doi = {10.1086/312674},
url = {https://dx.doi.org/10.1086/312674},
year = {2000},
month = {may},
publisher = {},
volume = {534},
number = {2},
pages = {L143},
author = {Burkert, Andreas},
title = {The Structure and Evolution of Weakly Self-interacting
Cold Dark Matter Halos},
journal = {The Astrophysical Journal}
}

@article{Cen_2001,
doi = {10.1086/318861},
url = {https://dx.doi.org/10.1086/318861},
year = {2001},
month = {jan},
publisher = {},
volume = {546},
number = {2},
pages = {L77},
author = {Cen, Renyue},
title = {Decaying Cold Dark Matter Model and Small-Scale Power},
journal = {The Astrophysical Journal}
}

@ARTICLE{2000PhRvL..85.1158H,
       author = {{Hu}, Wayne and {Barkana}, Rennan and {Gruzinov}, Andrei},
        title = "{Fuzzy Cold Dark Matter: The Wave Properties of Ultralight Particles}",
      journal = {\prl},
         year = 2000,
        month = aug,
       volume = {85},
       number = {6},
        pages = {1158-1161},
          doi = {10.1103/PhysRevLett.85.1158},
archivePrefix = {arXiv},
       eprint = {astro-ph/0003365},
 primaryClass = {astro-ph},
       adsurl = {https://ui.adsabs.harvard.edu/abs/2000PhRvL..85.1158H}
}

@ARTICLE{2012MNRAS.422..282R,
       author = {{Robles}, Victor H. and {Matos}, T.},
        title = "{Flat central density profile and constant dark matter surface density in galaxies from scalar field dark matter}",
      journal = {\mnras},
         year = 2012,
        month = may,
       volume = {422},
       number = {1},
        pages = {282-289},
          doi = {10.1111/j.1365-2966.2012.20603.x},
archivePrefix = {arXiv},
       eprint = {1201.3032},
 primaryClass = {astro-ph.CO},
       adsurl = {https://ui.adsabs.harvard.edu/abs/2012MNRAS.422..282R}
}

@ARTICLE{2016PhRvL.116d1302K,
       author = {{Kaplinghat}, Manoj and {Tulin}, Sean and {Yu}, Hai-Bo},
        title = "{Dark Matter Halos as Particle Colliders: Unified Solution to Small-Scale Structure Puzzles from Dwarfs to Clusters}",
      journal = {\prl},
         year = 2016,
        month = jan,
       volume = {116},
       number = {4},
          eid = {041302},
        pages = {041302},
          doi = {10.1103/PhysRevLett.116.041302},
archivePrefix = {arXiv},
       eprint = {1508.03339},
 primaryClass = {astro-ph.CO},
       adsurl = {https://ui.adsabs.harvard.edu/abs/2016PhRvL.116d1302K}
}

@ARTICLE{2018PhR...730....1T,
       author = {{Tulin}, Sean and {Yu}, Hai-Bo},
        title = "{Dark matter self-interactions and small scale structure}",
      journal = {\physrep},
         year = 2018,
        month = feb,
       volume = {730},
        pages = {1-57},
          doi = {10.1016/j.physrep.2017.11.004},
archivePrefix = {arXiv},
       eprint = {1705.02358},
 primaryClass = {hep-ph},
       adsurl = {https://ui.adsabs.harvard.edu/abs/2018PhR...730....1T}
}

@ARTICLE{2019PhRvX...9c1020R,
       author = {{Ren}, Tao and {Kwa}, Anna and {Kaplinghat}, Manoj and {Yu}, Hai-Bo},
        title = "{Reconciling the Diversity and Uniformity of Galactic Rotation Curves with Self-Interacting Dark Matter}",
      journal = {Physical Review X},
         year = 2019,
        month = jul,
       volume = {9},
       number = {3},
          eid = {031020},
        pages = {031020},
          doi = {10.1103/PhysRevX.9.031020},
archivePrefix = {arXiv},
       eprint = {1808.05695},
 primaryClass = {astro-ph.GA},
       adsurl = {https://ui.adsabs.harvard.edu/abs/2019PhRvX...9c1020R}
}

@ARTICLE{2000AJ....119.1579V,
       author = {{van den Bosch}, Frank C. and {Robertson}, Brant E. and {Dalcanton}, Julianne J. and {de Blok}, W.~J.~G.},
        title = "{Constraints on the Structure of Dark Matter Halos from the Rotation Curves of Low Surface Brightness Galaxies}",
      journal = {\aj},
         year = 2000,
        month = apr,
       volume = {119},
       number = {4},
        pages = {1579-1591},
          doi = {10.1086/301315},
archivePrefix = {arXiv},
       eprint = {astro-ph/9911372},
 primaryClass = {astro-ph},
       adsurl = {https://ui.adsabs.harvard.edu/abs/2000AJ....119.1579V}
}

@article{10.1093/mnras/stw3004,
    author = {Pineda, Juan C. B. and Hayward, Christopher C. and Springel, Volker and Mendes de Oliveira, Claudia},
    title = {Rotation curve fitting and its fatal attraction to cores in realistically simulated galaxy observations},
    journal = {Monthly Notices of the Royal Astronomical Society},
    volume = {466},
    number = {1},
    pages = {63-87},
    year = {2016},
    month = {11},
    issn = {0035-8711},
    doi = {10.1093/mnras/stw3004},
    url = {https://doi.org/10.1093/mnras/stw3004},
    eprint = {https://academic.oup.com/mnras/article-pdf/466/1/63/10865036/stw3004.pdf},
}

@ARTICLE{2019MNRAS.482..821O,
       author = {{Oman}, Kyle A. and {Marasco}, Antonino and {Navarro}, Julio F. and {Frenk}, Carlos S. and {Schaye}, Joop and {Ben{\'\i}tez-Llambay}, Alejandro},
        title = "{Non-circular motions and the diversity of dwarf galaxy rotation curves}",
      journal = {\mnras},
         year = 2019,
        month = jan,
       volume = {482},
       number = {1},
        pages = {821-847},
          doi = {10.1093/mnras/sty2687},
archivePrefix = {arXiv},
       eprint = {1706.07478},
 primaryClass = {astro-ph.GA},
       adsurl = {https://ui.adsabs.harvard.edu/abs/2019MNRAS.482..821O}
}

@ARTICLE{2023MNRAS.522.3318D,
       author = {{Downing}, Eleanor R. and {Oman}, Kyle A.},
        title = "{The many reasons that the rotation curves of low-mass galaxies can fail as tracers of their matter distributions}",
      journal = {\mnras},
         year = 2023,
        month = jul,
       volume = {522},
       number = {3},
        pages = {3318-3336},
          doi = {10.1093/mnras/stad868},
archivePrefix = {arXiv},
       eprint = {2301.05242},
 primaryClass = {astro-ph.GA},
       adsurl = {https://ui.adsabs.harvard.edu/abs/2023MNRAS.522.3318D}
}

@ARTICLE{2023MNRAS.521.1316R,
       author = {{Roper}, Finn A. and {Oman}, Kyle A. and {Frenk}, Carlos S. and {Ben{\'\i}tez-Llambay}, Alejandro and {Navarro}, Julio F. and {Santos-Santos}, Isabel M.~E.},
        title = "{The diversity of rotation curves of simulated galaxies with cusps and cores}",
      journal = {\mnras},
         year = 2023,
        month = may,
       volume = {521},
       number = {1},
        pages = {1316-1336},
          doi = {10.1093/mnras/stad549},
archivePrefix = {arXiv},
       eprint = {2203.16652},
 primaryClass = {astro-ph.GA},
       adsurl = {https://ui.adsabs.harvard.edu/abs/2023MNRAS.521.1316R}
}

@ARTICLE{2024arXiv240416247S,
       author = {{Sands}, Isabel S. and {Hopkins}, Philip F. and {Shen}, Xuejian and {Boylan-Kolchin}, Michael and {Bullock}, James and {Faucher-Giguere}, Claude-Andre and {Mercado}, Francisco J. and {Moreno}, Jorge and {Necib}, Lina and {Ou}, Xiaowei and {Wellons}, Sarah and {Wetzel}, Andrew},
        title = "{Confronting the Diversity Problem: The Limits of Galaxy Rotation Curves as a tool to Understand Dark Matter Profiles}",
      journal = {arXiv e-prints},
         year = 2024,
        month = apr,
          eid = {arXiv:2404.16247},
        pages = {arXiv:2404.16247},
          doi = {10.48550/arXiv.2404.16247},
archivePrefix = {arXiv},
       eprint = {2404.16247},
 primaryClass = {astro-ph.GA},
       adsurl = {https://ui.adsabs.harvard.edu/abs/2024arXiv240416247S}
}

@ARTICLE{2020MNRAS.491.4993K,
       author = {{Kurapati}, Sushma and {Chengalur}, Jayaram N. and {Kamphuis}, Peter and {Pustilnik}, Simon},
        title = "{Mass models of gas-rich void dwarf galaxies}",
      journal = {\mnras},
         year = 2020,
        month = feb,
       volume = {491},
       number = {4},
        pages = {4993-5014},
          doi = {10.1093/mnras/stz3334},
archivePrefix = {arXiv},
       eprint = {1911.11877},
 primaryClass = {astro-ph.GA},
       adsurl = {https://ui.adsabs.harvard.edu/abs/2020MNRAS.491.4993K}
}

@phdthesis{pISO_Begeman1987,
  author = {Begeman, K. G.},
  title = {{HI Rotation Curves of Spiral Galaxies}},
  school = {University of Groningen},
  year = {1987},
  address = {Groningen}
}

@ARTICLE{burkert1995ApJ...447L..25B,
       author = {{Burkert}, A.},
        title = "{The Structure of Dark Matter Halos in Dwarf Galaxies}",
      journal = {\apjl},
         year = 1995,
        month = jul,
       volume = {447},
        pages = {L25-L28},
          doi = {10.1086/309560},
archivePrefix = {arXiv},
       eprint = {astro-ph/9504041},
 primaryClass = {astro-ph},
       adsurl = {https://ui.adsabs.harvard.edu/abs/1995ApJ...447L..25B}
}

@ARTICLE{Einasto_1965TrAlm...5...87E,
       author = {{Einasto}, J.},
        title = "{On the Construction of a Composite Model for the Galaxy and on the Determination of the System of Galactic Parameters}",
      journal = {Trudy Astrofizicheskogo Instituta Alma-Ata},
         year = 1965,
        month = jan,
       volume = {5},
        pages = {87-100},
       adsurl = {https://ui.adsabs.harvard.edu/abs/1965TrAlm...5...87E}
}

@ARTICLE{Einasto_1969Afz.....5..137E,
       author = {{Einasto}, J.},
        title = "{The Andromeda galaxy M31. I. A preliminary model}",
      journal = {Astrofizika},
         year = 1969,
        month = feb,
       volume = {5},
        pages = {137-159},
       adsurl = {https://ui.adsabs.harvard.edu/abs/1969Afz.....5..137E}
}

@ARTICLE{2005MNRAS.358.1325C,
       author = {{Cardone}, V.~F. and {Piedipalumbo}, E. and {Tortora}, C.},
        title = "{Spherical galaxy models with power-law logarithmic slope}",
      journal = {\mnras},
         year = 2005,
        month = apr,
       volume = {358},
       number = {4},
        pages = {1325-1336},
          doi = {10.1111/j.1365-2966.2005.08834.x},
archivePrefix = {arXiv},
       eprint = {astro-ph/0501151},
 primaryClass = {astro-ph},
       adsurl = {https://ui.adsabs.harvard.edu/abs/2005MNRAS.358.1325C}
}

@ARTICLE{2011AJ....142..109C,
       author = {{Chemin}, Laurent and {de Blok}, W.~J.~G. and {Mamon}, Gary A.},
        title = "{Improved Modeling of the Mass Distribution of Disk Galaxies by the Einasto Halo Model}",
      journal = {\aj},
         year = 2011,
        month = oct,
       volume = {142},
       number = {4},
          eid = {109},
        pages = {109},
          doi = {10.1088/0004-6256/142/4/109},
archivePrefix = {arXiv},
       eprint = {1109.4247},
 primaryClass = {astro-ph.CO},
       adsurl = {https://ui.adsabs.harvard.edu/abs/2011AJ....142..109C}
}

@ARTICLE{2005MNRAS.362...95M,
       author = {{Mamon}, Gary A. and {{\L}okas}, Ewa L.},
        title = "{Dark matter in elliptical galaxies - I. Is the total mass density profile of the NFW form or even steeper?}",
      journal = {\mnras},
         year = 2005,
        month = sep,
       volume = {362},
       number = {1},
        pages = {95-109},
          doi = {10.1111/j.1365-2966.2005.09225.x},
archivePrefix = {arXiv},
       eprint = {astro-ph/0405466},
 primaryClass = {astro-ph},
       adsurl = {https://ui.adsabs.harvard.edu/abs/2005MNRAS.362...95M}
}

@ARTICLE{NAVARRO_EIN2004MNRAS.349.1039N,
       author = {{Navarro}, J.~F. and {Hayashi}, E. and {Power}, C. and {Jenkins}, A.~R. and {Frenk}, C.~S. and {White}, S.~D.~M. and {Springel}, V. and {Stadel}, J. and {Quinn}, T.~R.},
        title = "{The inner structure of {\ensuremath{\Lambda}}CDM haloes - III. Universality and asymptotic slopes}",
      journal = {\mnras},
         year = 2004,
        month = apr,
       volume = {349},
       number = {3},
        pages = {1039-1051},
          doi = {10.1111/j.1365-2966.2004.07586.x},
archivePrefix = {arXiv},
       eprint = {astro-ph/0311231},
 primaryClass = {astro-ph},
       adsurl = {https://ui.adsabs.harvard.edu/abs/2004MNRAS.349.1039N}
}

@ARTICLE{2006AJ....132.2685M,
       author = {{Merritt}, David and {Graham}, Alister W. and {Moore}, Ben and {Diemand}, J{\"u}rg and {Terzi{\'c}}, Bal{\v{s}}a},
        title = "{Empirical Models for Dark Matter Halos. I. Nonparametric Construction of Density Profiles and Comparison with Parametric Models}",
      journal = {\aj},
         year = 2006,
        month = dec,
       volume = {132},
       number = {6},
        pages = {2685-2700},
          doi = {10.1086/508988},
archivePrefix = {arXiv},
       eprint = {astro-ph/0509417},
 primaryClass = {astro-ph},
       adsurl = {https://ui.adsabs.harvard.edu/abs/2006AJ....132.2685M}
}

@ARTICLE{1983ApJ...270..365M,
       author = {{Milgrom}, M.},
        title = "{A modification of the Newtonian dynamics as a possible alternative to the hidden mass hypothesis.}",
      journal = {\apj},
         year = 1983,
        month = jul,
       volume = {270},
        pages = {365-370},
          doi = {10.1086/161130},
       adsurl = {https://ui.adsabs.harvard.edu/abs/1983ApJ...270..365M}
}

@ARTICLE{1983ApJ...270..371M,
       author = {{Milgrom}, M.},
        title = "{A modification of the Newtonian dynamics - Implications for galaxies.}",
      journal = {\apj},
         year = 1983,
        month = jul,
       volume = {270},
        pages = {371-383},
          doi = {10.1086/161131},
       adsurl = {https://ui.adsabs.harvard.edu/abs/1983ApJ...270..371M}
}

@ARTICLE{1974ApJ...193..309R,
       author = {{Rogstad}, D.~H. and {Lockhart}, I.~A. and {Wright}, M.~C.~H.},
        title = "{Aperture-synthesis observations of H I in the galaxy M83.}",
      journal = {\apj},
         year = 1974,
        month = oct,
       volume = {193},
        pages = {309-319},
          doi = {10.1086/153164},
       adsurl = {https://ui.adsabs.harvard.edu/abs/1974ApJ...193..309R}
}

@ARTICLE{2008MNRAS.390...71C,
       author = {{Cappellari}, Michele},
        title = "{Measuring the inclination and mass-to-light ratio of axisymmetric galaxies via anisotropic Jeans models of stellar kinematics}",
      journal = {\mnras},
         year = 2008,
        month = oct,
       volume = {390},
       number = {1},
        pages = {71-86},
          doi = {10.1111/j.1365-2966.2008.13754.x},
archivePrefix = {arXiv},
       eprint = {0806.0042},
 primaryClass = {astro-ph},
       adsurl = {https://ui.adsabs.harvard.edu/abs/2008MNRAS.390...71C}
}

@ARTICLE{2002MNRAS.333..400C,
       author = {{Cappellari}, Michele},
        title = "{Efficient multi-Gaussian expansion of galaxies}",
      journal = {\mnras},
         year = 2002,
        month = jun,
       volume = {333},
       number = {2},
        pages = {400-410},
          doi = {10.1046/j.1365-8711.2002.05412.x},
archivePrefix = {arXiv},
       eprint = {astro-ph/0201430},
 primaryClass = {astro-ph},
       adsurl = {https://ui.adsabs.harvard.edu/abs/2002MNRAS.333..400C}
}

@ARTICLE{2015MNRAS.451.3021D,
       author = {{Di Teodoro}, E.~M. and {Fraternali}, F.},
        title = "{$^{3D}$ BAROLO: a new 3D algorithm to derive rotation curves of galaxies}",
      journal = {\mnras},
         year = 2015,
        month = aug,
       volume = {451},
       number = {3},
        pages = {3021-3033},
          doi = {10.1093/mnras/stv1213},
archivePrefix = {arXiv},
       eprint = {1505.07834},
 primaryClass = {astro-ph.GA},
       adsurl = {https://ui.adsabs.harvard.edu/abs/2015MNRAS.451.3021D}
}

@ARTICLE{2015MNRAS.452.3139K,
       author = {{Kamphuis}, P. and {J{\'o}zsa}, G.~I.~G. and {Oh}, S. -. H. and {Spekkens}, K. and {Urbancic}, N. and {Serra}, P. and {Koribalski}, B.~S. and {Dettmar}, R. -J.},
        title = "{Automated kinematic modelling of warped galaxy discs in large H I surveys: 3D tilted-ring fitting of H I emission cubes}",
      journal = {\mnras},
         year = 2015,
        month = sep,
       volume = {452},
       number = {3},
        pages = {3139-3158},
          doi = {10.1093/mnras/stv1480},
archivePrefix = {arXiv},
       eprint = {1507.00413},
 primaryClass = {astro-ph.GA},
       adsurl = {https://ui.adsabs.harvard.edu/abs/2015MNRAS.452.3139K}
}

@ARTICLE{2021MNRAS.503.1753S,
       author = {{Sharma}, Gauri and {Salucci}, Paolo and {Harrison}, C.~M. and {van de Ven}, Glenn and {Lapi}, Andrea},
        title = "{Flat rotation curves of z {\ensuremath{\sim}} 1 star-forming galaxies}",
      journal = {\mnras},
         year = 2021,
        month = may,
       volume = {503},
       number = {2},
        pages = {1753-1772},
          doi = {10.1093/mnras/stab249},
archivePrefix = {arXiv},
       eprint = {2005.00279},
 primaryClass = {astro-ph.GA},
       adsurl = {https://ui.adsabs.harvard.edu/abs/2021MNRAS.503.1753S}
}

@article{Deg_SpekkensWALLABY, 
    title={WALLABY Pilot Survey: Public release of HI kinematic models for more than 100 galaxies from phase 1 of ASKAP pilot observations},
    volume={39}, 
    DOI={10.1017/pasa.2022.43}, 
    journal={Publications of the Astronomical Society of Australia}, author={Deg, N. and Spekkens, K. and Westmeier, T. and Reynolds, T. N. and Venkataraman, P. and Goliath, S. and Shen, A. X. and Halloran, R. and Bosma, A. and Catinella, B and et al.}, year={2022}, 
    pages={e059}
}

@ARTICLE{2017MNRAS.466.4159I,
       author = {{Iorio}, G. and {Fraternali}, F. and {Nipoti}, C. and {Di Teodoro}, E. and {Read}, J.~I. and {Battaglia}, G.},
        title = "{LITTLE THINGS in 3D: robust determination of the circular velocity of dwarf irregular galaxies}",
      journal = {\mnras},
         year = 2017,
        month = apr,
       volume = {466},
       number = {4},
        pages = {4159-4192},
          doi = {10.1093/mnras/stw3285},
archivePrefix = {arXiv},
       eprint = {1611.03865},
 primaryClass = {astro-ph.GA},
       adsurl = {https://ui.adsabs.harvard.edu/abs/2017MNRAS.466.4159I}
}

@ARTICLE{2017MNRAS.470.2410R,
       author = {{Rodrigues}, Davi C. and {del Popolo}, Antonino and {Marra}, Valerio and {de Oliveira}, Paulo L.~C.},
        title = "{Evidence against cuspy dark matter haloes in large galaxies}",
      journal = {\mnras},
         year = 2017,
        month = sep,
       volume = {470},
       number = {2},
        pages = {2410-2426},
          doi = {10.1093/mnras/stx1384},
archivePrefix = {arXiv},
       eprint = {1701.02698},
 primaryClass = {astro-ph.GA},
       adsurl = {https://ui.adsabs.harvard.edu/abs/2017MNRAS.470.2410R}
}

@ARTICLE{biswas2023,
       author = {{Biswas}, Prerana and {Kalinova}, Veselina and {Roy}, Nirupam and {Patra}, Narendra Nath and {Tyulneva}, Nadezda},
        title = "{The GMRT archive atomic gas survey - II. Mass modelling and dark matter halo properties across late-type spirals}",
      journal = {\mnras},
         year = 2023,
        month = oct,
       volume = {524},
       number = {4},
        pages = {6213-6228},
          doi = {10.1093/mnras/stad2285},
archivePrefix = {arXiv},
       eprint = {2307.13738},
 primaryClass = {astro-ph.GA},
       adsurl = {https://ui.adsabs.harvard.edu/abs/2023MNRAS.524.6213B}
}

@ARTICLE{biswas2022,
       author = {{Biswas}, Prerana and {Patra}, Narendra Nath and {Roy}, Nirupam and {Rashid}, Md},
        title = "{The GMRT archive atomic gas survey - I. Survey definition, methodology, and initial results from the pilot sample}",
      journal = {\mnras},
         year = 2022,
        month = jun,
       volume = {513},
       number = {1},
        pages = {168-185},
          doi = {10.1093/mnras/stac791},
archivePrefix = {arXiv},
       eprint = {2203.16584},
 primaryClass = {astro-ph.GA},
       adsurl = {https://ui.adsabs.harvard.edu/abs/2022MNRAS.513..168B}
}

@mastersthesis{Tyulneva2021,
  author  = {Tyulneva, N.},
  title   = {Title of the Thesis},
  school  = {Argelander-Institut für Astronomie, University of Bonn},
  year    = {2021},
  address = {Auf dem Hügel 71, D-53121 Bonn, Germany},
  type    = {Master's Thesis}
}

@ARTICLE{2017MNRAS.464.1903K,
       author = {{Kalinova}, Veselina and {van de Ven}, Glenn and {Lyubenova}, Mariya and {Falc{\'o}n-Barroso}, Jes{\'u}s and {Colombo}, Dario and {Rosolowsky}, Erik},
        title = "{The inner mass distribution of late-type spiral galaxies from <monospace>SAURON</monospace> stellar kinematic maps}",
      journal = {\mnras},
         year = 2017,
        month = jan,
       volume = {464},
       number = {2},
        pages = {1903-1922},
          doi = {10.1093/mnras/stw2448},
archivePrefix = {arXiv},
       eprint = {1609.08700},
 primaryClass = {astro-ph.GA},
       adsurl = {https://ui.adsabs.harvard.edu/abs/2017MNRAS.464.1903K}
}

@ARTICLE{2013PASP..125..306F,
       author = {{Foreman-Mackey}, Daniel and {Hogg}, David W. and {Lang}, Dustin and {Goodman}, Jonathan},
        title = "{emcee: The MCMC Hammer}",
      journal = {\pasp},
         year = 2013,
        month = mar,
       volume = {125},
       number = {925},
        pages = {306},
          doi = {10.1086/670067},
archivePrefix = {arXiv},
       eprint = {1202.3665},
 primaryClass = {astro-ph.IM},
       adsurl = {https://ui.adsabs.harvard.edu/abs/2013PASP..125..306F}
}

@ARTICLE{2014MNRAS.441.3359D,
       author = {{Dutton}, Aaron A. and {Macci{\`o}}, Andrea V.},
        title = "{Cold dark matter haloes in the Planck era: evolution of structural parameters for Einasto and NFW profiles}",
      journal = {\mnras},
         year = 2014,
        month = jul,
       volume = {441},
       number = {4},
        pages = {3359-3374},
          doi = {10.1093/mnras/stu742},
archivePrefix = {arXiv},
       eprint = {1402.7073},
 primaryClass = {astro-ph.CO},
       adsurl = {https://ui.adsabs.harvard.edu/abs/2014MNRAS.441.3359D}
}

@ARTICLE{2013MNRAS.428.3121M,
       author = {{Moster}, Benjamin P. and {Naab}, Thorsten and {White}, Simon D.~M.},
        title = "{Galactic star formation and accretion histories from matching galaxies to dark matter haloes}",
      journal = {\mnras},
         year = 2013,
        month = feb,
       volume = {428},
       number = {4},
        pages = {3121-3138},
          doi = {10.1093/mnras/sts261},
archivePrefix = {arXiv},
       eprint = {1205.5807},
 primaryClass = {astro-ph.CO},
       adsurl = {https://ui.adsabs.harvard.edu/abs/2013MNRAS.428.3121M}
}

@ARTICLE{2022ApJ...936L..11S,
       author = {{Spilker}, Justin S. and {Suess}, Katherine A. and {Setton}, David J. and {Bezanson}, Rachel and {Feldmann}, Robert and {Greene}, Jenny E. and {Kriek}, Mariska and {Lower}, Sidney and {Narayanan}, Desika and {Verrico}, Margaret},
        title = "{Star Formation Suppression by Tidal Removal of Cold Molecular Gas from an Intermediate-redshift Massive Post-starburst Galaxy}",
      journal = {\apjl},
         year = 2022,
        month = sep,
       volume = {936},
       number = {1},
          eid = {L11},
        pages = {L11},
          doi = {10.3847/2041-8213/ac75ea},
archivePrefix = {arXiv},
       eprint = {2208.13917},
 primaryClass = {astro-ph.GA},
       adsurl = {https://ui.adsabs.harvard.edu/abs/2022ApJ...936L..11S}
}

@ARTICLE{1994A&A...283..753D,
       author = {{Dottori}, H. and {Cepa}, J. and {Vilchez}, J. and {Barth}, C.~S.},
        title = "{CCD imaging of NGC 4861: morphology and brightness distribution.}",
      journal = {\aap},
         year = 1994,
        month = mar,
       volume = {283},
        pages = {753-758},
       adsurl = {https://ui.adsabs.harvard.edu/abs/1994A&A...283..753D}
}

@ARTICLE{2008MNRAS.391.1940M,
       author = {{Macci{\`o}}, Andrea V. and {Dutton}, Aaron A. and {van den Bosch}, Frank C.},
        title = "{Concentration, spin and shape of dark matter haloes as a function of the cosmological model: WMAP1, WMAP3 and WMAP5 results}",
      journal = {\mnras},
         year = 2008,
        month = dec,
       volume = {391},
       number = {4},
        pages = {1940-1954},
          doi = {10.1111/j.1365-2966.2008.14029.x},
archivePrefix = {arXiv},
       eprint = {0805.1926},
 primaryClass = {astro-ph},
       adsurl = {https://ui.adsabs.harvard.edu/abs/2008MNRAS.391.1940M}
}

@ARTICLE{2023MNRAS.518.6340D,
       author = {{Di Teodoro}, Enrico M. and {Posti}, Lorenzo and {Fall}, S. Michael and {Ogle}, Patrick M. and {Jarrett}, Thomas and {Appleton}, Philip N. and {Cluver}, Michelle E. and {Haynes}, Martha P. and {Lisenfeld}, Ute},
        title = "{Dark matter halos and scaling relations of extremely massive spiral galaxies from extended H I rotation curves}",
      journal = {\mnras},
         year = 2023,
        month = feb,
       volume = {518},
       number = {4},
        pages = {6340-6354},
          doi = {10.1093/mnras/stac3424},
archivePrefix = {arXiv},
       eprint = {2207.02906},
 primaryClass = {astro-ph.GA},
       adsurl = {https://ui.adsabs.harvard.edu/abs/2023MNRAS.518.6340D}
}

@ARTICLE{2022MNRAS.514.3329M,
       author = {{Mancera Pi{\~n}a}, Pavel E. and {Fraternali}, Filippo and {Oosterloo}, Tom and {Adams}, Elizabeth A.~K. and {di Teodoro}, Enrico and {Bacchini}, Cecilia and {Iorio}, Giuliano},
        title = "{The impact of gas disc flaring on rotation curve decomposition and revisiting baryonic and dark matter relations for nearby galaxies}",
      journal = {\mnras},
         year = 2022,
        month = aug,
       volume = {514},
       number = {3},
        pages = {3329-3348},
          doi = {10.1093/mnras/stac1508},
archivePrefix = {arXiv},
       eprint = {2205.12977},
 primaryClass = {astro-ph.GA},
       adsurl = {https://ui.adsabs.harvard.edu/abs/2022MNRAS.514.3329M}
}

@ARTICLE{2006MNRAS.373.1117H,
       author = {{Hayashi}, Eric and {Navarro}, Julio F.},
        title = "{Hiding cusps in cores: kinematics of disc galaxies in triaxial dark matter haloes}",
      journal = {\mnras},
         year = 2006,
        month = dec,
       volume = {373},
       number = {3},
        pages = {1117-1124},
          doi = {10.1111/j.1365-2966.2006.10927.x},
archivePrefix = {arXiv},
       eprint = {astro-ph/0608376},
 primaryClass = {astro-ph},
       adsurl = {https://ui.adsabs.harvard.edu/abs/2006MNRAS.373.1117H}
}

@ARTICLE{2008AJ....136.2648D,
       author = {{de Blok}, W.~J.~G. and {Walter}, F. and {Brinks}, E. and {Trachternach}, C. and {Oh}, S. -H. and {Kennicutt}, Jr., R.~C.},
        title = "{High-Resolution Rotation Curves and Galaxy Mass Models from THINGS}",
      journal = {\aj},
         year = 2008,
        month = dec,
       volume = {136},
       number = {6},
        pages = {2648-2719},
          doi = {10.1088/0004-6256/136/6/2648},
archivePrefix = {arXiv},
       eprint = {0810.2100},
 primaryClass = {astro-ph},
       adsurl = {https://ui.adsabs.harvard.edu/abs/2008AJ....136.2648D}
}

@ARTICLE{2011AJ....141..193O,
       author = {{Oh}, Se-Heon and {de Blok}, W.~J.~G. and {Brinks}, Elias and {Walter}, Fabian and {Kennicutt}, Jr., Robert C.},
        title = "{Dark and Luminous Matter in THINGS Dwarf Galaxies}",
      journal = {\aj},
         year = 2011,
        month = jun,
       volume = {141},
       number = {6},
          eid = {193},
        pages = {193},
          doi = {10.1088/0004-6256/141/6/193},
archivePrefix = {arXiv},
       eprint = {1011.0899},
 primaryClass = {astro-ph.CO},
       adsurl = {https://ui.adsabs.harvard.edu/abs/2011AJ....141..193O}
}

@ARTICLE{2016AJ....152..157L,
       author = {{Lelli}, Federico and {McGaugh}, Stacy S. and {Schombert}, James M.},
        title = "{SPARC: Mass Models for 175 Disk Galaxies with Spitzer Photometry and Accurate Rotation Curves}",
      journal = {\aj},
         year = 2016,
        month = dec,
       volume = {152},
       number = {6},
          eid = {157},
        pages = {157},
          doi = {10.3847/0004-6256/152/6/157},
archivePrefix = {arXiv},
       eprint = {1606.09251},
 primaryClass = {astro-ph.GA},
       adsurl = {https://ui.adsabs.harvard.edu/abs/2016AJ....152..157L}
}

@ARTICLE{2016ApJ...827L..19L,
       author = {{Lelli}, Federico and {McGaugh}, Stacy S. and {Schombert}, James M. and {Pawlowski}, Marcel S.},
        title = "{The Relation between Stellar and Dynamical Surface Densities in the Central Regions of Disk Galaxies}",
      journal = {\apjl},
         year = 2016,
        month = aug,
       volume = {827},
       number = {1},
          eid = {L19},
        pages = {L19},
          doi = {10.3847/2041-8205/827/1/L19},
archivePrefix = {arXiv},
       eprint = {1607.02145},
 primaryClass = {astro-ph.GA},
       adsurl = {https://ui.adsabs.harvard.edu/abs/2016ApJ...827L..19L}
}

@ARTICLE{2019MNRAS.484.3267L,
       author = {{Lelli}, Federico and {McGaugh}, Stacy S. and {Schombert}, James M. and {Desmond}, Harry and {Katz}, Harley},
        title = "{The baryonic Tully-Fisher relation for different velocity definitions and implications for galaxy angular momentum}",
      journal = {\mnras},
         year = 2019,
        month = apr,
       volume = {484},
       number = {3},
        pages = {3267-3278},
          doi = {10.1093/mnras/stz205},
archivePrefix = {arXiv},
       eprint = {1901.05966},
 primaryClass = {astro-ph.GA},
       adsurl = {https://ui.adsabs.harvard.edu/abs/2019MNRAS.484.3267L}
}

@ARTICLE{2019ApJ...886L..11L,
       author = {{Li}, Pengfei and {Lelli}, Federico and {McGaugh}, Stacy and {Pawlowski}, Marcel S. and {Zwaan}, Martin A. and {Schombert}, James},
        title = "{The Halo Mass Function of Late-type Galaxies from H I Kinematics}",
      journal = {\apjl},
         year = 2019,
        month = nov,
       volume = {886},
       number = {1},
          eid = {L11},
        pages = {L11},
          doi = {10.3847/2041-8213/ab53e6},
archivePrefix = {arXiv},
       eprint = {1911.00517},
 primaryClass = {astro-ph.GA},
       adsurl = {https://ui.adsabs.harvard.edu/abs/2019ApJ...886L..11L}
}

@ARTICLE{2020ApJS..247...31L,
       author = {{Li}, Pengfei and {Lelli}, Federico and {McGaugh}, Stacy and {Schombert}, James},
        title = "{A Comprehensive Catalog of Dark Matter Halo Models for SPARC Galaxies}",
      journal = {\apjs},
         year = 2020,
        month = mar,
       volume = {247},
       number = {1},
          eid = {31},
        pages = {31},
          doi = {10.3847/1538-4365/ab700e},
archivePrefix = {arXiv},
       eprint = {2001.10538},
 primaryClass = {astro-ph.GA},
       adsurl = {https://ui.adsabs.harvard.edu/abs/2020ApJS..247...31L}
}

@ARTICLE{2017MNRAS.466.1648K,
       author = {{Katz}, Harley and {Lelli}, Federico and {McGaugh}, Stacy S. and {Di Cintio}, Arianna and {Brook}, Chris B. and {Schombert}, James M.},
        title = "{Testing feedback-modified dark matter haloes with galaxy rotation curves: estimation of halo parameters and consistency with {\ensuremath{\Lambda}}CDM scaling relations}",
      journal = {\mnras},
         year = 2017,
        month = apr,
       volume = {466},
       number = {2},
        pages = {1648-1668},
          doi = {10.1093/mnras/stw3101},
archivePrefix = {arXiv},
       eprint = {1605.05971},
 primaryClass = {astro-ph.GA},
       adsurl = {https://ui.adsabs.harvard.edu/abs/2017MNRAS.466.1648K}
}

@ARTICLE{2014A&A...570A..13M,
       author = {{Makarov}, Dmitry and {Prugniel}, Philippe and {Terekhova}, Nataliya and {Courtois}, H{\'e}l{\`e}ne and {Vauglin}, Isabelle},
        title = "{HyperLEDA. III. The catalogue of extragalactic distances}",
      journal = {\aap},
         year = 2014,
        month = oct,
       volume = {570},
          eid = {A13},
        pages = {A13},
          doi = {10.1051/0004-6361/201423496},
archivePrefix = {arXiv},
       eprint = {1408.3476},
 primaryClass = {astro-ph.GA},
       adsurl = {https://ui.adsabs.harvard.edu/abs/2014A&A...570A..13M}
}

@ARTICLE{2014AJ....148...77M,
       author = {{McGaugh}, Stacy S. and {Schombert}, James M.},
        title = "{Color-Mass-to-light-ratio Relations for Disk Galaxies}",
      journal = {\aj},
         year = 2014,
        month = nov,
       volume = {148},
       number = {5},
          eid = {77},
        pages = {77},
          doi = {10.1088/0004-6256/148/5/77},
archivePrefix = {arXiv},
       eprint = {1407.1839},
 primaryClass = {astro-ph.GA},
       adsurl = {https://ui.adsabs.harvard.edu/abs/2014AJ....148...77M}
}

@ARTICLE{2014PASA...31...36S,
       author = {{Schombert}, James and {McGaugh}, Stacy},
        title = "{Stellar Populations and the Star Formation Histories of LSB Galaxies: III. Stellar Population Models}",
      journal = {\pasa},
         year = 2014,
        month = sep,
       volume = {31},
          eid = {e036},
        pages = {e036},
          doi = {10.1017/pasa.2014.32},
archivePrefix = {arXiv},
       eprint = {1407.6778},
 primaryClass = {astro-ph.GA},
       adsurl = {https://ui.adsabs.harvard.edu/abs/2014PASA...31...36S}
}

@ARTICLE{2014ApJ...788..144M,
       author = {{Meidt}, Sharon E. and {Schinnerer}, Eva and {van de Ven}, Glenn and {Zaritsky}, Dennis and {Peletier}, Reynier and {Knapen}, Johan H. and {Sheth}, Kartik and {Regan}, Michael and {Querejeta}, Miguel and {Mu{\~n}oz-Mateos}, Juan-Carlos and {Kim}, Taehyun and {Hinz}, Joannah L. and {Gil de Paz}, Armando and {Athanassoula}, E. and {Bosma}, Albert and {Buta}, Ronald J. and {Cisternas}, Mauricio and {Ho}, Luis C. and {Holwerda}, Benne and {Skibba}, Ramin and {Laurikainen}, E. and {Salo}, H. and {Gadotti}, D.~A. and {Laine}, Jarkko and {Erroz-Ferrer}, S. and {Comer{\'o}n}, S{\'e}bastien and {Men{\'e}ndez-Delmestre}, K. and {Seibert}, M. and {Mizusawa}, T.},
        title = "{Reconstructing the Stellar Mass Distributions of Galaxies Using S$^{4}$G IRAC 3.6 and 4.5 {\ensuremath{\mu}}m Images. II. The Conversion from Light to Mass}",
      journal = {\apj},
         year = 2014,
        month = jun,
       volume = {788},
       number = {2},
          eid = {144},
        pages = {144},
          doi = {10.1088/0004-637X/788/2/144},
archivePrefix = {arXiv},
       eprint = {1402.5210},
 primaryClass = {astro-ph.GA},
       adsurl = {https://ui.adsabs.harvard.edu/abs/2014ApJ...788..144M}
}

@ARTICLE{2016ApJ...832..198N,
       author = {{Norris}, Mark A. and {Van de Ven}, Glenn and {Schinnerer}, Eva and {Crain}, Robert A. and {Meidt}, Sharon and {Groves}, Brent and {Bower}, Richard G. and {Furlong}, Michelle and {Schaller}, Matthieu and {Schaye}, Joop and {Theuns}, Tom},
        title = "{Being WISE II: Reducing the Influence of Star formation History on the Mass-to-Light Ratio of Quiescent Galaxies}",
      journal = {\apj},
         year = 2016,
        month = dec,
       volume = {832},
       number = {2},
          eid = {198},
        pages = {198},
          doi = {10.3847/0004-637X/832/2/198},
archivePrefix = {arXiv},
       eprint = {1608.07584},
 primaryClass = {astro-ph.GA},
       adsurl = {https://ui.adsabs.harvard.edu/abs/2016ApJ...832..198N}
}

@article{10.1093/mnras/sty3223,
    author = {Schombert, James and McGaugh, Stacy and Lelli, Federico},
    title = {The mass-to-light ratios and the star formation histories of disc galaxies},
    journal = {Monthly Notices of the Royal Astronomical Society},
    volume = {483},
    number = {2},
    pages = {1496-1512},
    year = {2018},
    month = {12},
    issn = {0035-8711},
    doi = {10.1093/mnras/sty3223},
    url = {https://doi.org/10.1093/mnras/sty3223},
    eprint = {https://academic.oup.com/mnras/article-pdf/483/2/1496/27089556/sty3223.pdf},
}

@ARTICLE{2001ApJ...550..212B,
       author = {{Bell}, Eric F. and {de Jong}, Roelof S.},
        title = "2003{Stellar Mass-to-Light Ratios and the Tully-Fisher Relation}",
      journal = {\apj},
         year = 2001,
        month = mar,
       volume = {550},
       number = {1},
        pages = {212-229},
          doi = {10.1086/319728},
archivePrefix = {arXiv},
       eprint = {astro-ph/0011493},
 primaryClass = {astro-ph},
       adsurl = {https://ui.adsabs.harvard.edu/abs/2001ApJ...550..212B}
}

@ARTICLE{2003ApJS..149..289B,
       author = {{Bell}, Eric F. and {McIntosh}, Daniel H. and {Katz}, Neal and {Weinberg}, Martin D.},
        title = "{The Optical and Near-Infrared Properties of Galaxies. I. Luminosity and Stellar Mass Functions}",
      journal = {\apjs},
         year = 2003,
        month = dec,
       volume = {149},
       number = {2},
        pages = {289-312},
          doi = {10.1086/378847},
archivePrefix = {arXiv},
       eprint = {astro-ph/0302543},
 primaryClass = {astro-ph},
       adsurl = {https://ui.adsabs.harvard.edu/abs/2003ApJS..149..289B}
}

@ARTICLE{2009MNRAS.400.1181Z,
       author = {{Zibetti}, Stefano and {Charlot}, St{\'e}phane and {Rix}, Hans-Walter},
        title = "{Resolved stellar mass maps of galaxies - I. Method and implications for global mass estimates}",
      journal = {\mnras},
         year = 2009,
        month = dec,
       volume = {400},
       number = {3},
        pages = {1181-1198},
          doi = {10.1111/j.1365-2966.2009.15528.x},
archivePrefix = {arXiv},
       eprint = {0904.4252},
 primaryClass = {astro-ph.CO},
       adsurl = {https://ui.adsabs.harvard.edu/abs/2009MNRAS.400.1181Z}
}

@ARTICLE{2006ApJS..164...52S,
       author = {{Schmitt}, H.~R. and {Calzetti}, D. and {Armus}, L. and {Giavalisco}, M. and {Heckman}, T.~M. and {Kennicutt}, Jr., R.~C. and {Leitherer}, C. and {Meurer}, G.~R.},
        title = "{Multiwavelength Star Formation Indicators: Observations}",
      journal = {\apjs},
         year = 2006,
        month = may,
       volume = {164},
       number = {1},
        pages = {52-80},
          doi = {10.1086/501529},
archivePrefix = {arXiv},
       eprint = {astro-ph/0602063},
 primaryClass = {astro-ph},
       adsurl = {https://ui.adsabs.harvard.edu/abs/2006ApJS..164...52S}
}

@ARTICLE{2004MNRAS.355..728F,
       author = {{Fernandes}, I.~F. and {de Carvalho}, R. and {Contini}, T. and {Gal}, R.~R.},
        title = "{Massive star populations in Wolf-Rayet galaxies}",
      journal = {\mnras},
         year = 2004,
        month = dec,
       volume = {355},
       number = {3},
        pages = {728-746},
          doi = {10.1111/j.1365-2966.2004.08352.x},
archivePrefix = {arXiv},
       eprint = {astro-ph/0409114},
 primaryClass = {astro-ph},
       adsurl = {https://ui.adsabs.harvard.edu/abs/2004MNRAS.355..728F}
}

@ARTICLE{2023ApJS..267...44A,
       author = {{Almeida}, Andr{\'e}s and {Anderson}, Scott F. and {Argudo-Fern{\'a}ndez}, Maria and {Badenes}, Carles and {Barger}, Kat and {Barrera-Ballesteros}, Jorge K. and {Bender}, Chad F. and {Benitez}, Erika and {Besser}, Felipe and {Bird}, Jonathan C. and {Bizyaev}, Dmitry and {Blanton}, Michael R. and {Bochanski}, John and {Bovy}, Jo and {Brandt}, William Nielsen and {Brownstein}, Joel R. and {Buchner}, Johannes and {Bulbul}, Esra and {Burchett}, Joseph N. and {Cano D{\'\i}az}, Mariana and {Carlberg}, Joleen K. and {Casey}, Andrew R. and {Chandra}, Vedant and {Cherinka}, Brian and {Chiappini}, Cristina and {Coker}, Abigail A. and {Comparat}, Johan and {Conroy}, Charlie and {Contardo}, Gabriella and {Cortes}, Arlin and {Covey}, Kevin and {Crane}, Jeffrey D. and {Cunha}, Katia and {Dabbieri}, Collin and {Davidson}, James W. and {Davis}, Megan C. and {de Andrade Queiroz}, Anna Barbara and {De Lee}, Nathan and {M{\'e}ndez Delgado}, Jos{\'e} Eduardo and {Demasi}, Sebastian and {Di Mille}, Francesco and {Donor}, John and {Dow}, Peter and {Dwelly}, Tom and {Eracleous}, Mike and {Eriksen}, Jamey and {Fan}, Xiaohui and {Farr}, Emily and {Frederick}, Sara and {Fries}, Logan and {Frinchaboy}, Peter and {G{\"a}nsicke}, Boris T. and {Ge}, Junqiang and {Gonz{\'a}lez {\'A}vila}, Consuelo and {Grabowski}, Katie and {Grier}, Catherine and {Guiglion}, Guillaume and {Gupta}, Pramod and {Hall}, Patrick and {Hawkins}, Keith and {Hayes}, Christian R. and {Hermes}, J.~J. and {Hern{\'a}ndez-Garc{\'\i}a}, Lorena and {Hogg}, David W. and {Holtzman}, Jon A. and {Ibarra-Medel}, Hector Javier and {Ji}, Alexander and {Jofre}, Paula and {Johnson}, Jennifer A. and {Jones}, Amy M. and {Kinemuchi}, Karen and {Kluge}, Matthias and {Koekemoer}, Anton and {Kollmeier}, Juna A. and {Kounkel}, Marina and {Krishnarao}, Dhanesh and {Krumpe}, Mirko and {Lacerna}, Ivan and {Lago}, Paulo Jakson Assuncao and {Laporte}, Chervin and {Liu}, Chao and {Liu}, Ang and {Liu}, Xin and {Lopes}, Alexandre Roman and {Macktoobian}, Matin and {Majewski}, Steven R. and {Malanushenko}, Viktor and {Maoz}, Dan and {Masseron}, Thomas and {Masters}, Karen L. and {Matijevic}, Gal and {McBride}, Aidan and {Medan}, Ilija and {Merloni}, Andrea and {Morrison}, Sean and {Myers}, Natalie and {M{\'e}sz{\'a}ros}, Szabolcs and {Negrete}, C. Alenka and {Nidever}, David L. and {Nitschelm}, Christian and {Oravetz}, Daniel and {Oravetz}, Audrey and {Pan}, Kaike and {Peng}, Yingjie and {Pinsonneault}, Marc H. and {Pogge}, Rick and {Qiu}, Dan and {Ramirez}, Solange V. and {Rix}, Hans-Walter and {Fern{\'a}ndez Rosso}, Daniela and {Runnoe}, Jessie and {Salvato}, Mara and {Sanchez}, Sebastian F. and {Santana}, Felipe A. and {Saydjari}, Andrew and {Sayres}, Conor and {Schlaufman}, Kevin C. and {Schneider}, Donald P. and {Schwope}, Axel and {Serna}, Javier and {Shen}, Yue and {Sobeck}, Jennifer and {Song}, Ying-Yi and {Souto}, Diogo and {Spoo}, Taylor and {Stassun}, Keivan G. and {Steinmetz}, Matthias and {Straumit}, Ilya and {Stringfellow}, Guy and {S{\'a}nchez-Gallego}, Jos{\'e} and {Taghizadeh-Popp}, Manuchehr and {Tayar}, Jamie and {Thakar}, Ani and {Tissera}, Patricia B. and {Tkachenko}, Andrew and {Hernandez Toledo}, Hector and {Trakhtenbrot}, Benny and {Fern{\'a}ndez-Trincado}, Jos{\'e} G. and {Troup}, Nicholas and {Trump}, Jonathan R. and {Tuttle}, Sarah and {Ulloa}, Natalie and {Vazquez-Mata}, Jose Antonio and {Vera Alfaro}, Pablo and {Villanova}, Sandro and {Wachter}, Stefanie and {Weijmans}, Anne-Marie and {Wheeler}, Adam and {Wilson}, John and {Wojno}, Leigh and {Wolf}, Julien and {Xue}, Xiang-Xiang and {Ybarra}, Jason E. and {Zari}, Eleonora and {Zasowski}, Gail},
        title = "{The Eighteenth Data Release of the Sloan Digital Sky Surveys: Targeting and First Spectra from SDSS-V}",
      journal = {\apjs},
         year = 2023,
        month = aug,
       volume = {267},
       number = {2},
          eid = {44},
        pages = {44},
          doi = {10.3847/1538-4365/acda98},
archivePrefix = {arXiv},
       eprint = {2301.07688},
 primaryClass = {astro-ph.GA},
       adsurl = {https://ui.adsabs.harvard.edu/abs/2023ApJS..267...44A}
}

@ARTICLE{2010ApJ...717..379B,
       author = {{Behroozi}, Peter S. and {Conroy}, Charlie and {Wechsler}, Risa H.},
        title = "{A Comprehensive Analysis of Uncertainties Affecting the Stellar Mass-Halo Mass Relation for 0 < z < 4}",
      journal = {\apj},
         year = 2010,
        month = jul,
       volume = {717},
       number = {1},
        pages = {379-403},
          doi = {10.1088/0004-637X/717/1/379},
archivePrefix = {arXiv},
       eprint = {1001.0015},
 primaryClass = {astro-ph.CO},
       adsurl = {https://ui.adsabs.harvard.edu/abs/2010ApJ...717..379B}
}

@ARTICLE{2010ApJ...710..903M,
       author = {{Moster}, Benjamin P. and {Somerville}, Rachel S. and {Maulbetsch}, Christian and {van den Bosch}, Frank C. and {Macci{\`o}}, Andrea V. and {Naab}, Thorsten and {Oser}, Ludwig},
        title = "{Constraints on the Relationship between Stellar Mass and Halo Mass at Low and High Redshift}",
      journal = {\apj},
         year = 2010,
        month = feb,
       volume = {710},
       number = {2},
        pages = {903-923},
          doi = {10.1088/0004-637X/710/2/903},
archivePrefix = {arXiv},
       eprint = {0903.4682},
 primaryClass = {astro-ph.CO},
       adsurl = {https://ui.adsabs.harvard.edu/abs/2010ApJ...710..903M}
}

@ARTICLE{2025A&A...695L..23M,
       author = {{Marasco}, A. and {Fall}, S.~M. and {Di Teodoro}, E.~M. and {Mancera Pi{\~n}a}, P.~E.},
        title = "{Photometric versus dynamical stellar masses and their impact on scaling relations in nearby disc galaxies}",
      journal = {\aap},
         year = 2025,
        month = mar,
       volume = {695},
          eid = {L23},
        pages = {L23},
          doi = {10.1051/0004-6361/202553925},
archivePrefix = {arXiv},
       eprint = {2502.19472},
 primaryClass = {astro-ph.GA},
       adsurl = {https://ui.adsabs.harvard.edu/abs/2025A&A...695L..23M}
}

@BOOK{2008gady.book.....B,
       author = {{Binney}, James and {Tremaine}, Scott},
        title = "{Galactic Dynamics: Second Edition}",
         year = 2008,
       adsurl = {https://ui.adsabs.harvard.edu/abs/2008gady.book.....B}
}

@ARTICLE{2021A&A...649A.119P,
       author = {{Posti}, L. and {Fall}, S.~M.},
        title = "{Dynamical evidence for a morphology-dependent relation between the stellar and halo masses of galaxies}",
      journal = {\aap},
         year = 2021,
        month = may,
       volume = {649},
          eid = {A119},
        pages = {A119},
          doi = {10.1051/0004-6361/202040256},
archivePrefix = {arXiv},
       eprint = {2102.11282},
 primaryClass = {astro-ph.GA},
       adsurl = {https://ui.adsabs.harvard.edu/abs/2021A&A...649A.119P}
}

@ARTICLE{2019A&A...626A..56P,
       author = {{Posti}, Lorenzo and {Fraternali}, Filippo and {Marasco}, Antonino},
        title = "{Peak star formation efficiency and no missing baryons in massive spirals}",
      journal = {\aap},
         year = 2019,
        month = jun,
       volume = {626},
          eid = {A56},
        pages = {A56},
          doi = {10.1051/0004-6361/201935553},
archivePrefix = {arXiv},
       eprint = {1812.05099},
 primaryClass = {astro-ph.GA},
       adsurl = {https://ui.adsabs.harvard.edu/abs/2019A&A...626A..56P}
}

@ARTICLE{2017MNRAS.467.2019R,
       author = {{Read}, J.~I. and {Iorio}, G. and {Agertz}, O. and {Fraternali}, F.},
        title = "{The stellar mass-halo mass relation of isolated field dwarfs: a critical test of {\ensuremath{\Lambda}}CDM at the edge of galaxy formation}",
      journal = {\mnras},
         year = 2017,
        month = may,
       volume = {467},
       number = {2},
        pages = {2019-2038},
          doi = {10.1093/mnras/stx147},
archivePrefix = {arXiv},
       eprint = {1607.03127},
 primaryClass = {astro-ph.GA},
       adsurl = {https://ui.adsabs.harvard.edu/abs/2017MNRAS.467.2019R}
}

@INPROCEEDINGS{2024eas..conf.1858B,
       author = {{Ba{\~n}ares Hern{\'a}ndez}, Andr{\'e}s and {Mart{\'\i}n Camalich}, Jorge and {Castillo}, Andr{\'e}s and {Iorio}, Giuliano and {Mancera-Pi{\~n}a}, Pavel and {di Cintio}, Arianna},
        title = "{Confronting Dark Matter Models with Nearby Dwarf Irregular Galaxies}",
    booktitle = {EAS2024, European Astronomical Society Annual Meeting},
         year = 2024,
        month = jul,
          eid = {1858},
        pages = {1858},
       adsurl = {https://ui.adsabs.harvard.edu/abs/2024eas..conf.1858B}
}

@article{ refId0,
	author = {{Bañares-Hernández, Andrés} and {Castillo, Andrés} and {Martin Camalich, Jorge} and {Iorio, Giuliano}},
	title = {Confronting fuzzy dark matter with the rotation curves of nearby dwarf irregular galaxies},
	DOI= "10.1051/0004-6361/202346686",
	url= "https://doi.org/10.1051/0004-6361/202346686",
	journal = {A&A},
	year = 2023,
	volume = 676,
	pages = "A63",
}

@ARTICLE{2001MNRAS.321..559B,
       author = {{Bullock}, J.~S. and {Kolatt}, T.~S. and {Sigad}, Y. and {Somerville}, R.~S. and {Kravtsov}, A.~V. and {Klypin}, A.~A. and {Primack}, J.~R. and {Dekel}, A.},
        title = "{Profiles of dark haloes: evolution, scatter and environment}",
      journal = {\mnras},
         year = 2001,
        month = mar,
       volume = {321},
       number = {3},
        pages = {559-575},
          doi = {10.1046/j.1365-8711.2001.04068.x},
archivePrefix = {arXiv},
       eprint = {astro-ph/9908159},
 primaryClass = {astro-ph},
       adsurl = {https://ui.adsabs.harvard.edu/abs/2001MNRAS.321..559B}
}

@ARTICLE{2003ApJ...597L...9Z,
       author = {{Zhao}, D.~H. and {Jing}, Y.~P. and {Mo}, H.~J. and {B{\"o}rner}, G.},
        title = "{Mass and Redshift Dependence of Dark Halo Structure}",
      journal = {\apjl},
         year = 2003,
        month = nov,
       volume = {597},
       number = {1},
        pages = {L9-L12},
          doi = {10.1086/379734},
archivePrefix = {arXiv},
       eprint = {astro-ph/0309375},
 primaryClass = {astro-ph},
       adsurl = {https://ui.adsabs.harvard.edu/abs/2003ApJ...597L...9Z}
}

@ARTICLE{2006MNRAS.368.1931L,
       author = {{Lu}, Yu and {Mo}, H.~J. and {Katz}, Neal and {Weinberg}, Martin D.},
        title = "{On the origin of cold dark matter halo density profiles}",
      journal = {\mnras},
         year = 2006,
        month = jun,
       volume = {368},
       number = {4},
        pages = {1931-1940},
          doi = {10.1111/j.1365-2966.2006.10270.x},
archivePrefix = {arXiv},
       eprint = {astro-ph/0508624},
 primaryClass = {astro-ph},
       adsurl = {https://ui.adsabs.harvard.edu/abs/2006MNRAS.368.1931L}
}

@ARTICLE{2016MNRAS.457.4340K,
       author = {{Klypin}, Anatoly and {Yepes}, Gustavo and {Gottl{\"o}ber}, Stefan and {Prada}, Francisco and {He{\ss}}, Steffen},
        title = "{MultiDark simulations: the story of dark matter halo concentrations and density profiles}",
      journal = {\mnras},
         year = 2016,
        month = apr,
       volume = {457},
       number = {4},
        pages = {4340-4359},
          doi = {10.1093/mnras/stw248},
archivePrefix = {arXiv},
       eprint = {1411.4001},
 primaryClass = {astro-ph.CO},
       adsurl = {https://ui.adsabs.harvard.edu/abs/2016MNRAS.457.4340K}
}

@ARTICLE{2015MNRAS.452.1217C,
       author = {{Correa}, Camila A. and {Wyithe}, J. Stuart B. and {Schaye}, Joop and {Duffy}, Alan R.},
        title = "{The accretion history of dark matter haloes - III. A physical model for the concentration-mass relation}",
      journal = {\mnras},
         year = 2015,
        month = sep,
       volume = {452},
       number = {2},
        pages = {1217-1232},
          doi = {10.1093/mnras/stv1363},
archivePrefix = {arXiv},
       eprint = {1502.00391},
 primaryClass = {astro-ph.CO},
       adsurl = {https://ui.adsabs.harvard.edu/abs/2015MNRAS.452.1217C}
}

@ARTICLE{2015ApJ...799..108D,
       author = {{Diemer}, Benedikt and {Kravtsov}, Andrey V.},
        title = "{A Universal Model for Halo Concentrations}",
      journal = {\apj},
         year = 2015,
        month = jan,
       volume = {799},
       number = {1},
          eid = {108},
        pages = {108},
          doi = {10.1088/0004-637X/799/1/108},
archivePrefix = {arXiv},
       eprint = {1407.4730},
 primaryClass = {astro-ph.CO},
       adsurl = {https://ui.adsabs.harvard.edu/abs/2015ApJ...799..108D}
}

@ARTICLE{1997PASA...14...11S,
       author = {{Sackett}, Penny D.},
        title = "{Structure of Dark Halos: Model-independent Information from HI Rotation Curves}",
      journal = {\pasa},
         year = 1997,
        month = apr,
       volume = {14},
       number = {1},
        pages = {11-14},
          doi = {10.1071/AS97011},
       adsurl = {https://ui.adsabs.harvard.edu/abs/1997PASA...14...11S}
}

@ARTICLE{2013MNRAS.429.3068T,
       author = {{Teyssier}, Romain and {Pontzen}, Andrew and {Dubois}, Yohan and {Read}, Justin I.},
        title = "{Cusp-core transformations in dwarf galaxies: observational predictions}",
      journal = {\mnras},
         year = 2013,
        month = mar,
       volume = {429},
       number = {4},
        pages = {3068-3078},
          doi = {10.1093/mnras/sts563},
archivePrefix = {arXiv},
       eprint = {1206.4895},
 primaryClass = {astro-ph.CO},
       adsurl = {https://ui.adsabs.harvard.edu/abs/2013MNRAS.429.3068T}
}

@ARTICLE{2019MNRAS.488.2387B,
       author = {{Ben{\'\i}tez-Llambay}, Alejandro and {Frenk}, Carlos S. and {Ludlow}, Aaron D. and {Navarro}, Julio F.},
        title = "{Baryon-induced dark matter cores in the EAGLE simulations}",
      journal = {\mnras},
         year = 2019,
        month = sep,
       volume = {488},
       number = {2},
        pages = {2387-2404},
          doi = {10.1093/mnras/stz1890},
archivePrefix = {arXiv},
       eprint = {1810.04186},
 primaryClass = {astro-ph.GA},
       adsurl = {https://ui.adsabs.harvard.edu/abs/2019MNRAS.488.2387B}
}

@ARTICLE{2024MNRAS.528..693O,
       author = {{Ou}, Xiaowei and {Eilers}, Anna-Christina and {Necib}, Lina and {Frebel}, Anna},
        title = "{The dark matter profile of the Milky Way inferred from its circular velocity curve}",
      journal = {\mnras},
         year = 2024,
        month = feb,
       volume = {528},
       number = {1},
        pages = {693-710},
          doi = {10.1093/mnras/stae034},
archivePrefix = {arXiv},
       eprint = {2303.12838},
 primaryClass = {astro-ph.GA},
       adsurl = {https://ui.adsabs.harvard.edu/abs/2024MNRAS.528..693O}
}

@ARTICLE{2003MNRAS.339..243J,
       author = {{Jimenez}, Raul and {Verde}, Licia and {Oh}, S. Peng},
        title = "{Dark halo properties from rotation curves}",
      journal = {\mnras},
         year = 2003,
        month = feb,
       volume = {339},
       number = {1},
        pages = {243-259},
          doi = {10.1046/j.1365-8711.2003.06165.x},
archivePrefix = {arXiv},
       eprint = {astro-ph/0201352},
 primaryClass = {astro-ph},
       adsurl = {https://ui.adsabs.harvard.edu/abs/2003MNRAS.339..243J}
}

@ARTICLE{2003MNRAS.341...33K,
       author = {{Kauffmann}, Guinevere and {Heckman}, Timothy M. and {White}, Simon D.~M. and {Charlot}, St{\'e}phane and {Tremonti}, Christy and {Brinchmann}, Jarle and {Bruzual}, Gustavo and {Peng}, Eric W. and {Seibert}, Mark and {Bernardi}, Mariangela and {Blanton}, Michael and {Brinkmann}, Jon and {Castander}, Francisco and {Cs{\'a}bai}, Istvan and {Fukugita}, Masataka and {Ivezic}, Zeljko and {Munn}, Jeffrey A. and {Nichol}, Robert C. and {Padmanabhan}, Nikhil and {Thakar}, Aniruddha R. and {Weinberg}, David H. and {York}, Donald},
        title = "{Stellar masses and star formation histories for {}10$^{5}$ galaxies from the Sloan Digital Sky Survey}",
      journal = {\mnras},
         year = 2003,
        month = may,
       volume = {341},
       number = {1},
        pages = {33-53},
          doi = {10.1046/j.1365-8711.2003.06291.x},
archivePrefix = {arXiv},
       eprint = {astro-ph/0204055},
 primaryClass = {astro-ph},
       adsurl = {https://ui.adsabs.harvard.edu/abs/2003MNRAS.341...33K}
}

@article{10.1093/mnras/sty2968,
    author = {Li, Pengfei and Lelli, Federico and McGaugh, Stacy S and Starkman, Nathaniel and Schombert, James M},
    title = {A constant characteristic volume density of dark matter haloes from SPARC rotation curve fits},
    journal = {Monthly Notices of the Royal Astronomical Society},
    volume = {482},
    number = {4},
    pages = {5106-5124},
    year = {2018},
    month = {11},
    issn = {0035-8711},
    doi = {10.1093/mnras/sty2968},
    url = {https://doi.org/10.1093/mnras/sty2968},
    eprint = {https://academic.oup.com/mnras/article-pdf/482/4/5106/26899055/sty2968.pdf},
}

@ARTICLE{2016MNRAS.460.1214L,
       author = {{Ludlow}, Aaron D. and {Bose}, Sownak and {Angulo}, Ra{\'u}l E. and {Wang}, Lan and {Hellwing}, Wojciech A. and {Navarro}, Julio F. and {Cole}, Shaun and {Frenk}, Carlos S.},
        title = "{The mass-concentration-redshift relation of cold and warm dark matter haloes}",
      journal = {\mnras},
         year = 2016,
        month = aug,
       volume = {460},
       number = {2},
        pages = {1214-1232},
          doi = {10.1093/mnras/stw1046},
archivePrefix = {arXiv},
       eprint = {1601.02624},
 primaryClass = {astro-ph.CO},
       adsurl = {https://ui.adsabs.harvard.edu/abs/2016MNRAS.460.1214L}
}

@ARTICLE{2017MNRAS.465L..84L,
       author = {{Ludlow}, Aaron D. and {Angulo}, Ra{\'u}l E.},
        title = "{Einasto profiles and the dark matter power spectrum}",
      journal = {\mnras},
         year = 2017,
        month = feb,
       volume = {465},
       number = {1},
        pages = {L84-L88},
          doi = {10.1093/mnrasl/slw216},
archivePrefix = {arXiv},
       eprint = {1610.04620},
 primaryClass = {astro-ph.CO},
       adsurl = {https://ui.adsabs.harvard.edu/abs/2017MNRAS.465L..84L}
}

@ARTICLE{2008MNRAS.387..536G,
       author = {{Gao}, Liang and {Navarro}, Julio F. and {Cole}, Shaun and {Frenk}, Carlos S. and {White}, Simon D.~M. and {Springel}, Volker and {Jenkins}, Adrian and {Neto}, Angelo F.},
        title = "{The redshift dependence of the structure of massive {\ensuremath{\Lambda}} cold dark matter haloes}",
      journal = {\mnras},
         year = 2008,
        month = jun,
       volume = {387},
       number = {2},
        pages = {536-544},
          doi = {10.1111/j.1365-2966.2008.13277.x},
archivePrefix = {arXiv},
       eprint = {0711.0746},
 primaryClass = {astro-ph},
       adsurl = {https://ui.adsabs.harvard.edu/abs/2008MNRAS.387..536G}
}

@ARTICLE{2016AJ....151...94F,
       author = {{Frank}, B.~S. and {de Blok}, W.~J.~G. and {Walter}, F. and {Leroy}, A. and {Carignan}, C.},
        title = "{The Impact of Molecular Gas on Mass Models of Nearby Galaxies}",
      journal = {\aj},
         year = 2016,
        month = apr,
       volume = {151},
       number = {4},
          eid = {94},
        pages = {94},
          doi = {10.3847/0004-6256/151/4/94},
archivePrefix = {arXiv},
       eprint = {1512.01367},
 primaryClass = {astro-ph.GA},
       adsurl = {https://ui.adsabs.harvard.edu/abs/2016AJ....151...94F}
}

@ARTICLE{2019A&A...621A.120G,
       author = {{Garc{\'\i}a-Benito}, R. and {Gonz{\'a}lez Delgado}, R.~M. and {P{\'e}rez}, E. and {Cid Fernandes}, R. and {S{\'a}nchez}, S.~F. and {de Amorim}, A.~L.},
        title = "{Spatially resolved mass-to-light from the CALIFA survey. Mass-to-light ratio vs. color relations}",
      journal = {\aap},
         year = 2019,
        month = jan,
       volume = {621},
          eid = {A120},
        pages = {A120},
          doi = {10.1051/0004-6361/201833993},
archivePrefix = {arXiv},
       eprint = {1811.08431},
 primaryClass = {astro-ph.GA},
       adsurl = {https://ui.adsabs.harvard.edu/abs/2019A&A...621A.120G}
}

@ARTICLE{2015ApJS..219....4S,
       author = {{Salo}, Heikki and {Laurikainen}, Eija and {Laine}, Jarkko and {Comer{\'o}n}, Sebastien and {Gadotti}, Dimitri A. and {Buta}, Ron and {Sheth}, Kartik and {Zaritsky}, Dennis and {Ho}, Luis and {Knapen}, Johan and {Athanassoula}, E. and {Bosma}, Albert and {Laine}, Seppo and {Cisternas}, Mauricio and {Kim}, Taehyun and {Mu{\~n}oz-Mateos}, Juan Carlos and {Regan}, Michael and {Hinz}, Joannah L. and {Gil de Paz}, Armando and {Menendez-Delmestre}, Karin and {Mizusawa}, Trisha and {Erroz-Ferrer}, Santiago and {Meidt}, Sharon E. and {Querejeta}, Miguel},
        title = "{The Spitzer Survey of Stellar Structure in Galaxies (S$^{4}$G): Multi-component Decomposition Strategies and Data Release}",
      journal = {\apjs},
         year = 2015,
        month = jul,
       volume = {219},
       number = {1},
          eid = {4},
        pages = {4},
          doi = {10.1088/0067-0049/219/1/4},
archivePrefix = {arXiv},
       eprint = {1503.06550},
 primaryClass = {astro-ph.GA},
       adsurl = {https://ui.adsabs.harvard.edu/abs/2015ApJS..219....4S}
}

@article{Schombert_2022,
doi = {10.3847/1538-3881/ac5249},
url = {https://doi.org/10.3847/1538-3881/ac5249},
year = {2022},
month = {mar},
publisher = {The American Astronomical Society},
volume = {163},
number = {4},
pages = {154},
author = {Schombert, James and McGaugh, Stacy and Lelli, Federico},
title = {Stellar Mass-to-light Ratios: Composite Bulge+Disk Models and the Baryonic Tully–Fisher Relation},
journal = {The Astronomical Journal}
}

@ARTICLE{2019MNRAS.486.1995S,
       author = {{Staudaher}, Shawn M. and {Dale}, Daniel A. and {van Zee}, Liese},
        title = "{The Extended Disc Galaxy Exploration Science Survey: description and surface brightness profile properties}",
      journal = {\mnras},
         year = 2019,
        month = jun,
       volume = {486},
       number = {2},
        pages = {1995-2010},
          doi = {10.1093/mnras/stz935},
archivePrefix = {arXiv},
       eprint = {1904.00050},
 primaryClass = {astro-ph.GA},
       adsurl = {https://ui.adsabs.harvard.edu/abs/2019MNRAS.486.1995S}
}




\appendix

\section{Scaled Dark Matter Density Plot with Fiducial Value}
\label{A1}

\begin{figure*}  
 \centering
    \begin{subfigure}{0.95\textwidth}
        
        \includegraphics[width=\linewidth]{Results_img/galaxy_colorbar_larger_font_and_title_gap.png}
    \end{subfigure}

    \centering
    \begin{subfigure}{0.49\textwidth}
        \centering
        \includegraphics[width=\textwidth]{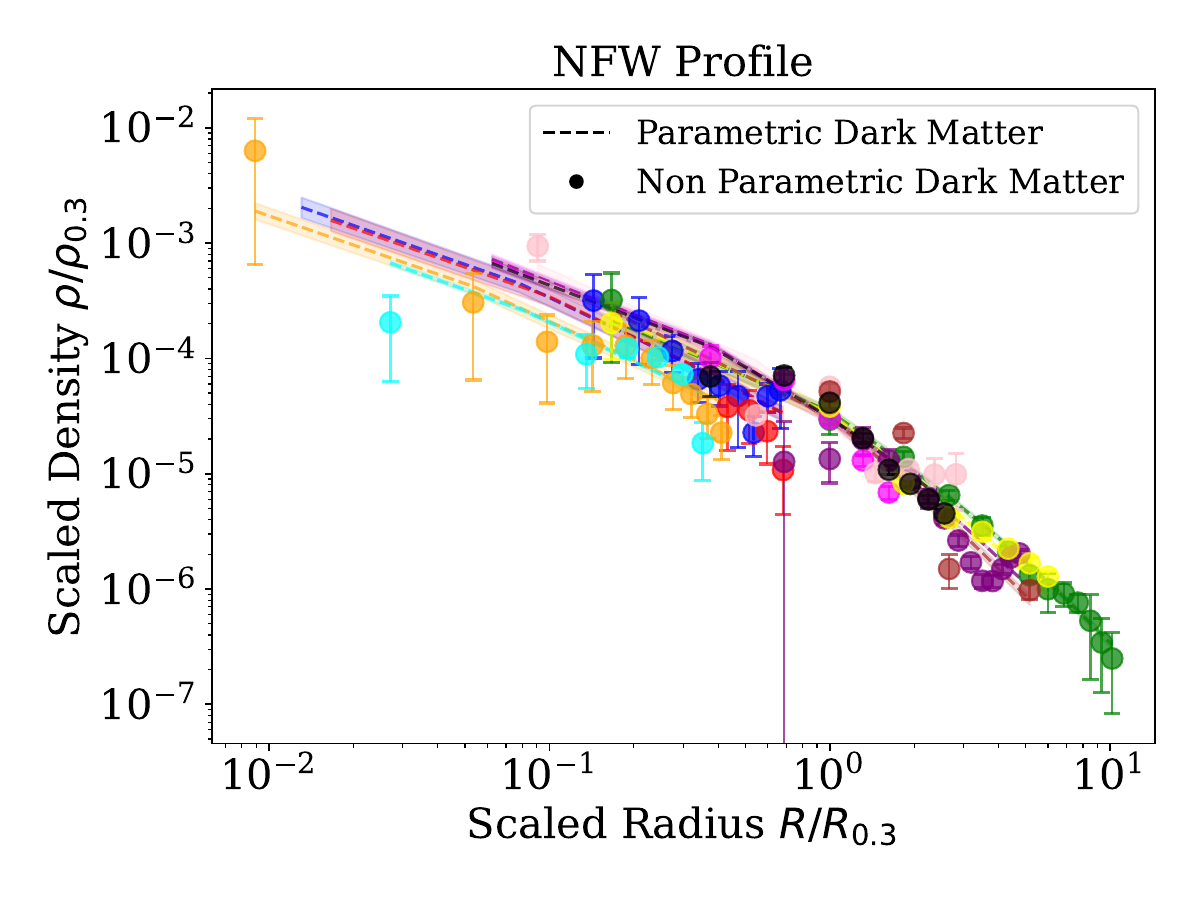}
    \end{subfigure}
    \begin{subfigure}{0.49\textwidth}
        \centering
        \includegraphics[width=\textwidth]{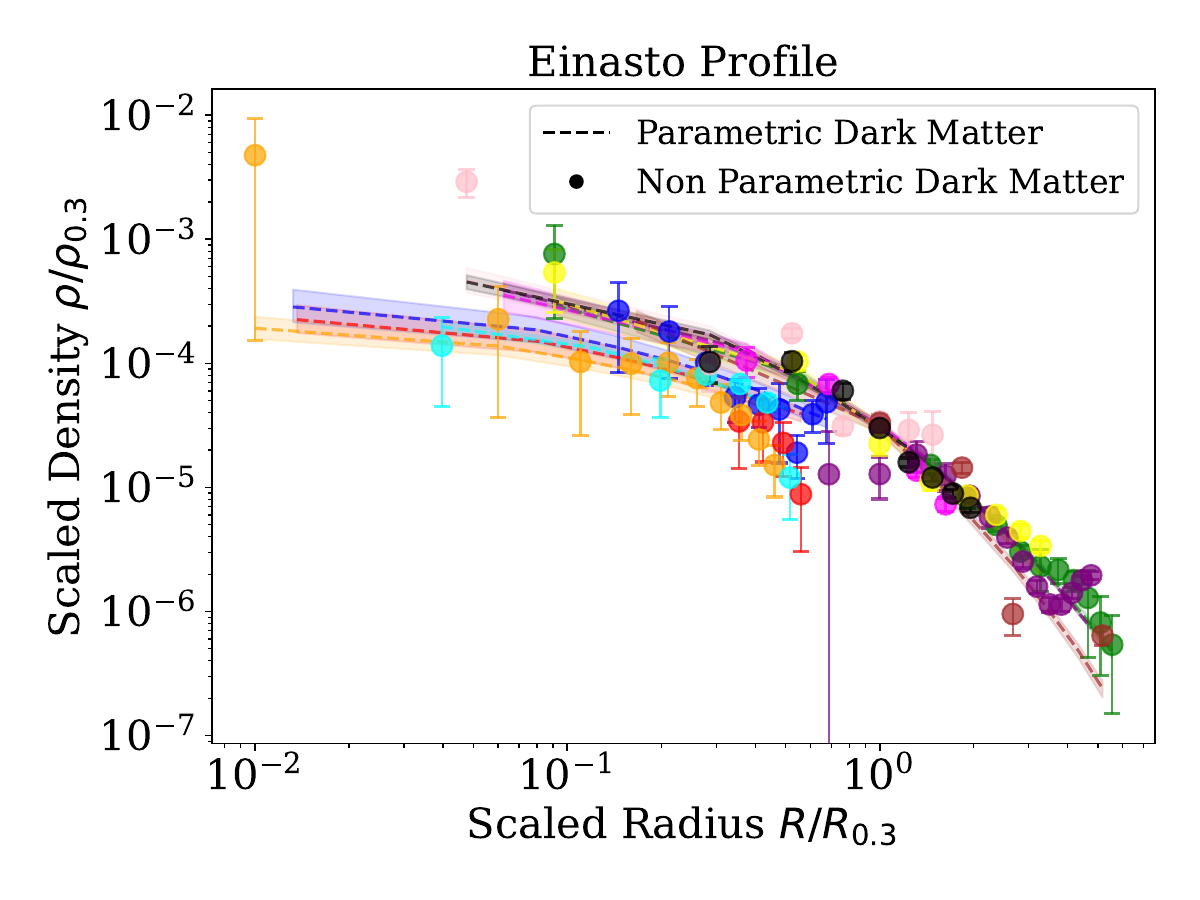}
    \end{subfigure}

    \vspace{0.05cm} 

    \begin{subfigure}{0.49\textwidth}
        \centering
        \includegraphics[width=\textwidth]{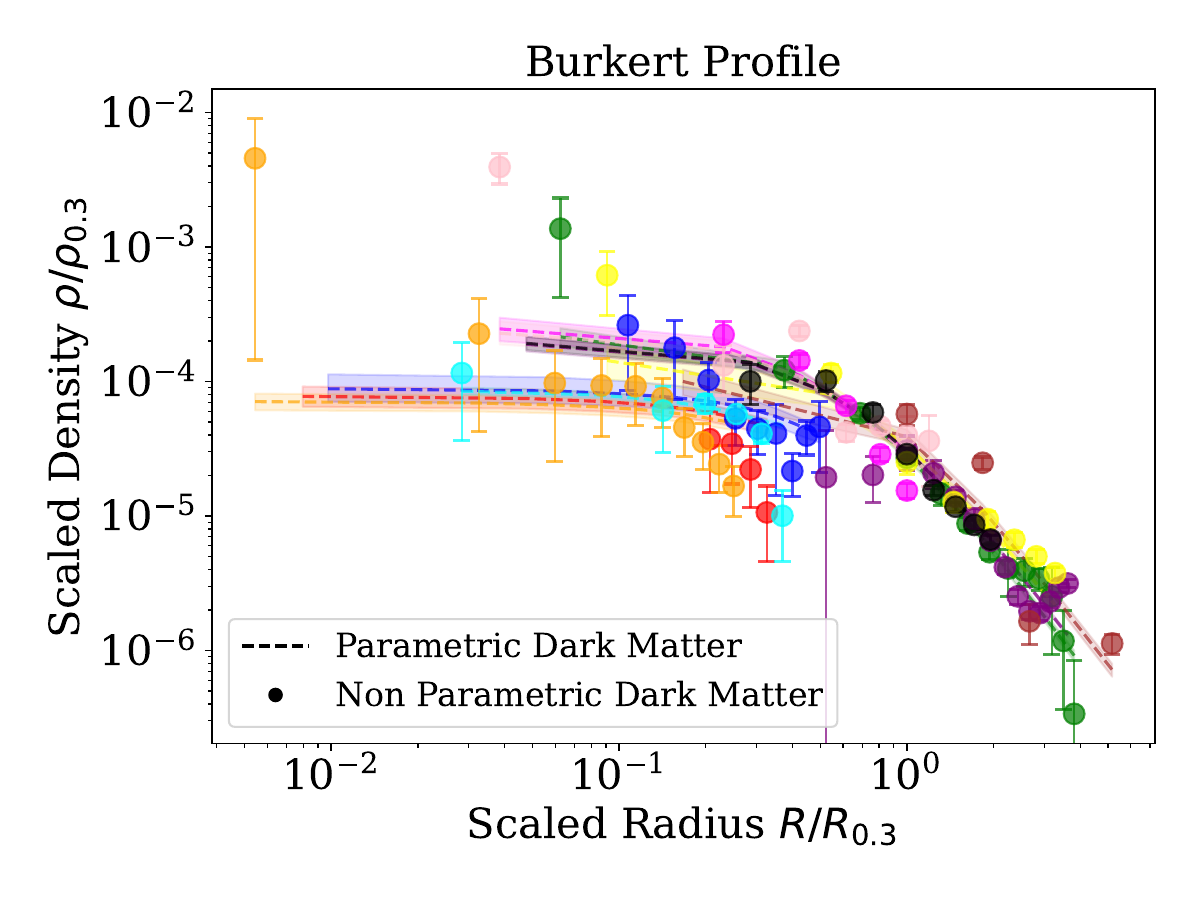}
    \end{subfigure}
    \begin{subfigure}{0.49\textwidth}
        \centering
        \includegraphics[width=\textwidth]{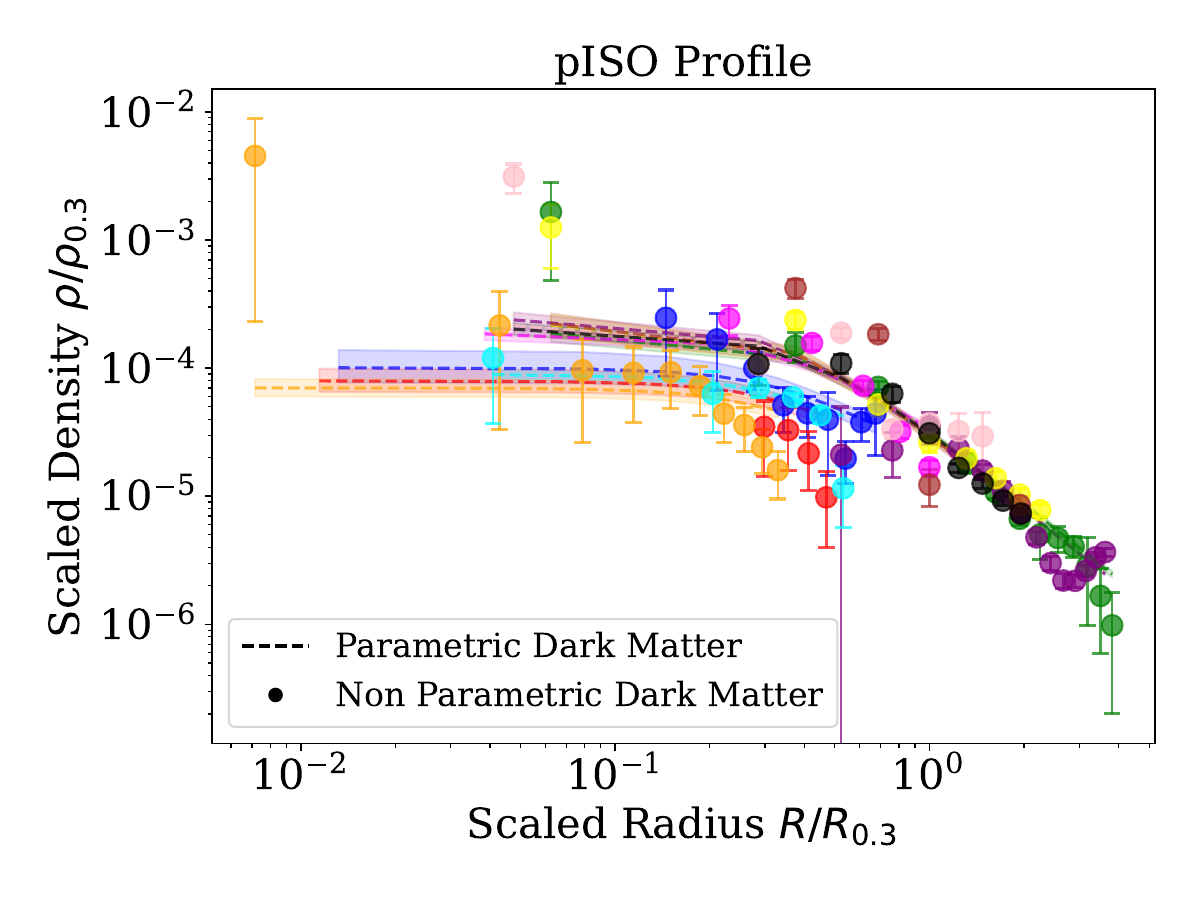}
    \end{subfigure}
    \caption{Scaled dark matter density profiles: parametric (dashed lines; see Section~\ref{subsubsec: parametric dm}) and non-parametric (circles; see Section~\ref{subsubsec:non para dm}). Each panel corresponds to a different dark matter halo model, with different colours representing different galaxies.}
    \label{fig:scaled_den_app}
\end{figure*} 

We repeated the non-parametric analysis using the stellar velocity contribution, $V_{star}$, using a fiducial stellar mass-to-light ratio at 3.6\,$\mu$m of $0.6$ with an associated uncertainty of 0.1\,dex, following \citet{2014ApJ...788..144M}. This constant value provides a robust estimate for old stellar populations with minimal bias from dust emission or young stellar populations. For most galaxies, we applied this value directly to the Spitzer IRAC 3.6\,$\mu$m images to convert surface brightness to stellar mass surface density, followed by Multi-Gaussian Expansion (MGE) fitting and Jeans Anisotropic Modelling (JAM) to obtain $V_{star}$.

Two galaxies in our sample NGC7610 and NGC7292 do not have 3.6\,$\mu$m data. For these, we derived the stellar mass-to-light ratio using the color-based prescription from \citet{2003ApJS..149..289B}:
\begin{equation}
\log_{10}(M/L)_r = a_r + b_r\,(g - r),
\end{equation}
where $a_r$ and $b_r$ are the coefficients for r band taken from Table A7 in \citet{2003ApJS..149..289B}, and $(g - r)$ colors were taken from SDSS~DR18 photometry \citep{2023ApJS..267...44A}.The resulting $M/L$ values were then applied to construct the stellar mass distribution and compute $V_{star}$ using the same MGE+JAM procedure as for the rest of the sample. Then the stellar velocity component, together with the previously derived gas velocity component \citep{biswas2023}, was combined to form the baryonic contribution. This baryonic contribution was then subtracted from the observed rotation curve to obtain the dark matter velocity component. Using this dark matter component, we constructed the scaled density distribution following equation~\ref{eq:scl_den} (see Section~\ref{subsubsec:Scaled Density}).

Figure~\ref{fig:scaled_den_app} presents the scaled dark matter density profiles for four different halo models. In each panel, the dashed lines represent the parametric dark matter densities from MCMC-based mass modelling (see Section~\ref{subsubsec: parametric dm}), while the circles show the non-parametric dark matter densities derived directly from the data (see Section~\ref{subsubsec:non para dm}). The overall trends remain consistent with our earlier results: the outer regions exhibit good agreement between the parametric and non-parametric profiles, whereas cored profiles continue to show systematic deviations in the inner regions.

\section{Impact of bulge--disc $M/L$ variations}
\label{B}
    \begin{table}
    \centering
    \caption{Disk and bulge light (or mass) fractions \citep{2015ApJS..219....4S}, NGC4861 from \citet{2019MNRAS.486.1995S}}
    \label{tab:disk_bulge_fractions}
    \begin{tabular}{lcc}
    \hline
    Name & Disk & Bulge \\
    \hline
    NGC0784 & 0.998 & 0.002 \\
    NGC1156 & -- & -- \\
    NGC3027 & 0.887 & 0.113 \\
    NGC3359 & 0.816 & 0.184 \\
    NGC4068 & 0.637 & 0.363 \\
    NGC4861 & 1.00  & --   \\
    NGC7292 & --    & --   \\
    NGC7497 & 0.909 & 0.091 \\
    NGC7610 & --    & --   \\
    NGC7741 & 0.849 & 0.151 \\
    NGC7800 & 0.790 & 0.210 \\
    \hline
    \end{tabular}
    \end{table}

In the main analysis, we adopt a single stellar mass-to-light ratio, $M/L$, for the full stellar component of each galaxy. However, as discussed in Sec.~\ref{subsubsec: parametric dm}, $M/L$ may differ between structural components. Therefore, We have done mass modelling for one of the galaxies (NGC4068, the galaxy with the highest bulge fraction in our sample) with different $M/L$ for the bulge and disc.

To separate the stellar contribution into bulge and disc components at the level of the circular velocity calculation, we first decomposed the observed surface brightness distribution into two structural components using the \texttt{bulge\_disk=True} in MGE analysis. Each component is described by its own set of Gaussian amplitudes ($\Sigma_j$), dispersions ($\sigma_j$), and observed axial ratios ($q_{\mathrm{obs},j}$). We have fittted the MGE model surface brightness profile with a Sérsic (bulge) plus an exponential (disc) model and used the intersection radius of the two profiles to define a transition between the components. The stellar circular velocity contributions of the bulge and disc were computed independently by scaling the surface brightness amplitudes of the Gaussians assigned to each component by their respective mass-to-light ratios, $(M/L)_{\rm bulge}$ and $(M/L)_{\rm disc}$. The total stellar circular velocity was then constructed by adding the bulge and disc contributions in quadrature,

\begin{equation}
V_{\star}(R) = \sqrt{V_{\rm bulge}^2(R) + V_{\rm disc}^2(R)}.
\end{equation}

Finally, the full model rotation curve was obtained by combining the stellar, gaseous, and dark-matter halo components in quadrature,
\begin{equation}
V_{\rm dyn}(R) = \sqrt{V_{\rm gas}^2(R) + V_{\star}^2(R) + V_{\rm halo}^2(R)}.
\end{equation}

Table~\ref{tab:disk_bulge_fractions} represents the bulge-to-total ($B/T$) light ratio for our galaxies, $B/T$ is not available in literature for three galaxies (NGC1156, NGC7292, NGC7610). We considered NGC4068, which has the highest bulge fraction in our sample (B/T $=0.363$, Table.~\ref{tab:disk_bulge_fractions}). For this galaxy we presented here the mass modelling in two configurations: (i) a single stellar $M/L$ applied to all MGE Gaussians (our fiducial setup), and (ii) different $M/L$ to MGE Gaussians coming from bulge and disc, $M/L_{\mathrm{bulge}}$ and $M/L_{\mathrm{disc}}$ parameters, both fitted with MCMC. 

\begin{figure*}
    \centering

    \begin{subfigure}[b]{0.45\textwidth}
        \centering
        \includegraphics[width=\textwidth]{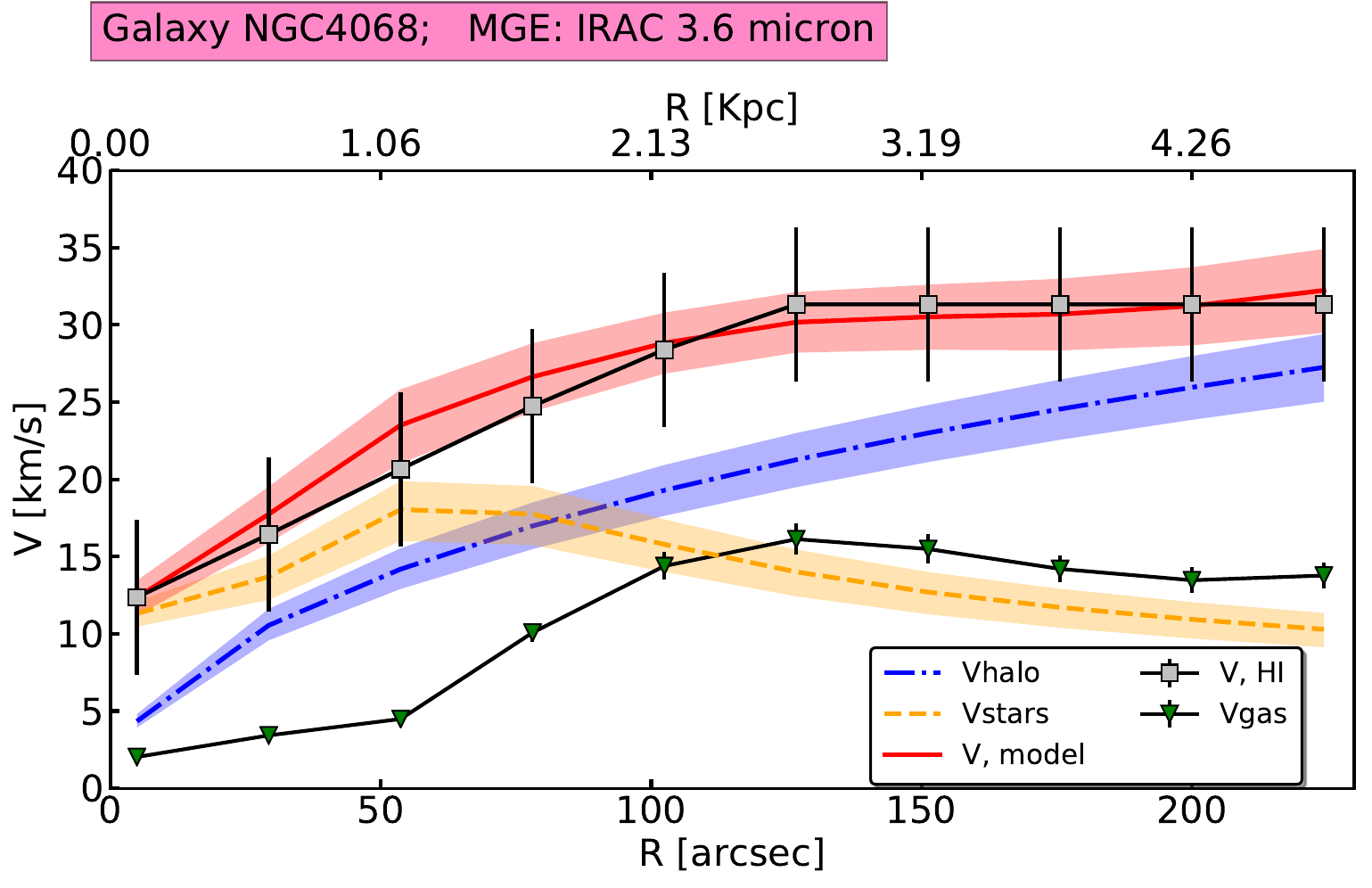}
        \subcaption*{\textbf{NGC4068: Single $M/L$}}
    \end{subfigure}
    \begin{subfigure}[b]{0.45\textwidth}
        \centering
        \includegraphics[width=\textwidth]{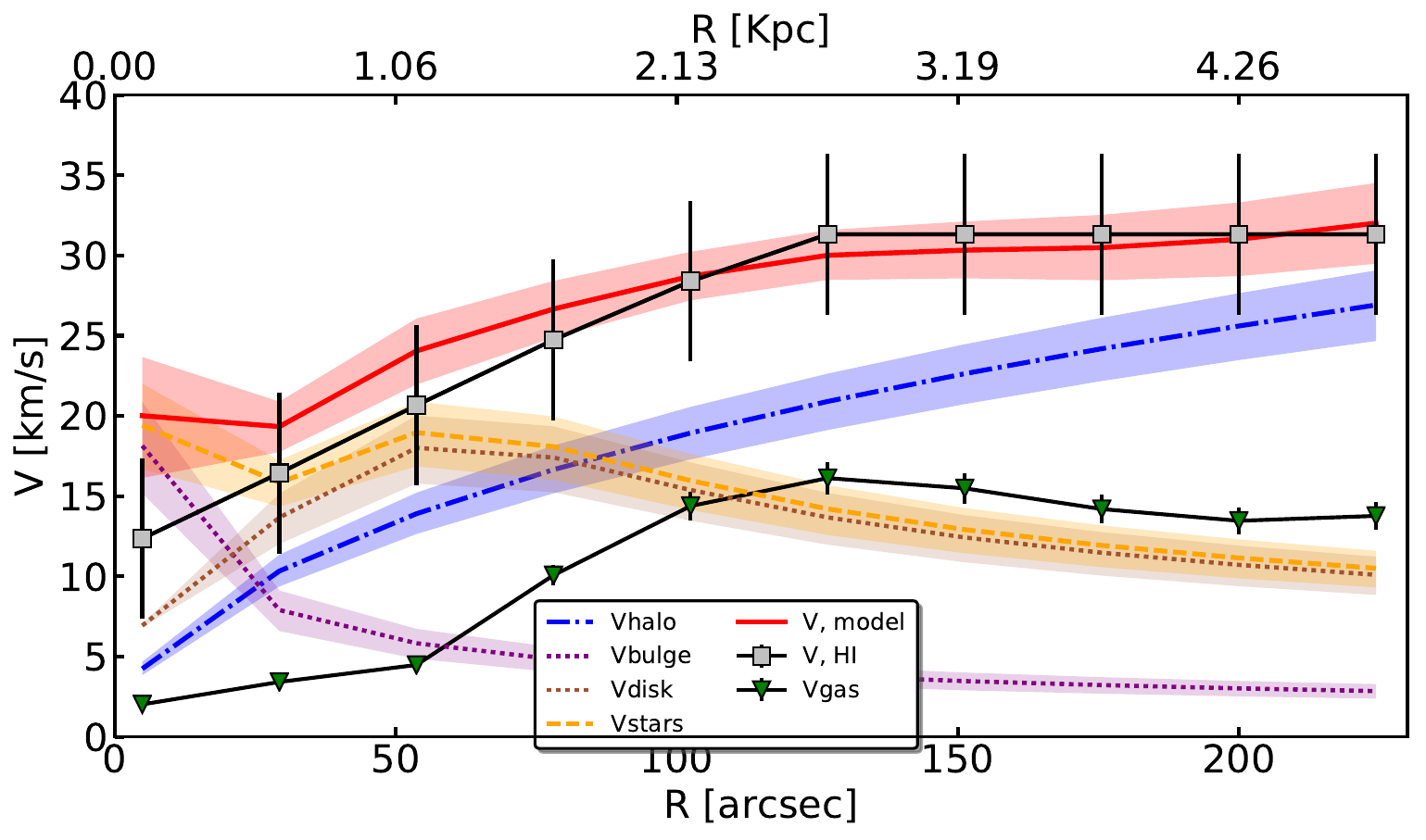}
        \subcaption*{\textbf{NGC4068: Different $M/L$}}
    \end{subfigure}

    \vspace{0.7em}

    \begin{subfigure}[b]{0.3\textwidth}
        \centering
        \includegraphics[width=\textwidth]{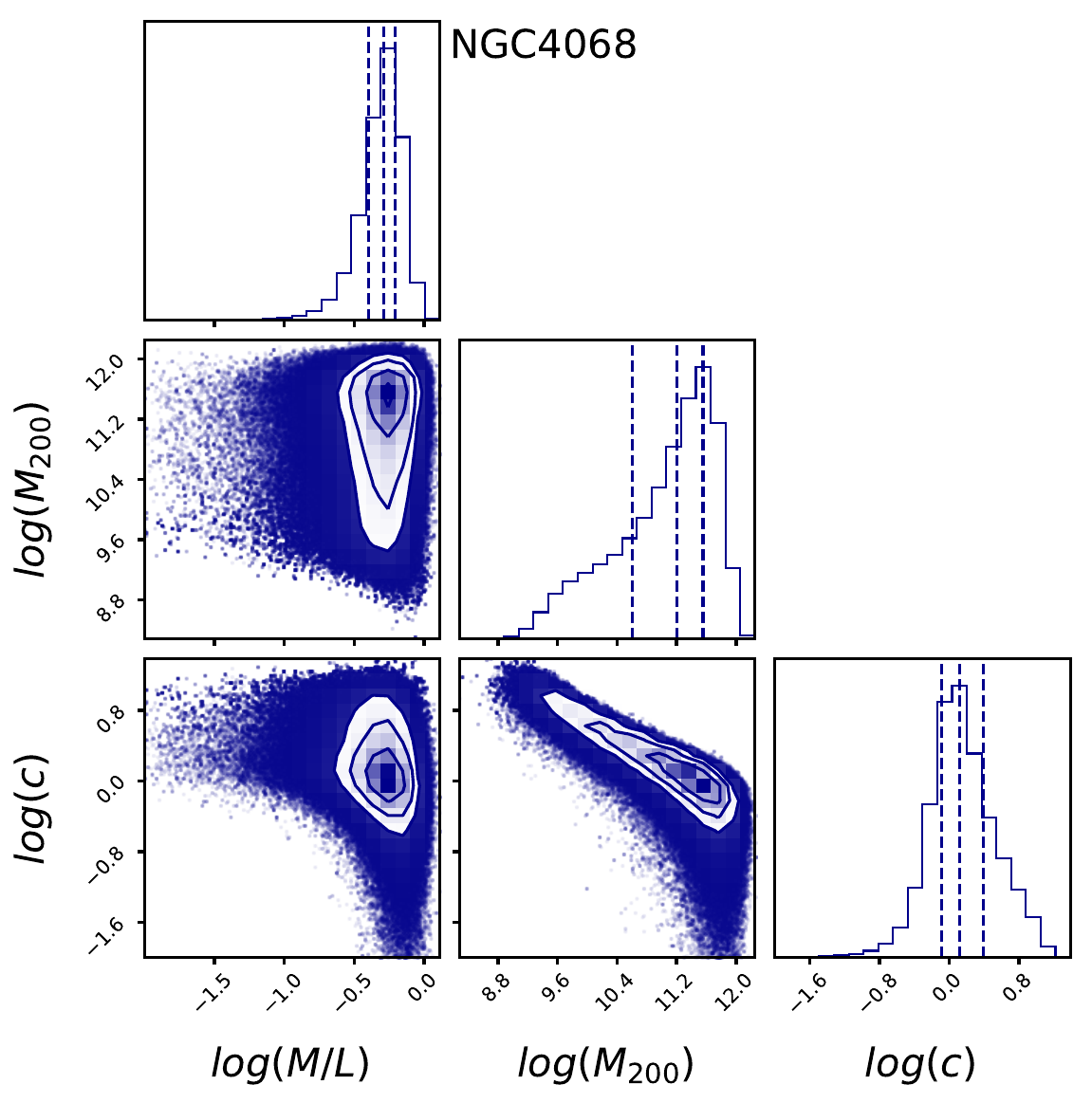}
    \end{subfigure}
    \hspace*{0.10\textwidth}  
    \begin{subfigure}[b]{0.3\textwidth}
        \centering
        \includegraphics[width=\textwidth]{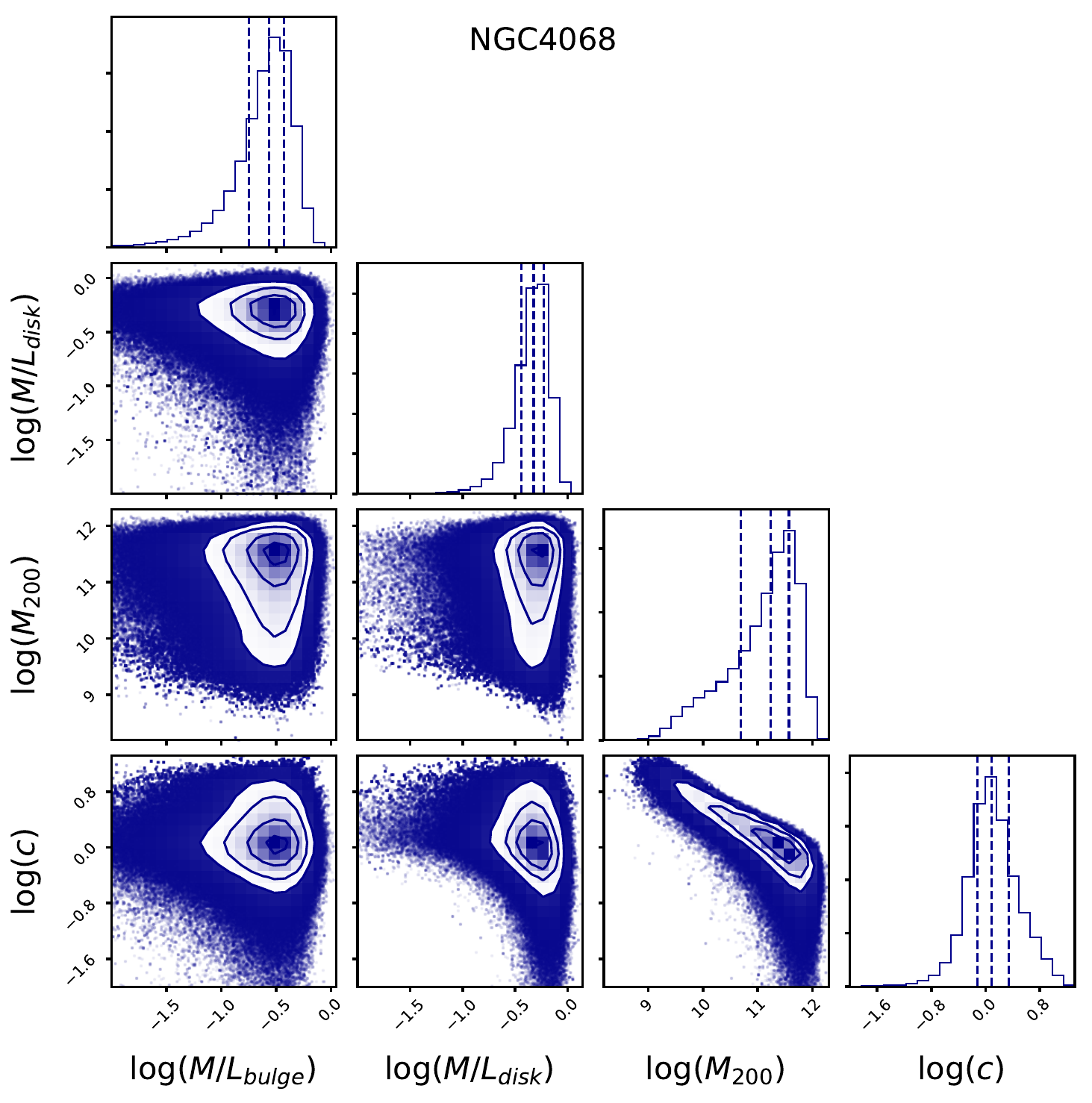}
    \end{subfigure}

    \vspace{1.0em}

    \caption{Comparison of single-$M/L$ (left) and bulge+disc $M/L$ (right) mass models for NGC4068; the upper panels show the modelled rotation curves and data, while the lower panels show the posterior distributions of the fitted parameters.}
    \label{fig:app_mltest}
\end{figure*}

The resulting best-fitting parameters are summarised in Table~\ref{tab:ml_merged_4row}, and the corresponding rotation curves and posterior distributions are shown in Fig.~\ref{fig:app_mltest}. Allowing different bulge and disc $M/L$ values produces only small changes in the derived rotation curves and halo parameters: the best-fitting $M_{200}$, $M/L$ and concentration, and total stellar mass (see Table~\ref{tab:ml_merged_4row}) all remain consistent within the error bars.

These tests indicate that, even for the galaxy with the largest B/T in our sample, adopting a single global stellar $M/L$ does not significantly bias our main conclusions on halo masses and concentrations. Given the relatively small bulge fractions of most galaxies in the pilot sample, the single-$M/L$ approach provides an adequate and internally consistent description of the stellar contribution to the rotation curves. The MGE method is sufficiently flexible to reproduce the surface brightness distributions of realistic, multi component galaxies; however, enabling \texttt{bulge\_disk=True} restricts the photometry to a two-component representation, which can reduce the generality of the MGE description for complex morphologies. We will compare different modelling approaches for the full GARCIA-II sample, including MGE with a single $M/L$, bulge--disc decompositions with distinct $M/L$ values, and with a radially varying $(M/L)$ in forthcoming paper.

\begin{table*}
\centering
\caption{Best-fit parameters for the single-$M/L$ and two-component (disk+bulge) $M/L$ mass models, and the inferred stellar mass.}
\label{tab:ml_merged_4row}
\setlength{\tabcolsep}{5pt}
\renewcommand{\arraystretch}{2}
\scriptsize
\begin{tabular}{llcccc}
\toprule
Name & Model & $M/L$ & $M_{200}\,[M_\odot]$ & $M_\star\,[M_\odot]$ & $C$ \\
\midrule
\multirow{2}{*}{NGC4068}
& Single $M/L$
& $0.51 \pm 0.16$
& $1.6^{+3.1}_{-1.4}\times10^{11}$
& $(1.2 \pm 0.4)\times10^{8}$
& $1.3^{+2.1}_{-0.7}$ \\
& Different $M/L$
& $(M/L)_{\rm d}=0.47^{+0.17}_{-0.16}$; $(M/L)_{\rm b}=0.27^{+0.15}_{-0.13}$
& $1.73^{+3.0}_{-1.5}\times10^{11}$
& $(1.17 \pm 0.39)\times10^{8}$
& $1.23^{+1.79}_{-0.64}$ \\
\bottomrule
\end{tabular}
\end{table*}


\bsp	
\label{lastpage}
\end{document}